\documentclass{article}
\usepackage{bookstyle}
\usepackage{graphicx}
\usepackage{cite}
\def\gsim{\mbox{\raisebox{-.6ex}{~$\stackrel{>}{\sim}$~}}}
\def\lsim{\mbox{\raisebox{-.6ex}{~$\stackrel{<}{\sim}$~}}}
\def\dsl{\hbox{\phantom{.}/\kern-.4600em$\partial$}}
\def\qsl{\hbox{\phantom{.}/\kern-.4600em$q$}}
\def\psl{\hbox{\phantom{.}/\kern-.4600em$p$}}
\def\beq{\begin{equation}}
\def\eeq{\end{equation}}
\def\beqa{\begin{eqnarray}}
\def\eeqa{\end{eqnarray}}
\begin{document}
\author{James M.\ Cline}
\address{McGill University, Dept.\ of Physics, Montreal, Qc H3A 2T8,
Canada}
\title{Baryogenesis}
%
\frontmatter
\maketitle    
\mainmatter

\section{Observational evidence for the BAU}

In everyday life it is obvious that there is more matter than
antimatter.  In nature we see antimatter mainly in cosmic rays,
{\it e.g.,} $\bar p$, whose flux expressed as a ratio to that of
protons, is
\begin{equation}
	{\bar p\over p} \sim 10^{-4}
\end{equation}
consistent with no ambient $\bar p$'s, only  $\bar p$'s produced
through high energy collisions with ordinary matter.  On earth, we
need to work very hard to produce and keep $e^+$ (as in the former
LEP experiment) or $\bar p$ (as at the Tevatron).  

We can characterize the asymmetry between matter and antimatter in 
terms of the baryon-to-photon ratio
\begin{equation}
	\eta \equiv {n_B - n_{\bar B}\over n_\gamma}
\end{equation}
where
\begin{eqnarray}
	n_B &=& \hbox{number density of baryons}\nonumber\\
	n_{\bar B} &=& \hbox{number density of antibaryons}\nonumber\\
	n_{\gamma} &=& \hbox{number density of photons}\nonumber\\
		&\equiv& {\zeta(3)\over \pi^2} g_* T^3,
	\qquad g_* = \hbox{2 spin polarizations}
\end{eqnarray}
and $\zeta(3) = 1.20206\dots$  This is a useful measure because it
remains constant with the expansion of the universe, at least at
late times.  But at early times, and high temperatures, 
many heavy particles were in thermal equilibrium, which later
annihilated to produce more photons but not baryons.  In this case,
the entropy density $s$ is a better quantity to compare the baryon
density to, and it is convenient to consider
\begin{equation}
	{n_B - n_{\bar B}\over s} = {1\over 7.04}\, \eta
\label{eq1.4}
\end{equation}
where the conversion factor $1/7$ is valid in the present universe,
since the epoch when neutrinos went out of equilibrium and positrons
annihilated.

Historically, $\eta$ was determined using big bang nucleosynthesis.
Abundances of $\,^3$He, $\,^4$He, D,  $\,^6$Li and $\,^7$Li are
sensitive to the value of $\eta$; see the Particle Data Group
\cite{PDG} for a review.  The theoretical predictions and experimental
measurements are summarized in figure 1, similar to a figure
in \cite{Cyburt}.  The boxes represent the regions which are
consistent with experimental determinations, showing that different
measurements obtain somewhat different values.  The smallest error
bars are for the deuterium (D/H) abundance, giving
\begin{equation}
	\eta = 10^{-10}\times\left\{\begin{array}{l}
	6.28\pm0.35\cr 5.92\pm 0.56\end{array}\right.
\end{equation}
for the two experimental determinations which are used.  These values
are consistent with the $\,^4$He abundance, though marginally
inconsistent with $\,^7$Li (the ``Lithium problem'').

In the last few years, the Cosmic Microwave Background has given
us an independent way of measuring the baryon asymmetry \cite{Hu}.
The relative sizes of the Doppler peaks of the temperature anisotropy
are sensitive to $\eta$, as illustrated in fig.\ 2. The Wilkinson
Microwave Anisotropy Probe first-year data fixed the combination
$\Omega_b h^2 = 0.0224\pm 0.0009$, corresponding to \cite{Cyburt}
\begin{equation}
	\eta = (6.14\pm0.25)\times 10^{-10} 
\end{equation}
which is more accurate than the BBN determination.  Apart from 
the Lithium problem, this value is seen to be in good agreement with the
BBN value.  The CMB-allowed range is shown as the vertical band in
figure 1.

We can thus be confident in our knowledge of the baryon asymmetry.
The question is, why is it not zero?  This would be a quite natural
value; a priori, one might expect the big bang---or in our more modern
understanding, reheating following inflation---to produce equal numbers
of particles and antiparticles.

\centerline{
\includegraphics[width=0.9\textwidth]{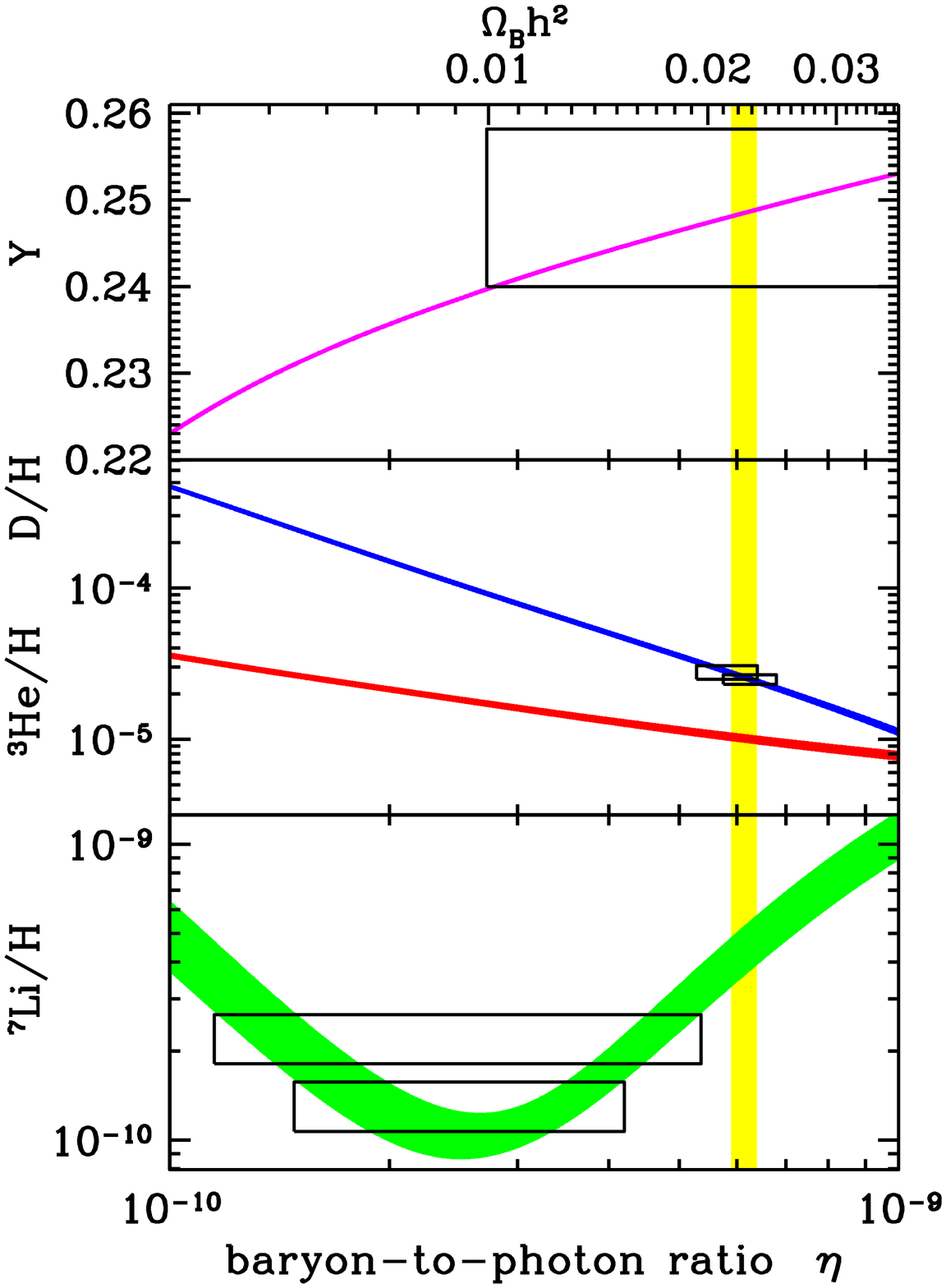}}
\centerline{Fig.\ 1. Primordial abundances versus $\eta$, courtesy of
R.\ Cyburt}
\setcounter{figure}{1}

\begin{figure}[h]
\centerline{
\includegraphics[width=0.9\textwidth]{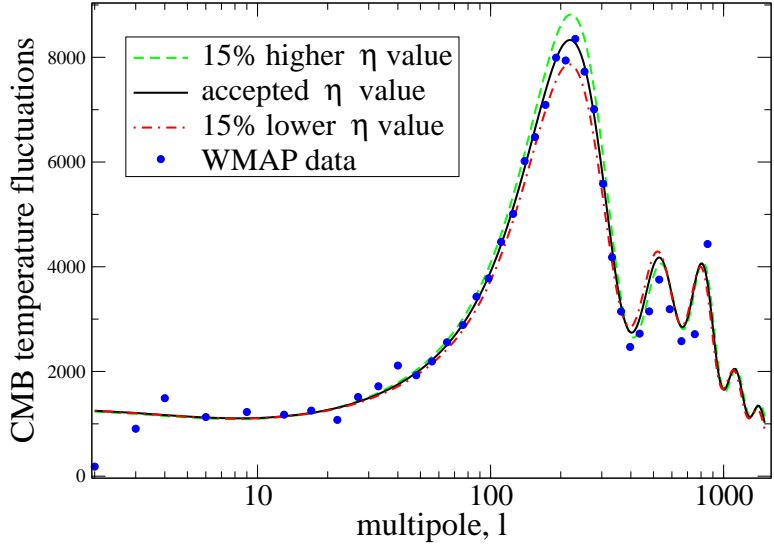}}
\caption{Dependence of the CMB Doppler peaks on $\eta$.}
\label{fig2}
\end{figure}

One might wonder whether the universe could be baryon-symmetric on
very large scales, and separated into regions which are either
dominated by baryons or antibaryons.  However we know that even in
the least dense regions of the universe there are hydrogen gas
clouds, so one would expect to see an excess of gamma rays in the
regions between baryon and antibaryon dominated regions, due to
annihilations.  These are not seen, indicating that such patches
should be as large as the presently observable universe.  There seems
to be no plausible way of separating baryons and antibaryons from
each other on such large scales.

It is interesting to note that in a homogeneous, baryon-symmetric
universe, there would still be a few baryons and antibaryons left
since annihilations aren't perfectly efficient.  But the freeze-out
abundance is
\begin{equation}
	{n_B\over n_\gamma} = {n_{\bar B}\over n_\gamma}
	\approx 10^{-20}
\end{equation}
(see ref.\ \cite{KT}, p.\ 159), which is far too small for the BBN
or CMB.

In the early days of big bang cosmology, the baryon asymmetry was 
considered to be an initial condition, but in the context of inflation
this idea is no longer tenable.  Any baryon asymmetry existing before
inflation would be diluted to a negligible value during inflation, due
to the production of entropy during reheating.

It is impressive that A.\ Sakharov realized the need for dynamically
creating the baryon asymmetry in 1967 \cite{Sak}, more than a decade before inflation
was invented.  The idea was not initially taken seriously; in fact it
was not referenced again, with respect to the idea of baryogenesis,
until 1979 \cite{Barr}.  Now it has 1040 citations (encouragement to
those of us who are still waiting for our most interesting papers to
be noticed!).  It was only with the advent of grand unified theories,
which contained the necessary ingredients for baryogenesis, that
interest in the subject started to grow dramatically.  I have
documented the rate of activity in the field from 1967 until now in
figure 3.

\begin{figure}[t]
\centerline{
\includegraphics[width=0.9\textwidth]{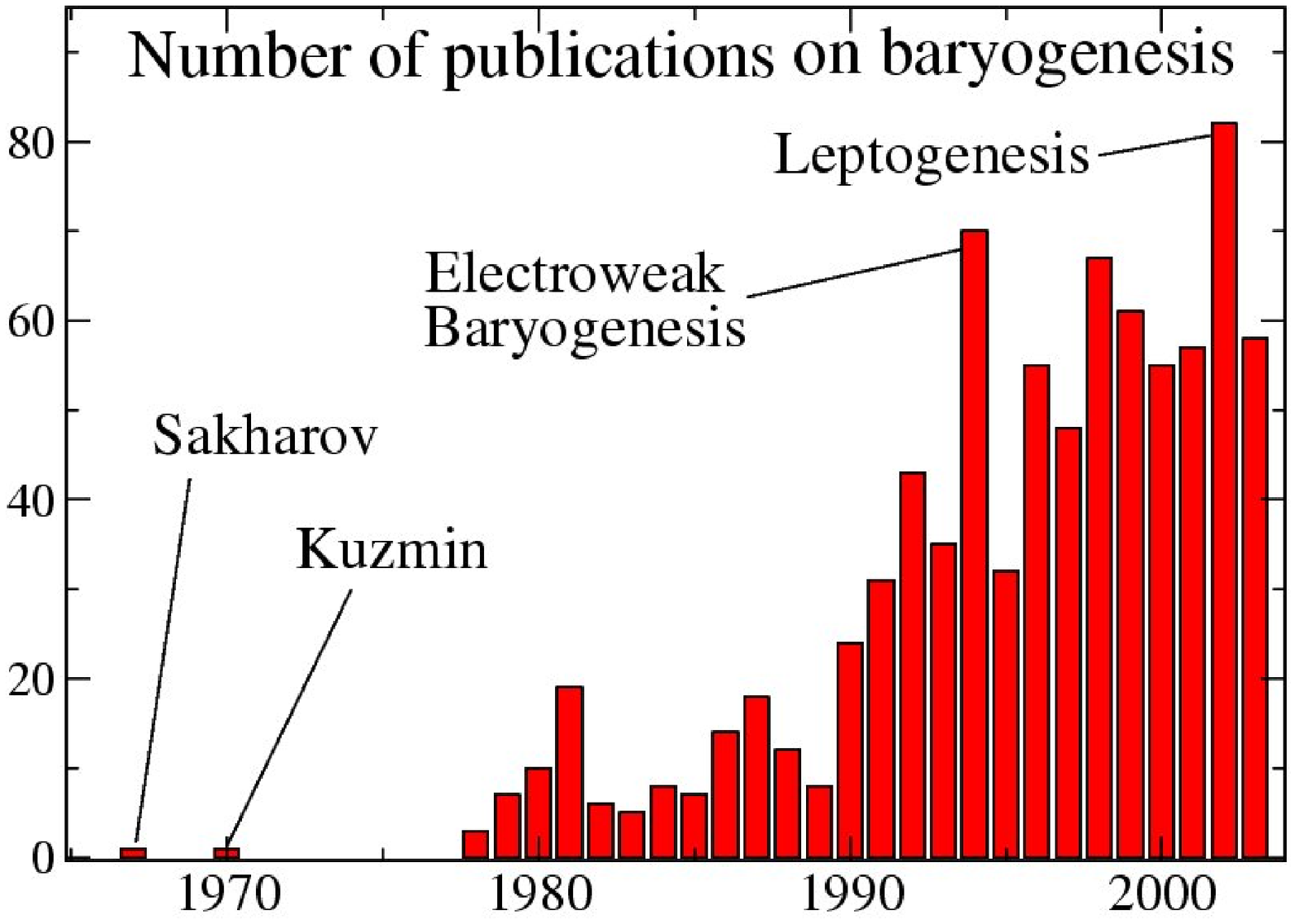}}
\caption{Number of publications on baryogenesis as a function of time.}
\label{fig3}
\end{figure}

\section{Sakharov's Conditions for Baryogenesis}

It is traditional to start any discussion of baryogenesis with the
list of three necessary ingredients needed to create a baryon
asymmetry where none previously existed:

\begin{enumerate}
\item B violation
\item Loss of thermal equilibrium
\item C, CP violation
\end{enumerate}

Although these principles have come to be attributed to Sakharov, he
did not enunciate them as clearly in his three-page paper as one
might have been led to think, especially the second point.  (Sakharov
describes a scenario where a universe which was initially contracting
and with equal and opposite baryon asymmetry to that existing today
goes through a bounce at the singularity and  reverses the magnitude
of its baryon asymmetry.)  It is easy to see why these conditions
are necessary.  The need for B (baryon) violation is obvious.  Let's
consider some examples of B violation.

\subsection{B violation}

In the standard model, B is violated by the triangle anomaly, which
spoils conservation of the left-handed baryon $+$ lepton current,
\begin{equation}
	\partial_\mu J^\mu_{B_L+L_L} = 
	{3 g^2\over 32\pi^2}\epsilon_{\alpha\beta\gamma\delta}
	W_a^{\alpha\beta} W_a^{\gamma\delta}
\label{eq2.1}
\end{equation}
where $W_a^{\alpha\beta}$ is the SU(2) field strength.  As we will
discuss in more detail in section 4, this leads to the nonperturbative
sphaleron process pictured in fig. 4.  It involves 9 left-handed
(SU(2) doublet) quarks, 3 from each generation, and 3 left-handed
leptons, one from each generation.  It violates B and L by 3 units
each,
\begin{equation}
	\Delta B = \Delta L = \pm 3
\end{equation}

\begin{figure}[h]
\centerline{
\includegraphics[width=0.9\textwidth]{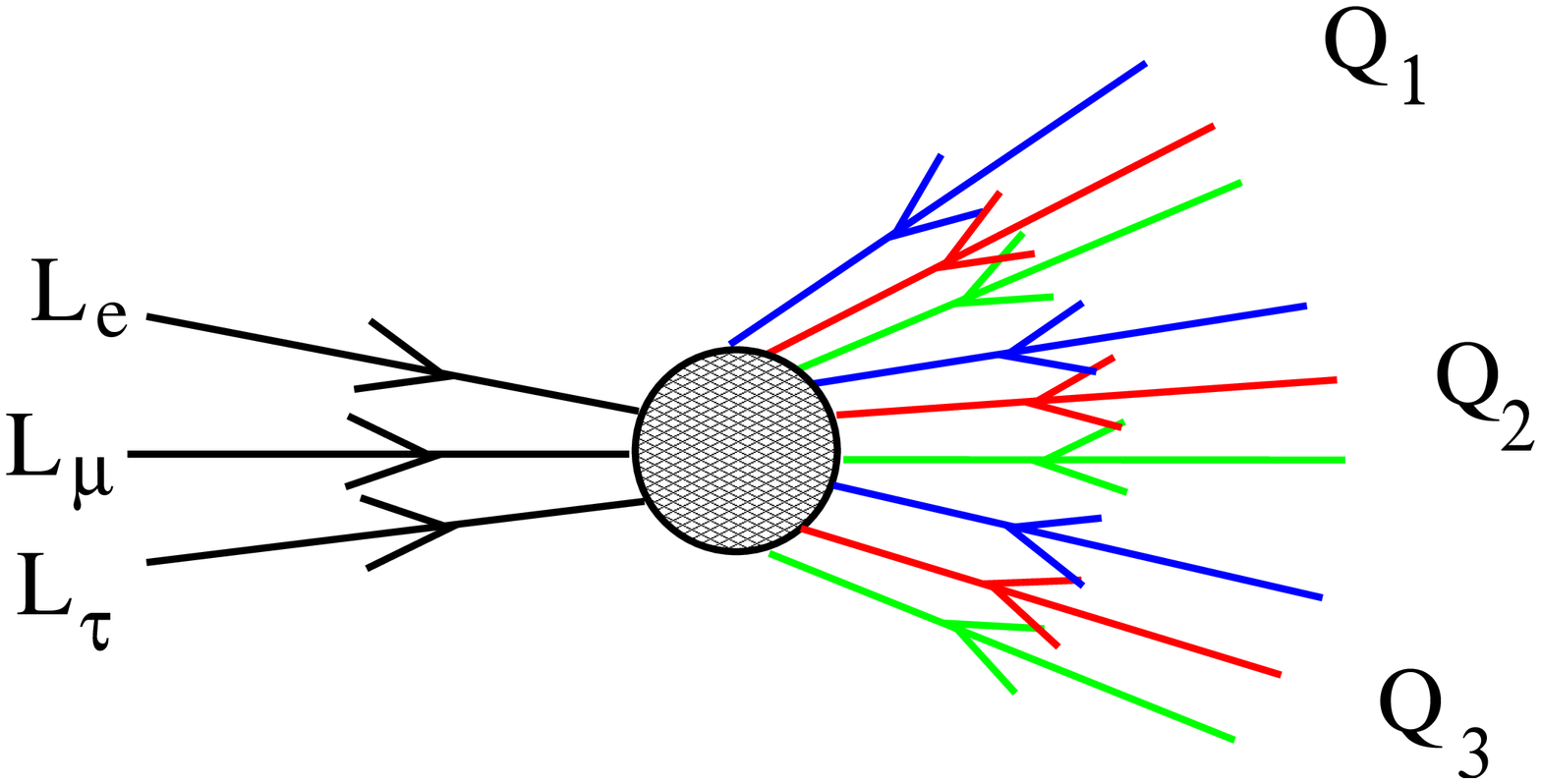}}
\caption{The sphaleron.}
\label{fig4}
\end{figure}

In grand unified theories, like SU(5), there are heavy gauge bosons
$X^\mu$ and heavy Higgs bosons $Y$ with couplings to quarks and
leptons of the form
\begin{equation}
	X q q, \quad X  \bar q \bar l
\end{equation}
and similarly for $Y$. The simultaneous existence of these two 
interactions imply that there is no consistent assignment of baryon
number to $X^\mu$.  Hence B is violated.

In supersymmetric models, one might choose not to impose R-parity.
Then one can introduce dimension-4 B-violating operators, like
\begin{equation}
	\tilde t^a_R b^b_R s^c_R \epsilon_{abc}
\label{rp}
\end{equation}
which couple the right-handed top squark to right-handed quarks.
One usually imposes R-parity to forbid proton decay, but the
interaction (\ref{rp}) could be tolerated if for some reason lepton
number was still conserved, since then the decay channels $p\to
\pi^+ \nu$ and $p\to \pi^0 e^+$ would be blocked.

\subsection{Loss of thermal equilibrium}

The second condition of Sakharov (as I am numbering them), loss of 
thermal equilibrium, is also easy to understand.  Consider a
hypothetical process
\begin{equation}
	X \to Y + B
\label{proc}
\end{equation}
where $X$ represents some initial state with vanishing baryon number,
$Y$ is a final state, also with $B=0$, and $B$ represents the excess
baryons produced.  If this process is in thermal equilibrium, then by
definition the rate for the inverse process, $Y+B\to X$, is equal
to the rate for (\ref{proc}):
\begin{equation}
	\Gamma(Y+B\to X) =\Gamma(X \to Y + B)
\end{equation}
No net baryon asymmetry can be produced since the inverse process
destroys B as fast as (\ref{proc}) creates it.

The classic example is out-of-equilibrium decays, where $X$ is a heavy
particle, such that $M_X > T$ at the time of decay, $\tau = 1/\Gamma$.
In this case, the energy of the final state $Y+B$ is of order $T$,
and there is no phase space for the inverse decay: $Y+B$ does not have
enough energy to create a heavy $X$ boson.  The rate for $Y+B\to X$
is Boltzmann-suppressed, $\Gamma(Y+B\to X)\sim e^{-M_X/T}$.

\subsection{C, CP violation}

The most subtle of Sakharov's requirements is C and CP violation.
Consider again some process $X\to Y+B$, and suppose that C (charge
conjugation) is a symmetry.  Then the rate for the C-conjugate
process, $\bar X\to \bar Y+\bar B$, is the same:
\begin{equation}
	\Gamma(\bar X\to \bar Y+\bar B) =\Gamma(X \to Y + B)
\end{equation}
The net rate of baryon production goes like the difference of these
rates,
\begin{equation}
	{dB\over dt} \propto \Gamma(\bar X\to \bar Y+\bar B) -\Gamma(X \to Y + B)
\end{equation}
and so vanishes in the case where C is a symmetry.

However, even if C is violated, this is not enough.  We also need
CP violation.  To see this, consider an example where X decays into
two left-handed or two right-handed quarks,
\begin{equation}
	X\to q_L\, q_L,\qquad X\to q_R\, q_R
\end{equation}
Under CP,
\begin{equation}
	CP:\quad q_L\to \bar q_R
\end{equation}
where $\bar q_R$ is the antiparticle of $q_R$, which is {\it
left}-handed (in my notation, L and R are really keeping track of
whether the fermion is an SU(2) doublet or singlet), whereas under
C,
\begin{equation}
	C:\quad q_L\to \bar q_L
\end{equation}
Therefore, even though $C$ violation implies that
\begin{equation}
	\Gamma(X\to q_L q_L) \neq \Gamma(\bar X \to \bar q_L \bar q_L)
\end{equation}
CP conservation would imply
\begin{equation}
	\Gamma(X\to q_L q_L) = \Gamma(\bar X \to \bar q_R \bar q_R)
\end{equation}
and also
\begin{equation}
	\Gamma(X\to q_R q_R) = \Gamma(\bar X \to \bar q_L \bar q_L)
\end{equation}
Then we would have
\begin{equation}
\Gamma(X\to q_L q_L) +\Gamma(X\to q_R q_R) = \Gamma(\bar X \to \bar
q_R \bar q_R)+\Gamma(\bar X \to \bar q_L \bar q_L)
\end{equation}
As long as the initial state has equal numbers of $X$ and $\bar X$,
we end up with no net asymmetry in quarks.  The best we can get is
an asymmetry between left- and right-handed quarks, but this is not
a baryon asymmetry.

Let us recall exactly how C, P and CP operate on scalar fields,
spinors, and vector fields.  For complex scalars,
\begin{eqnarray}
	C:&& \phi\to\phi^*\\
	P:&& \phi(t,\vec x) \to \pm \phi(t,-\vec x)\nonumber\\
	CP:&& \phi(t,\vec x) \to \pm \phi^*(t,-\vec x)\nonumber
\end{eqnarray}
For fermions,
\begin{eqnarray}
	C:&& \psi_L\to i\sigma_2\psi_R^*,\quad
	\psi_R\to -i\sigma_2\psi_L^*,\quad \psi\to i\gamma^2\psi^*
	\\
	P:&& \psi_L\to\psi_R(t,-\vec x),\quad
	\psi_R\to \psi_L(t,-\vec x),\quad 
	\psi\to \gamma^0\psi(t,-\vec x)\nonumber\\
	CP:&& \psi_L\to i\sigma_2\psi_R^*(t,-\vec x),\quad
	\psi_R\to -i\sigma_2\psi_L^*(t,-\vec x),
	\quad \psi\to i\gamma^2\psi^*(t,-\vec x)\nonumber
\end{eqnarray}
For vectors,
\begin{eqnarray}
	C:&& A^\mu\to-A^\mu\\
	P:&& A^\mu(t,\vec x) \to (A^0,-\vec A)(t,-\vec x)\nonumber\\
	CP:&& A^\mu(t,\vec x) \to (-A^0,\vec A)(t,-\vec x)\nonumber
\end{eqnarray}

To test your alertness, it is amusing to consider the following
puzzle.  Going back to the example of $X\to qq$, let us suppose that
C and CP are both violated and that
\begin{equation}
	\Gamma(X\to qq) \neq \Gamma(\bar X\to \bar q\bar q)
\end{equation}
where now we ignore the distinction between $q_R$ and $q_L$.  An
astute reader might object  that if all $X$'s decay into $qq$ and
all $\bar X$'s decay into $\bar q\bar q$, with $n_X = n_{\bar X}$
initially, then eventually we will still have equal numbers of $q$ and
$\bar q$, even though there was temporarily an excess.  To avoid this
outcome, there must exist at least one competing channel,
$X\to Y$, $\bar X\to \bar Y$, such that
\begin{equation}
	\Gamma(X\to Y) \neq \Gamma(\bar X\to \bar Y)
\end{equation}
and with the property that $Y$ has a different baryon number than
$qq$.  One can then see that a baryon asymmetry will develop by the
time all $X$'s have decayed.  Why is it then, that we do not have
an additional fourth requirement, that a competing decay channel with
the right properties exists?  The reason is that this is guaranteed
by the CPT theorem, combined with the requirement of B violation.
CPT assures us that the total rates of decay for $X$ and $\bar X$
are equal, which in the present example means
\begin{equation}
	\Gamma(X\to qq)+\Gamma(X\to Y)  
	= \Gamma(\bar X\to \bar q\bar q)+\Gamma(\bar X\to \bar Y)
\end{equation}
The stipulation that B is violated tells us that $Y$ has different
baryon number than $qq$.  Otherwise we could consistently assign the
baryon number $2/3$ to $X$ and there would be no B violation.

\subsection{Examples of CP violation}
We now consider the conditions under which a theory has CP violation.
Generally CP violation exists if there are complex phases in 
couplings in the Lagrangian which can't be removed by field
redefinitions.  Let's start with scalar theories.  For example
\begin{eqnarray}
	{\cal L} &=& |\partial\phi|^2 - m^2|\phi|^2 - \lambda|\phi|^4
	\nonumber\\
	&-&(\mu^2\phi^2 + g\phi^4 + {\rm h.c.})
\end{eqnarray}
where the top line is CP conserving, and the bottom line is
(potentially) CP-violating.  This can be seen by applying the CP
transformation $\phi\to\phi^*$ and observing that the bottom line
is not invariant:
\begin{equation}
	{\cal L}_{CPV} \to - (\mu^2{\phi^*}^2 + g{\phi^*}^4
	+ {\rm h.c.})
\end{equation}
Of course, this lack of invariance only occurs if $\mu^2$ and $g$
are not both real.  Parametrize
\begin{equation}
	\mu^2 = |\mu|^2 e^{i\phi_\mu},\qquad 
	g = |g| e^{i\phi_g}
\end{equation}
We can perform a field refefinition to get rid of one of these phases,
{\it e.g.} $\phi_\mu$, by letting
\begin{equation}
	\phi\to e^{-i\phi_\mu/2}\phi
\end{equation}
so that
\begin{equation}
	{\cal L}_{CPV} = - (|\mu^2|{\phi}^2 + |g| 
	e^{i(\phi_g-2\phi_\mu)}{\phi}^4
	+ {\rm h.c.})
\end{equation}
Now under CP, only the $\phi^4$ term changes.  We can see that it is
only the combination
\begin{equation}
	\phi_{\rm inv} = \phi_g - 2\phi_\mu = \hbox{"invariant
phase"}
\end{equation}
which violates CP, and which is independent of field redefinitions.
If $\phi_{\rm inv}=0$, the theory is CP-conserving.  We can also write
\begin{equation}
	\phi_{\rm inv} = \arg\left(g\over \mu^4\right)
\end{equation}
No physical result can depend on field redefinitions, since these
are arbitrary.  Therefore any physical effect of CP violation must
manifest itself through the combination $\phi_{\rm inv}$.  This can
sometimes provide a valuable check on calculations.

An equivalent way of thinking about the freedom to do field
redefinitions is the following.  One can also define the CP
transformation with a phase $\phi \to e^{i\alpha}\phi^*$.  As long as
there exists any value of $\alpha$ for which this transformation is a
symmetry, we can say that this is the real CP transformation, and CP
is conserved.  

Next let's consider an example with fermions.  We can put a complex
phase into the mass of a fermion (here we consider Dirac),
\begin{eqnarray}
	{\cal L} &=& \bar\psi\left(i\dsl
	-m(\cos\theta+i\sin\theta\gamma_5)\right)\psi\nonumber\\
	 &=& \bar\psi\left(i\dsl
	-me^{i\theta\gamma_5}\right)\psi
\end{eqnarray}
[In Weyl components, the mass term has the form 
$m(\bar\psi_L e^{i\theta}\psi_R + \bar\psi_R e^{-i\theta}\psi_L)$.]
Under CP, 
\begin{eqnarray}
	\psi &\to& i\gamma^2\gamma^0\psi^*,\quad
	\psi^* \to -i(-\gamma^2)\gamma^0\psi,\nonumber\\
	\psi^\dagger &\to& \psi^{T}\gamma^0(-\gamma^2)(-i),\nonumber\\
	\psi^T &\to& \psi^{\dagger} \gamma^0 \gamma^2 i = 
	\bar\psi \gamma^2 i
\end{eqnarray}
Also, by transposing the fields, $\bar\psi e^{i\theta\gamma_5}\psi = 
-\psi^T e^{i\theta\gamma_5}\gamma^0\psi^*$.  Putting these results
together, we find that the complex mass term transforms under CP
as
\begin{eqnarray}
	\bar\psi e^{i\theta\gamma_5}\psi &\to&
	-\bar\psi \gamma^2 i e^{i\theta\gamma_5}\gamma^0 i\gamma^2
	\gamma^0\psi\nonumber\\
	&=& - \bar\psi \gamma^2
	e^{i\theta\gamma_5}\gamma^2\psi\nonumber\\
	&=&	\bar\psi  e^{-i\theta\gamma_5}\psi
\end{eqnarray}
So ostensibly $\theta$ is CP-violating.  However, we can remove this
phase using the chiral field redefinition
\begin{equation}
	\psi \to e^{-i(\theta/2)\gamma_5}\psi
\end{equation}
To get an unremovable phase, we need to consider a more complicated
theory.  For example, with two Dirac fermions we can have
\begin{equation}
{\cal L} = \sum_{i=1}^2\bar\psi_i(i\dsl - m_i e^{i\theta_i\gamma_5})
	\psi_i + \mu(\bar\psi_1 e^{i\alpha\gamma_5}\psi_2 
	+{\rm h.c.})
\end{equation}
After field redefinitions we find that
\begin{equation}
{\cal L} \to \sum_{i=1}^2\bar\psi_i(i\dsl - m_i)
	\psi_i + \mu(\bar\psi_1 e^{i(\alpha-\theta_1/2-\theta_2/2)\gamma_5}\psi_2 
	+{\rm h.c.})
\end{equation}
so the invariant phase is $\alpha - \frac12(\theta_1+\theta_2)$.

It is also useful to think about this example in terms of 2-component
Weyl spinors, where the mass term takes the form
\begin{equation}
	\sum_{i=1}^2 \psi_{L,i}^\dagger e^{i\theta_i}\psi_{R,i}
	+ \mu \psi_{L,1}^\dagger e^{i\alpha}\psi_{R,2}
	+ {\rm h.c.}
\end{equation}
One might have thought that there are 4 field redefinitions,
corresponding to independent rephasings of $\psi_{L,i}$ and
$\psi_{R,i}$, so there would be enough freedom to remove all the
phases.  However, it is only the axial vector transformations,
where $L$ and $R$ components are rotated by equal and opposite
phases, that can be used to remove $\theta_i$.  Half of the
available field redefinitions are useless as far as removing 
phases from the mass terms.  We might more accurately state
the general rule as ``the number of invariant phases equals the
total number of phases in the Lagrangian parameters, minus the
number of {\it relevant} field redefinitions which can be done.''

A third example of CP violation is the famous $\theta$ term in QCD,
\begin{equation}
	\theta_{QCD} F^a_{\mu\nu} \tilde F^{a,\mu\nu} =
	\frac12 \theta_{QCD} \epsilon^{\mu\nu\alpha\beta}
	F^a_{\mu\nu} F^{a}_{\alpha\beta}
	= -2\theta_{QCD} \vec E^a\cdot \vec B^a
\end{equation}
This combination is P-odd and C-even, thus odd under CP.  Its
smallness is the strong CP problem of QCD: if $\theta_{QCD} 
\gsim 10^{-10}$, the electric dipole moment of the neutron would
exceed experimental limits.  The EDM operator is
\begin{eqnarray}
{\rm EDM} = 	-{i\over 2}d_n \bar n \sigma_{\mu\nu}\gamma_5 F^{\mu\nu} n
	&=& d_n\left( n^\dagger_L \vec\sigma\cdot\vec E n_R + 
			n^\dagger_R \vec\sigma\cdot\vec E n_L
	\right.\nonumber\\
	&&\left. + i(n^\dagger_L \vec\sigma\cdot\vec B n_R  - 
	n^\dagger_R \vec\sigma\cdot\vec B n_L)\right)
\end{eqnarray}
which is mostly $\vec\sigma\cdot \vec E$, since there is a cancellation
between left and right components in the bottom line. Contrast this
with the magnetic dipole moment operator,
\begin{eqnarray}
{\rm MDM} = 	{1\over 2}\mu_n \bar n \sigma_{\mu\nu}F^{\mu\nu} n
	&=& \mu_n\left( n^\dagger_L \vec\sigma\cdot\vec B n_R + 
			n^\dagger_R \vec\sigma\cdot\vec B n_L
	\right.\nonumber\\
	&&\left. + i(n^\dagger_L \vec\sigma\cdot\vec E n_R  - 
	n^\dagger_R \vec\sigma\cdot\vec E n_L)\right)
\end{eqnarray}
which is mostly  $\vec\sigma\cdot \vec B$.  To verify the CP properties
of these operators, notice that under CP, $\vec\sigma \to
\vec\sigma$, $L\leftrightarrow R$, $\vec E\to - \vec E$, $\vec B\to
\vec B$.  EDM's will play an important role in constraining models of
baryogenesis, since the phases needed for one can also appear in the
other.  The current limit on the neutron EDM is \cite{Baker}
\beq
	|d_n| < 3\times 10^{-26} \,e{\rm cm}
\eeq
In 1979 it was estimated that \cite{Crewther}
\beq
	d_n = 5\times 10^{-16} \theta_{QCD}  \,e{\rm cm}
\eeq
leading to the stringent bound on $\theta_{QCD}$.

Other particle EDM's provide competitive constraints on CP violation.
The electron EDM is constrained \cite{Regan},
\beq
 	|d_e| < 1.6 \times 10^{-27} \,e\cdot{\rm cm}
\eeq
as well those of Thallium and Mercury atoms \cite{Romalis},
\beqa
 	|d_{Tl}| &<& 9 \times 10^{-25} \,e{\rm cm}\nonumber\\
	|d_{Hg}| &<& 2 \times 10^{-28} \,e{\rm cm}
\eeqa

\subsection{More history} I conclude this section on ``Sakharov's
Laws'' with a bit more about the early history of baryogenesis. The
earliest papers on baryogenesis from the era when GUT's were invented
were not at first aware of Sakharov's contribution.  Yoshimura was
the first to attempt to get baryogenesis from the SU(5) theory, in
1978 \cite{Yoshimura78}.  However, this initial attempt was flawed
because it used only scatterings, not decays, and therefore did not
incorporate the loss of thermal equilibrium.  This defect was pointed
out by Barr \cite{Barr}, following work of Toussaint {\it et al.}
\cite{Toussaint}, and corrected
by Dimopoulos and Susskind \cite{Dimopoulos}, who were the first to use the
out-of-equilibrium decay mechanism.  None of these authors knew about
the Sakharov paper.  Weinberg \cite{Weinberg} and Yoshimura 
\cite{Yoshimura79} quickly  followed
Dimopoulos and Susskind with more quantitative calculations of the
baryon asymmetry in GUT models.  

The first paper to attribute to Sakharov the need for going out of
equilibrium, by Ignatiev {\it et al.,} \cite{Ignatiev} was not in the context of
GUTS. They constructed a lower-energy model of baryogenesis, having
all the necessary ingredients, by a modification of the electroweak
theory.  Their idea was not as popular as GUT baryogenesis, but it
did make people aware of Sakharov's early paper.  

\section{Example: GUT baryogenesis}
I will now illustrate the use of Sakharov's required ingredients in
the SU(5) GUT theory \cite{NanWein}.  This is no longer an attractive
theory for baryogenesis because it requires a higher reheating
temperature from inflation than is desired, from the point of view of
avoiding unwanted relics, like monopoles and heavy gravitinos. 
Nevertheless, it beautifully illustrates the principles, and it is
also useful for understanding leptogenesis, since the two approaches
are mathematically quite similar.

\begin{figure}[h]
\centerline{
\includegraphics[width=1.2\textwidth]{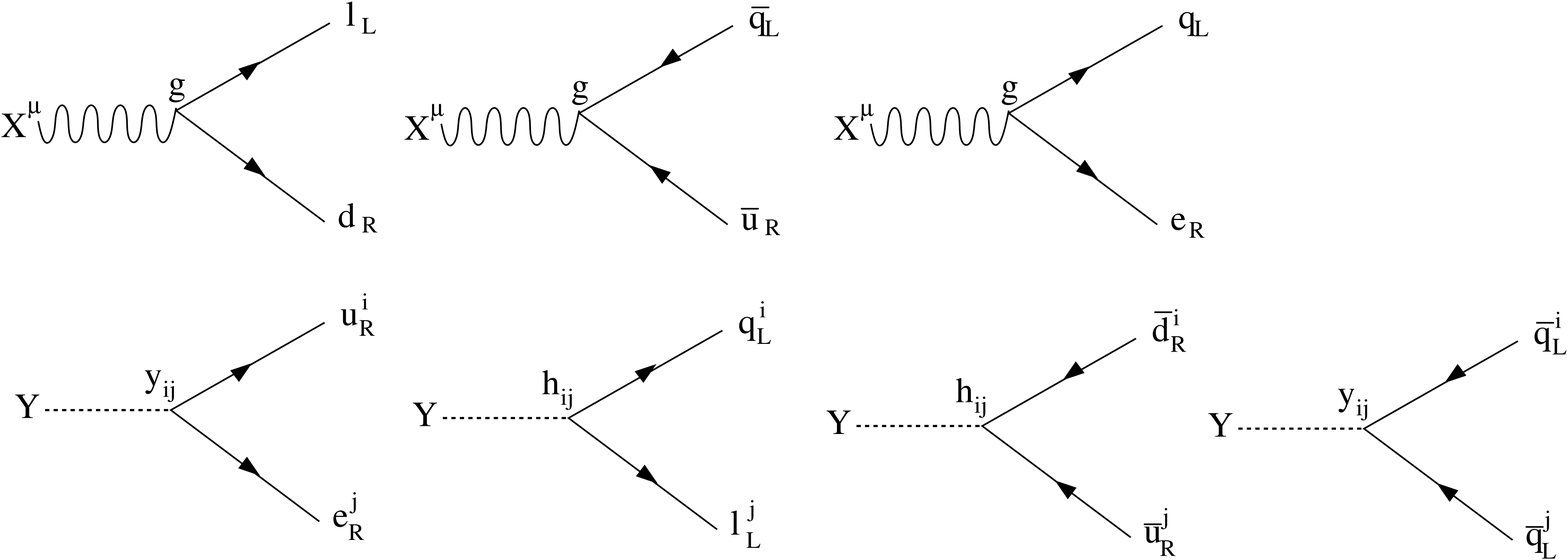}}
\caption{GUT couplings}
\label{fig5}
\end{figure}

In SU(5) there exist gauge bosons $X^\mu$ whose SU(3), SU(2) and
U(1)$_y$ quantum numbers are $(3,2,-\frac56)$, as well as Higgs bosons $Y$
with quantum numbers $(3,1,\frac13)$, and whose couplings are similar to
those of the vectors.  The couplings to quarks and leptons are shown
in figure \ref{fig5}.   We see that the requirement of B violation is
satisfied in this theory, and the CP-violating decays of any of these scalars can lead to
a baryon asymmetry.  To get CP violation, we need the matrix Yukawa couplings $h_{ij},
y_{ij}$ to be complex, where $i,j$ are generation indices.  It will
become apparent what the invariant phases are.  

Let's consider the requirement for $X^\mu$ or $Y$ to decay out of
equilibrium.  The decay rate is
\beq
	\Gamma_D \cong \alpha m N / \gamma
\eeq
where $\alpha = g^2/(4\pi)$ or $\sim (y+h)^2/(4\pi)$, for $X^\mu$ or $Y$,
resepctively, $m$ is the mass, $N$ is the number of decay channels, and the Lorentz 
gamma factor can be roughly estimated at high temperature as
\beq
	\gamma = {\langle E\rangle \over m} \cong
	(1 = 9 T^2/m^2)^{1/2}
\eeq
considering that $\langle E\rangle\cong 3 T$ for highly relativistic
particles.  The $\gamma$ factor will actually not be necessary for us,
because we are interested in cases where the particle decays at low
temperatures compared to its mass, so that the decays will occur out
of equilibrium.  Thus we can set $\gamma = 1$.  The age of the
universe is $\tau \equiv 1/H =  1/\Gamma_D$ at the time of the decay,
so we set
\beqa	
	\Gamma_D &=& H \cong \sqrt{g_*} {T^2\over M_p} = \alpha m N
	\nonumber\\ &\longrightarrow& (\alpha N m
	M_p/\sqrt{g_*})^{1/2} \ll m
\eeqa
where $g_*$ is the number of relativistic degrees of freedom at the
given temperature.  Therefore we need $\alpha \ll (m/M_p)
(\sqrt{g_*}/N)$.  Let us first consider the
decays of the $X$ gauge boson.  Since it couples to everything, the
number of decay channels $N$ is of the same order as $g_*$, and we get
\beq
	{g^2\over 4\pi}\ll {m\over M_p\sqrt{g_*}} 
\label{ooe}
\eeq
But unification occurs at the scale $m\cong 10^{16}$ GeV, for values
of $g^2$ such that $\alpha \cong 1/25$, so this condition cannot be
fulfilled.  On the other hand, for the Higgs bosons, $Y$, we get the
analogous bound
\beq
	{h^2\hbox{\ or\ }y^2\over 4\pi}\ll {m\over M_p\sqrt{g_*}} 
\eeq
which can be satisfied by taking small Yukawa couplings.  Therefore
the Higgs bosons are the promising candidates for decaying out of
equilibrium.

\begin{figure}[h]
\centerline{
\includegraphics[width=1.2\textwidth]{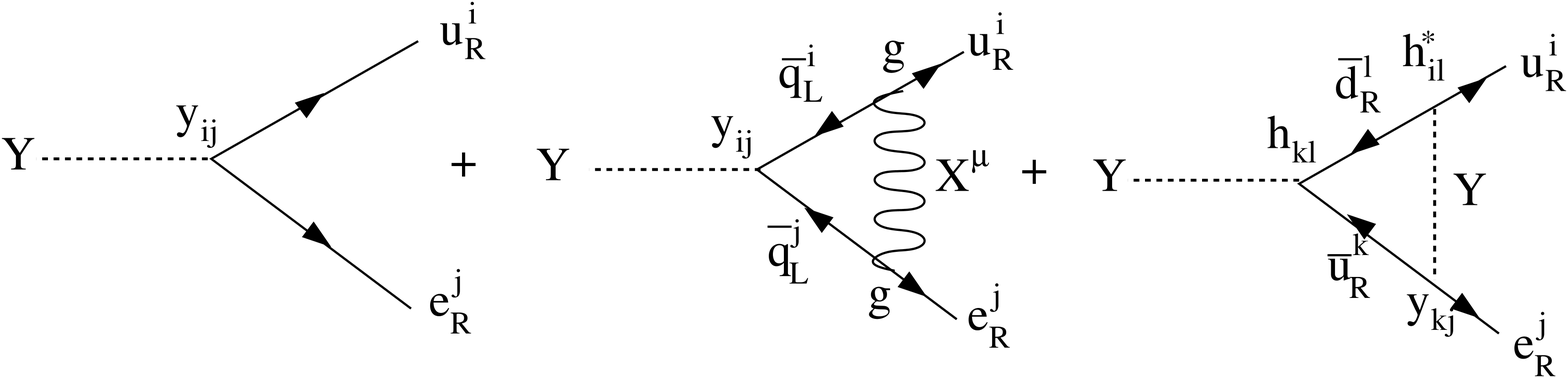}}
\caption{Tree plus 1-loop contributions to
$Y\to e_R u_R$}
\label{fig6}
\end{figure}

For clarity, I will now focus on a particular decay channel,
$Y\to u_R e_R$, whose interaction Lagrangian is
\beq
	y_{ij} Y^a \bar u_{R\, a,i} e^c_{R,j} + {\rm h.c.}
\eeq
At tree level, we cannot generate any difference between the squared
matrix elements, and we find that
\beq
	\left|{\cal M}_{Y\to e_R u_R}\right|^2 = 
	\left|{\cal M}_{\bar Y\to \bar e_R \bar u_R}\right|^2
\eeq
The complex phases are irrelevant at tree level.  To get a
CP-violating effect, we need interference between the tree amplitude
and loop corrections, as shown in figure \ref{fig6}. 

The one-loop diagrams develop imaginary parts
which interfere with the phase of the tree diagram in a CP-violating
manner.  We can write the amplitude as
\beq
	-i{\cal M} = -i\left(y_{ij} + y_{ij} F_X\left(M_X^2/p^2\right) 
	+ (h^* H^T y)_{ij} F_Y\left(M_X^2/p^2\right)\right)
\eeq
where $p$ is the 4-momentum of the decaying $Y$ boson, 
\beqa
	F_Y &=& i^5 \int{d^{4}q\over (2\pi)^4} \, {1\over q^2-M^2_Y}\,
	{1\over \qsl + \frac12\psl + i\epsilon}\,
	{1\over \qsl - \frac12\psl + i\epsilon}\nonumber\\
	&\equiv& R_Y + i I_Y
\eeqa
and $F_X$ is similar, but with $M_Y\to M_X$ and $\gamma_\mu$ factors
in between the propagators.  (It will turn out that the $X$ exchange
diagram does not contribute to the CP-violating interference.)  In the
second line we have indicated that the loop integral has real and
imaginary parts.  The latter arise as a consequence of unitarity,
\beq
	-i\left[{\cal M}(p_i\to p_f) - {\cal M}^*(p_f\to p_i)\right]
	= \sum_a d\Pi_a {\cal M}(p_i\to p_a){\cal M}^*(p_f\to p_a)
\eeq
where the sum is over all possible on-shell intermediate states and $d\Pi_a$ is
the Lorentz-invariant phase space measure.  
If ${\cal M}^*(p_f\to p_i) = {\cal M}^*(p_i\to p_f)$, then the
left-hand-side is the imaginary part of ${\cal M}$.  The
right-hand-side implies that an imaginary part develops if
intermediate states can go on shell:\\
\centerline{\includegraphics[width=1\textwidth]{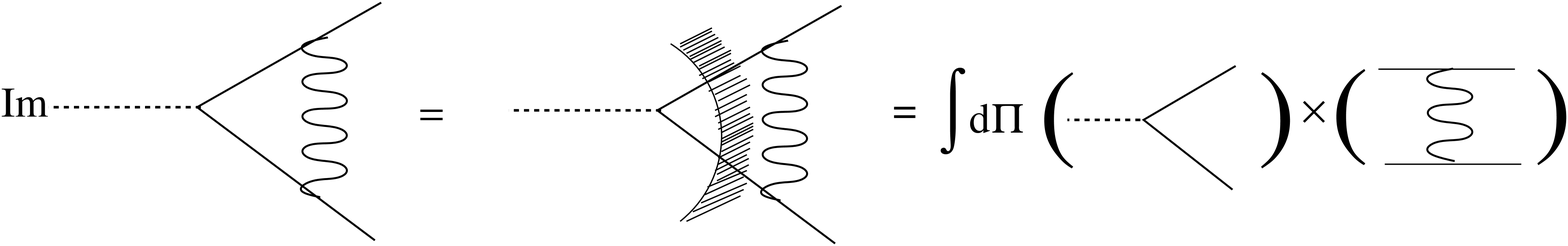}}
This will be the case as long as the decaying particle is heavier
than the intermediate states.

To see how a rate asymmetry comes about, schematically we can write
\beqa
	{\cal M}(Y\to eu) &=& y + y F_X +\tilde y F_Y \nonumber\\
	{\cal M}(\bar Y\to \bar e\bar u) &=& 
	y^* + y^* F_X +\tilde y^* F_Y 
\eeqa
where $\tilde y \equiv h^* h^T y$.  Then the difference between the
probability of decay of $Y$ and $\bar Y$ goes like
\beqa
	\left|{\cal M}_{Y\to eu}\right|^2 - 
	\left|{\cal M}_{\bar Y\to \bar e\bar u}\right|^2 &=&
	\left[y^*(1+F_X^*) +\tilde y^* F_Y^*\right]
	\left[y(1+F_X) +\tilde y F_Y\right] \nonumber\\
	&&- 
	\left[y(1+F_X^*) +\tilde y F_Y^*\right]
	\left[y^*(1+F_X) +\tilde y^* F_Y\right]\nonumber\\
	&=& y^*\tilde y F_Y + \tilde y^* y F_Y^* - y \tilde y^* F_Y
		- \tilde y y^* F_Y^*\nonumber\\
	&=& [\tilde y y^* -\tilde y^* y] 2i I_Y\nonumber\\
	&=& -4{\rm Im}[\tilde y y^*] I_Y\nonumber\\
	&=& -4{\rm Im}\,{\rm tr}(h^* h^T y y^\dagger) I_y
\eeqa
(You can check that the $F_X$ term does not contribute to this
difference, since $g$ is real.)  Now, unfortunately, 
it is easy to see using the properties of the trace
that ${\rm tr}(h^* H^T y y^\dagger)$ is purely real,
so this vanishes.  However, there is a simple generalization which
allows us to get a nonvanishing result.  Suppose there are two or more
$Y$ bosons with different Yukawa matrices; then we get
\beq
	\left|{\cal M}_{Y_A\to eu}\right|^2 - 
	\left|{\cal M}_{\bar Y_A\to \bar e\bar u}\right|^2 
	= -4\sum_B {\rm Im}{\,\rm tr}(h_B^* h_A^T y_B y_A^\dagger) 
	I_y\left(M^2_B\over M^2_A\right)
\eeq
which no longer vanishes.  The function $I_Y$ can be found in
\cite{NanWein}, who estimate it to be of order $10^{-2}-10^{-3}$ 
for typical values of the parameters.

Let us now proceed to estimate the baryon asymmetry.  Define the
rate asymmetry
\beq
	r_A = {\left|{\cal M}_{Y_A\to eu}\right|^2 - 
	\left|{\cal M}_{\bar Y_A\to \bar e\bar u}\right|^2 \over
	\sum_f \left|{\cal M}_{Y_A\to f}\right|^2}
\eeq
which is the fraction of decays that produce a baryon excess.
The sum in the denominator is over all possible final states.
The resulting baryon density is
\beq
	n_B - n_{\bar B} = \sum_A n_{Y_A} r_A B_q
\eeq
where $B_q$ is the amount of baryon number produced in each decay
($B_q = 1/3$ for the channel we have been considering).  Of course,
we should also add the corresponding contributions from all the
competing $B$-violating decay channels, as was emphasized in the 
previous section.  For simplicity we have shown how it works for
one particular decay channel.  Up to factors of order unity, this
channel by itself gives a good estimate of the baryon asymmetry.  

To compute the baryon asymmetry, it is convenient to take the ratio
of baryons to entropy first, since as we have noted in section 1, entropy density
changes in the same way as baryon density as the universe expands.
\beqa
	{n_B - n_{\bar B}\over s} &=& {\frac13 n_Y \sum_A r_A\over
	{4\pi^2\over 45} g_* T^2}
	= {90\zeta(3)\over 4\pi^4 g_*}\sum_A r_A
\eeqa
where we used $n_Y = (2\zeta(3)/\pi^2)T^3\times 3$, the final factor
of 3 counting the color states of $Y$.  The number of degrees of
freedom is
\beqa
	g_* &=& \left(\sum_{\rm bosons} + \frac78\sum_{\rm
	fermions}\right)(\hbox{spin states})\nonumber\\
	&=&\left\{\begin{array}{ll}106.75 & \hbox{standard model}\\
	160.75 & {\rm SU(5)}\end{array}\right\}\times (\sim 2\hbox{\ if SUSY})
\eeqa
(these numbers can be found in the PDG Big Bang Cosmology review
\cite{PDG}).  We can now use (\ref{eq1.4}) to convert this to $\eta$,
by multiplying by a factor of 7.  Suppose we have SUSY SU(5); then
\beq
	\eta \cong {1.2\over 320} \sum r_A \cong 10^{-2} \sum r_A
\eeq
Thus we only need $\sum r_A \cong 6\times 10^{-8}$, a quite small
value, easy to arrange since
\beq
	r_A \sim {y^2 h^2\over y^2 + h^2}\epsilon I_y
\eeq
where $\epsilon$ is the CP violating phase, and we noted that 
$I_Y\sim 10^{-2}-10^{-3}$.  According to our estimate of the size of
$y$ or $h$ needed to decay out of equilibrium, $y^2\sim 4\pi\times
(10^{-4}-10^{-6})$.  We can easily make $\eta$ large enough, or even
too large.

\subsection{Washout Processes}
The foregoing was the first quantitative estimate of the baryon asymmetry, 
but it is missing an important ingredient which reduces $\eta$:
B-violating rescattering and inverse decay processes. 
 We have ignored diagrams like\\
\centerline{\includegraphics[width=1\textwidth]{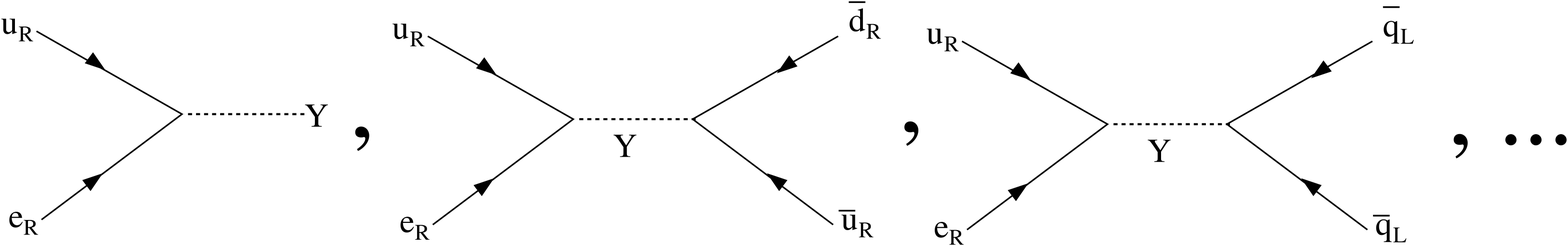}}
The first of these, inverse decay, was assumed to be negligible by
our assumption of out-of-equilibrium decay, but more generally we
could include this process and quantify the degree to which it is
out of equilibrium.  Any of the washout processes shown would cause
the baryon asymmetry to relax to zero if they came into equilibrium.

To quantify these effects we need the Boltzmann equation for the
evolution of  baryon number.  This was first carried out by Kolb
and Wolfram \cite{KW}, who were at that time postdoc and graduate student,
respectively, at Caltech.  The form of the Boltzmann equations is
\beqa
	{dn_{u_R}\over dt} + 3Hn_{u_R} &\cong& \int d\Pi_i \left(
	f_Y|{\cal M}(Y\to ue)|^2 - f_u f_e |{\cal M}(ue\to Y)|^2
	\right)\\
	&+& \int d\Pi_i\left[ f_{q_L} f_{q_L} |{\cal M}(qq\to ue)|^2
	- f_u f_e |{\cal M}(ue\to qq)|^2 \right] + \dots\nonumber
\eeqa
where $f_x$ is the distribution function for particle $x$, 
$\int d\Pi_i$ is the phase space integral over all final
and initial particles, 
\beq
	\prod_i d\Pi_i \equiv \left[\prod_i \left(
	{d^4 p_i\over(2\pi)^3} \delta(p_i^2-m_i^2)\right)\right]
	\times (2\pi)^4 \delta^{(4)}\left(\sum_{\rm final} p_a - 
	\sum_{\rm initial} p_a\right)
\eeq
and I have ignored final-state Pauli-blocking or Bose-enhancement
terms, $(1\pm f)$.   We should include similar equations for any other
particles which are relevant for the baryon asymmetry ($X^\mu$, $Y$,
quarks, leptons), and solve the coupled equations numerically to
determine the final density of baryons.  Depending on the strength of
the scattering, the washout effect can be important.  We can
estimate its significance without solving the Boltzmann equations
numerically.  

\begin{figure}[h]
\centerline{
\includegraphics[width=0.7\textwidth]{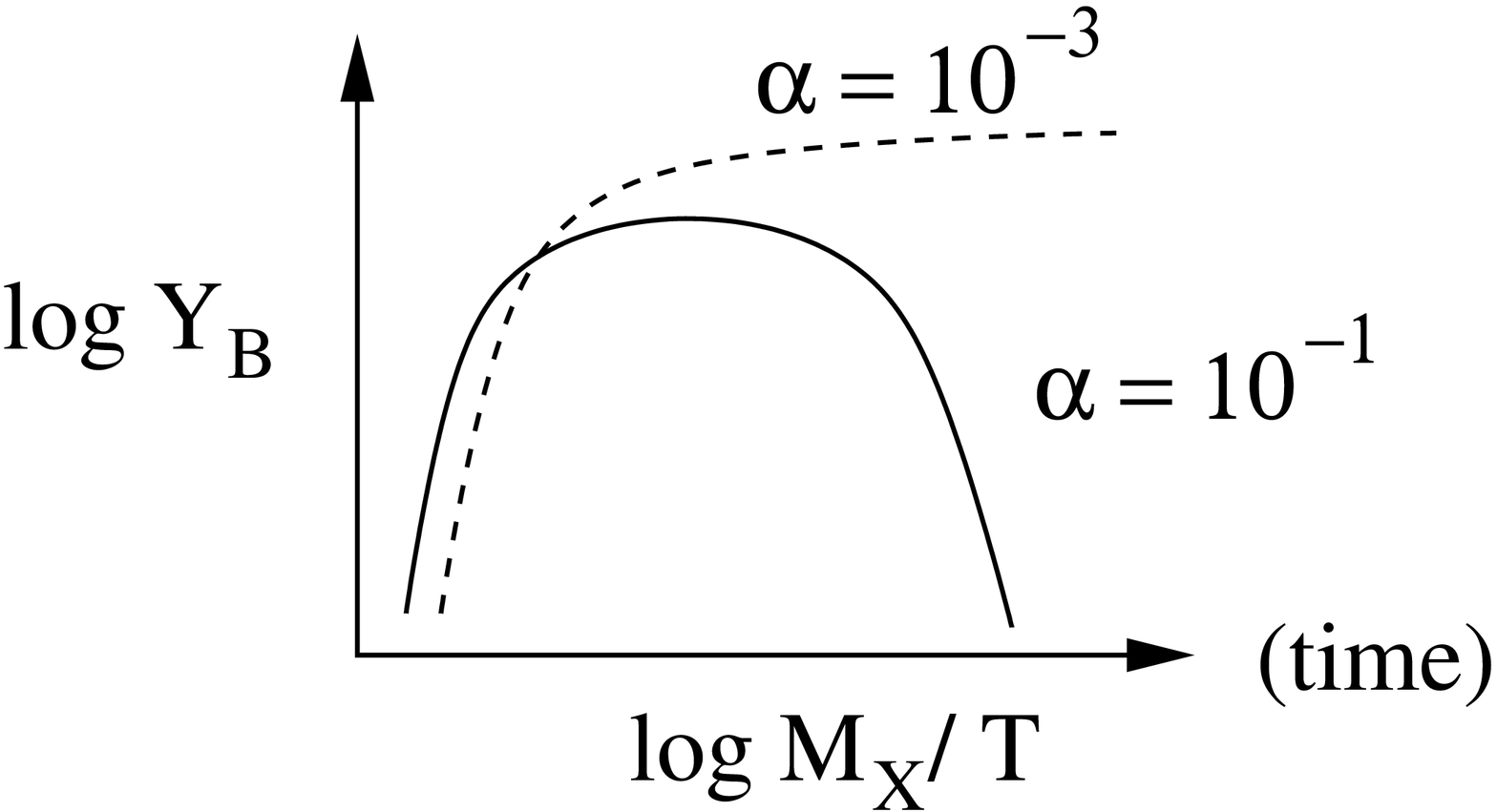}}
\caption{Evolution of baryon asymmetry with with B-violating
rescattering processes included.}
\label{fig7}
\end{figure}

Define the abundance $Y_i \equiv n_i/n_\gamma$ and consider a
simplified model where $X\to bb,$ $\bar b \bar b$ (baryons or
antibaryons), with
\beqa
	\Gamma_{X\to bb} &\equiv& \frac12(1+\epsilon)
\Gamma_X\nonumber\\
	\Gamma_{X\to \bar b\bar b} &\equiv& \frac12(1-\epsilon) \Gamma_X\nonumber\\
	\Gamma_{\bar X\to b b} &\equiv& \frac12(1-\bar\epsilon) \Gamma_X\nonumber\\
	\Gamma_{\bar X\to \bar b\bar b} &\equiv& \frac12(1+\bar\epsilon) \Gamma_X
\eeqa
Kolb and Wolfram found that the time rate of change of the baryon
abundance (which we could also have called $\dot\eta$) is
\beq
	\dot Y_B = \Gamma_X(\epsilon-\bar\epsilon)\left[
	Y_X - Y_{\rm eq}\right] - 2 Y_B\left[ \Gamma_X Y_{X,{\rm eq}}
	+ \Gamma_{bb\leftrightarrow\bar b\bar b}\right]
\eeq
On the right-hand-side, the first term in brackets represents baryon production 
due to CP-violating decays, while the second term is due to the
washout induced by inverse decays and scatterings.  The scattering
term is defined by the thermal average
\beq
	\Gamma_{bb\leftrightarrow\bar b\bar b} = n_\gamma \left(
	\langle v \sigma_{bb\to\bar b\bar b}\rangle + 
	\langle v \sigma_{\bar b\bar b\to bb}\rangle \right)
\eeq
In the treatment of Nanopoulos and Weinberg, the effect of 
$\Gamma_{bb\leftrightarrow\bar b\bar b}$ was neglected, but we can
estimate it using dimensional analysis as 
\beq
	\Gamma_{bb\leftrightarrow\bar b\bar b} \sim
	{\alpha^2 T^5\over (M_X^2 + 9 T^2)^2}
\eeq
(considering $s$-channel exchange of the $X$ boson), again using $\langle
E\rangle \sim 3 T$.  Then if we imagine the initial production of
the baryon asymmetry happens quickly (by the time $t\equiv 0$), followed
by gradual relaxation, then
\beq
	Y_B(t)\sim Y_B(0) \,
	 e^{-\int \Gamma_{bb\leftrightarrow\bar b\bar b}dt}
\eeq
Using $H\sim 1/t \sim g_* T^2/M_p$, $dt\sim (M_p/g_*)(dT/T^3)$, 
\beq
	\int\Gamma dt \sim {M_p\over g_*}\int \Gamma{dT\over T^3}
	\cong {\pi\alpha^2 M_p\over 100 M_X}
\eeq
so the maximum BAU gets diluted by a factor $\sim e^{-(\pi\alpha^2
M_p/(100 M_X)}$.  We thus need $\alpha^2\lsim 30 M_X/M_P$, hence
$\alpha \lsim 0.05$, using $M_X/M_P \sim 10^{-4}$.  This agrees with
the quantitative findings of Kolb and Turner, schematically shown in
figure \ref{fig7}.  Notice that this value of $\alpha$ is consistent
with the unification value, so GUT baryogenesis works.

This concludes our discussion of GUT baryogenesis.  As noted earlier,
it is disfavored because it needs a very high reheat temperature after
inflation, which could introduce monopoles or gravitinos as relics
that would overclose the universe.  Another weakness, shared by many
theories of baryogenesis, is a lack of testability.  There is no way
to distinguish whether baryogenesis happened via GUT's or some other
mechanism.  We now turn to the possibility of baryogenesis at lower
temperatures, with greater potential for testability in the
laboratory.

\section{B and CP violation in the Standard Model}
During the development of GUT baryogenesis, it was not known that the
standard model had a usable source of B violation.   In 1976, 't
Hooft showed that the triangle anomaly violates baryon number through
a nonperturbative effect \cite{'tHooft}.  In (\ref{eq2.1}) we saw that the baryon
current is not conserved in the presence of external $W$ boson field
strengths.  However this violation of B is never manifested in any
perturbative process.  It is associated with the vacuum structure of
SU(N) gauge theories with spontaneously broken symmetry.  To explain
this, we need to introduce the concept of Chern-Simons number, 
\beq
	N_{CS} = \int d^{\,3} x\, K^0
\eeq
where the current $K^\mu$ is given by
\beq
    K^\mu = {g^2\over 32\pi^2} \epsilon^{\mu\nu\alpha\beta}\left(
	F^a_{\nu\alpha} A^a_\beta - {g\over 3}\epsilon_{abc}
	A^a_\nu A^b_\alpha A^c_\beta \right)
\eeq
(note: there are a number of references \cite{KM,AM} in which the factor ${g/3}$
is incorrectly given as $2g/3$; this seems to be an error that
propagated without being checked).  This current has the property that
\beq
	\partial_\mu K^\mu = {g^2\over 32\pi^2} F^a_{\mu\nu}
	\widetilde  F^{a,\mu\nu}
\eeq
Chern-Simons number has a topological nature which can be seen when
considering configurations which are pure gauge at some initial and
final times, $t_0$ and $t_1$.  It can be shown that 
\beq
	N_{CS}(t_1) - N_{CS}(t_0) = \int_{t_0}^{t_1} dt \int d^{\,3} x
	\,\partial_\mu K^\mu = \nu
\eeq
an integer, which is a winding number.  The gauge field is a map
from the physical space to the manifold of the gauge group.
If we consider an SU(2) subgroup of SU(N), and the boundary of 4D
space compactified on ball, both manifolds are 3-spheres, and the
map can have nontrivial homotopy.  

\begin{figure}[h]
\centerline{
\includegraphics[width=1\textwidth]{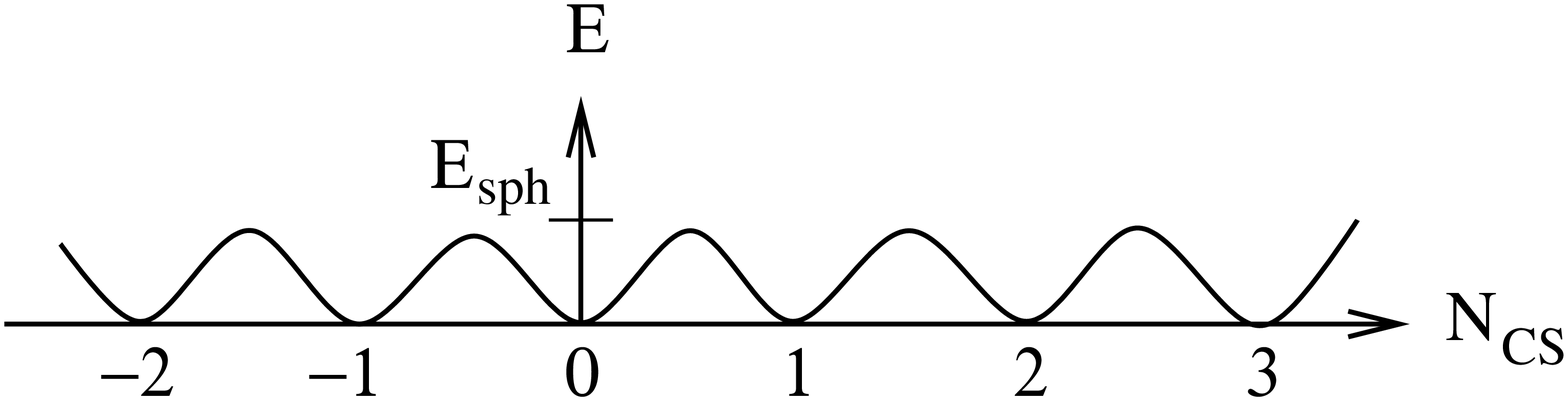}}
\caption{Energy of gauge field configurations as a function of
Chern-Simons number.}
\label{fig8}
\end{figure}

We are interested in the vacuum structure of SU(2) gauge theory.
Consider a family of static gauge field configurations with
continuously varying $N_{CS}$.  Those configuration with integer
values turn out to be pure gauge everywhere, hence with vanishing
field strength and zero energy.  But to interpolate between two
such configurations, one must pass through other configurations
whose field strength and energy are nonvanishing.  The energy versus
Chern-Simons number has the form shown in figure \ref{fig8}.

Each minimum is a valid perturbative vacuum state of the theory.
They are called $n$-vacua.  The height of the energy barrier is
\beq
	E_{\rm sph} = f\left(\lambda\over g^2\right) 
	{4\pi v\over g} \cong {8\pi v\over g} = {2 M_W\over\alpha_W}
	f\left(\lambda\over g^2\right) 
\eeq
where $v=174$ GeV is the Higgs field VEV, $\lambda$ is the Higgs
quartic coupling, $\alpha_W = g^2/{4\pi}\cong 1/30$, and the function
$f$ ranges between $f(0) = 1.56$ and $f(\infty) = 2.72$.

't Hooft discovered that tunneling occurs between $n$-vacua through
field configurations now called instantons (Physical Review tried to
suppress that name at first, so the term ``pseudoparticle'' appears in
the early literature, but it soon succumbed to the much more popular
name.)  The relevance of this for B violation is through the relation
between the divergence of the left-handed baryon plus lepton current
to the divergence of $K^\mu$:
\beq
	\partial_\mu J^\mu_{B+L} = N_f \partial_\mu K^\mu
\eeq
where $N_f =3$ is the number of families.  Integrating this relation
over space and time, the spatial divergence integrates to zero and 
we are left with
\beq
	3 {d\over dt} N_{CS} = {d\over dt} B = {d\over dt} L
\eeq
Hence each instanton transition violates B and L by 3 units
each---there is spontaneous production of 9 quarks and 3 leptons, with
each generation represented equally.  

However, the tunneling amplitude is proportional to
\beq
	{\cal A}\sim e^{-8\pi^2/g^2} \sim 10^{-173}
\eeq
which is so small as to never have happened during the lifetime of the
universe.  For this reason, anomalous B violation in the SM was not at
first considered relevant for baryogenesis.  But in 1985, Kuzmin,
Rubakov and Shaposhnikov realized that at high temperatures, these
transitions would become unsuppressed, due to the availability of
thermal energy to hop over the barrier, instead of tunneling through it
\cite{KRS}.
This occurs when $T\gsim 100$ GeV.  The finite-$T$ transitions are
known as sphaleron processes, a term coined by Klinkhamer and Manton
from the Greek, meaning ``ready to fall:'' it is the field
configuration of Higgs and $W^\mu$ which sits at the top of the energy
barrier between the $n$-vacua \cite{KM}.  A thermal transition between $n$-vacua
must pass through a configuration which is close to the sphaleron,
unless $T\gg E_{\rm sph}$.  The sphaleron is a static, saddle point
solution to the field equations.

Evaluating the path integral for a sphaleron transition
semiclassically, one finds that the amplitude goes like
\beq
	{\cal A}\sim e^{-E{\rm sph}/T}
\eeq
To find the actual rate of transitions, one must do a more detailed
calculation including the fluctuation determinant around the
saddle point \cite{AM}.  Khlebnikov and Shaposhnikov obtained the rate
per unit volume of sphaleron transitions 
\beq
	{\Gamma\over V} = {\rm const} \left({E_{\rm sph}\over
T}\right)^3\left({m_W(T)\over T}\right)^4 T^4  e^{-E{\rm sph}/T}
\eeq
where the constant is a number of order unity and $m_W(T)$ is the
temperature dependent mass of the $W$ boson.  But this formula is only
valid at temperatures $T< E_{\rm sph}$.  In fact, at $T\gsim m_W$, 
electroweak symmetry breaking has not yet occurred, and $m_W(T) = 0$.
The Higgs VEV vanishes in this symmetry-restored phase, and $E_{\rm
sph} = 0$.  There is no longer any barrier between the $n$-vacua at 
these high temperatures.

The rate of sphaleron interactions above the electroweak phase
transition (EWPT) cannot be computed
analytically; lattice computations are required.  For many years
it was believed that the parametric dependence was of the form
\beq
	{\Gamma\over V} = c \alpha_w^4 T^4
\label{gv}
\eeq
with $c\sim 1$ and $\alpha_w = g_2^4/(4\pi)\cong 1/30$.  This is based
on the idea that the transverse gauge bosons acquire a magnetic
thermal mass of order $g^2 T$, which is the only relevant scale in the
problem, and therefore determines ({\ref{gv}) by dimensional analysis.
The origin of this scale can be readily understood \cite{guy} by the 
following argument.  We are looking for field configurations with
$N_{CS}\sim 1$, so using the definition of Chern-Simons number,
\beq
	\int d^{\,3}x( g^2 F A \hbox{\ or \ } g^3 A^3)\sim 1
\eeq
but the energy of the configuration is determined by the temperature,
\beq
	\int d^{\,3}x\, F^2 \sim T
\eeq
We can therefore estimate for a configuration of size $R$ that
\beqa
	R^3\left( g^2 {A^2\over R^2} \hbox{\ or\ } g^3 A^3 \right)
	&\sim& 1 \nonumber\\
	R^3\, {A^2\over R^2} &\sim& T
\eeqa
which fixes the size of $A$ and $R$, giving $R\sim 1/(g^2 T)$.
Hence the estimate that $\Gamma/V \sim (g^2 T)^4$.  However it was
later shown \cite{ASY} that this is not quite right---the time scale
for sphaleron transitions is actually $g^4 T$, not $g^2 T$, giving
$\Gamma/V \sim \alpha_w^5 T^4$. Careful measurements in lattice
gauge theory  fixed the dimensionless coefficient to be \cite{MHM}
 $\Gamma/V = (29\pm 6)\alpha_w^5 T^4$, and more recently \cite{BMR}
\beq
	{\Gamma\over V} = (25.4\pm 2.0) \alpha_w^5 T^4
	= (1.06\pm 0.08)\times 10^{-6} T^4
\eeq
Ironically the prefactor is such that the old $\alpha_w^4 T^4$
estimate was good using $c=1$.

We can now determine when sphalerons were in thermal equilibrium in
the early universe.  To compute a rate we must choose a relevant
volume.  We can take the thermal volume, $1/T^3$, which is the average
space occupied by a particle in the thermal bath.  Then
\beq
	\Gamma = 10^{-6} T
\eeq
which must be compared to the Hubble rate, $H\sim \sqrt{g_*} T^2/M_p$.
At very high temperatures, sphalerons are {\it out} of equilibrium,
as illustrated in figure \ref{fig9}.
They come into equilibrium when
\beqa
	\Gamma &=& H \quad\Rightarrow\quad 10^{-6} T = \sqrt{g_*}{T^2\over
	M_p}\nonumber\\
	T &=& 10^{-6} g_*^{-1/2} M_p\sim 10^{-5}M_p\sim 10^{13}{\rm GeV}
\eeqa
Hence GUT baryogenesis is affected by sphalerons, after the fact.
Of course when $T$ falls below the EWPT temperature $\sim 100$ GeV,
the sphaleron rate again falls below $H$.

\begin{figure}[h]
\centerline{
\includegraphics[width=0.75\textwidth]{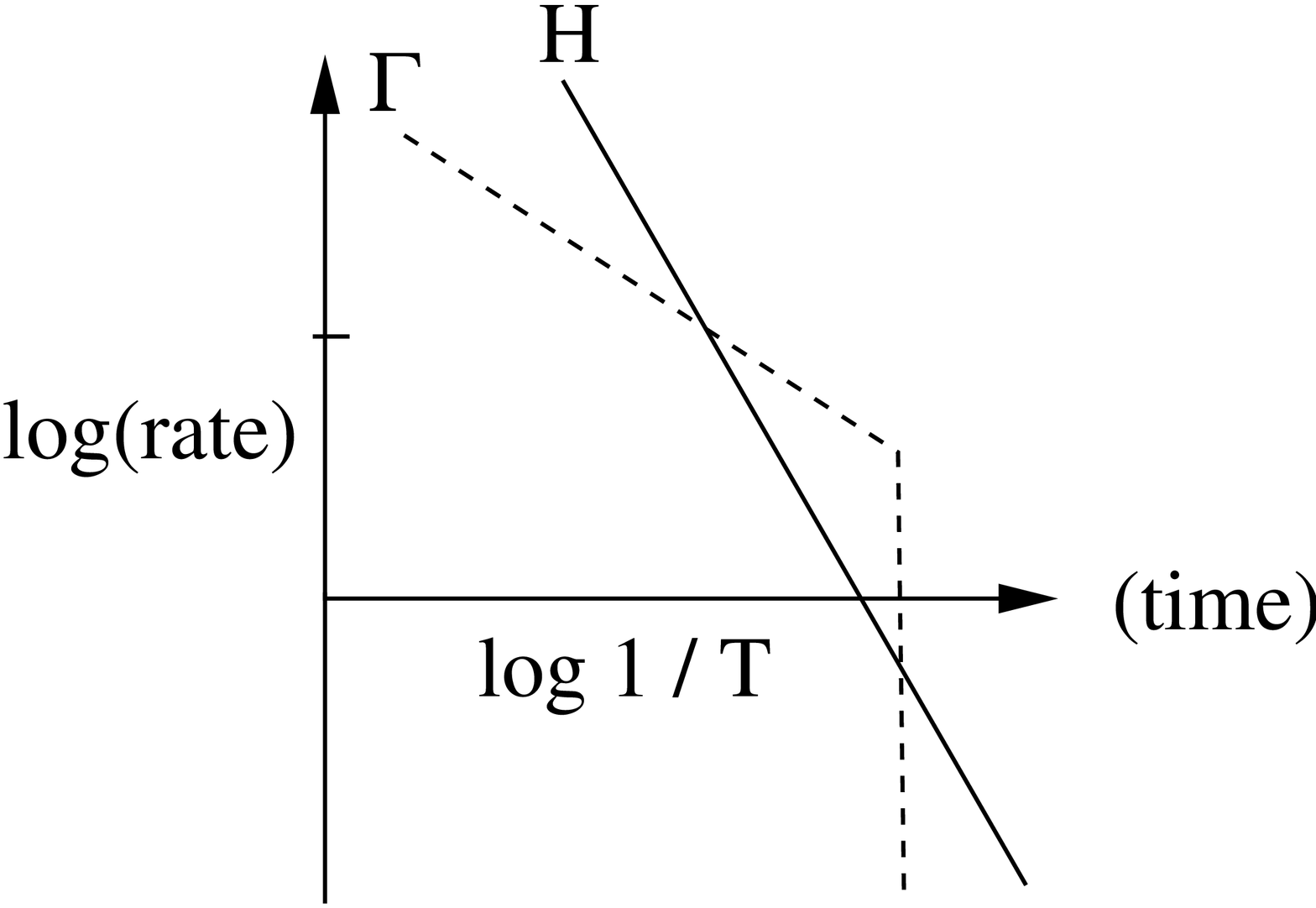}}
\caption{Sphaleron rate and Hubble rate versus time.  The sharp
drop in the sphaleron rate occurs at the EWPT.}
\label{fig9}
\end{figure}

\subsection{CP violation in the SM}
Since the SM provides such a strong source of B violation, it is
natural to wonder whether baryogenesis is possible within the SM.
We must investigate whether the other two criteria of Sakharov can
be fulfilled.

It is well known that CP violation exists in the CKM matrix of the 
SM,
\beq
	V_{CKM} = \left(\begin{array}{lll} V_{ud}& V_{us} & V_{ub}\\
	V_{cd} & V_{cs} & V_{cb} \\
	V_{td} & V_{ts} & V_{tb} \end{array}\right) = 
\left(\begin{array}{l|l|l} c_1 & -s_1 c_3 & - s_1 s_3\\
\hline
s_1 c_2 & {c_1 c_2 c_3\atop - s_2 s_3e^{i\delta}} &
{c_1 c_2 s_3 \atop + s_2 c_3 e^{i\delta}}\\
\hline
s_1 s_2 & {c_1 s_2 c_3\atop + c_2 s_3 e^{i\delta}} &
	{c_2 s_2 s_3 \atop - c_2 c_3 e^{i\delta}} \end{array}\right)
\eeq
This is the original parametrization, whereas in the Wolfenstein parametrization, $V_{ub}$ and $V_{td}$ contain the
CP-violating phase.  Where the phase resides in the CKM matrix can be changed by field redefinitions.
How do we express the invariant phase?  There is no unique answer:
it depends upon which physical process is manifesting the CP
violation.  However C.\ Jarlskog tried to address the question in a
rather general way \cite{Jarlskog}.  She showed that one possible
invariant is the combination
\beqa
	J &=& (m^2_t - m^2_c)(m^2_t-m^2_u)(m^2_c-m^2_u)\nonumber\\
	&& 
	\quad(m^2_b - m^2_s)(m^2_b-m^2_d)(m^2_s-m^2_d)\,K\nonumber\\
	\hbox{where\ }K &=& s_1^2 s_2 s_3 c_1 c_2 c_3 \sin\delta
	\nonumber\\
	&=& {\rm Im}\, V_{ii} V_{jj} V^*_{ij} V^*_{ji}
\hbox{\ for\ } {i\neq j}
\label{jeq}
\eeqa
This is derived by computing the determinant of the 
commutator of the up- and down-type quark mass matrices (squared):
\beq
	J = {\rm det}\left[ m^2_u,\, m^2_d\right]
\eeq
and therefore is invariant under rotations of the fields.

The form of $J$ has been used to argue that CP violation within the
standard model cannot be large enough for baryogenesis.  One should
find a dimensionless measure of the strength of CP violation, using
the relevant temperature of the universe, which must be at least
of order 100 GeV for sphalerons to be effective.  One then finds that
\beq
	{J\over (100 {\rm\ GeV})^{12}} \sim 10^{-20}
\label{jeq2}
\eeq
which is much too small to account for $\eta\sim 10^{-10}$.

One might doubt whether this argument is really robust.  For example,
in the original PRL of Jarlskog \cite{Jarlskog}, $J$ was defined
in terms of the linear quark mass matrices, which yielded a formula
like (\ref{jeq}), but with linear rather than squared mass
differences.  In the subsequent paper it was argued that actually
the sign of a fermion mass has no absolute physical significance, so
any physical quantity should depend on squares of masses.  However,
this seems to have no bearing on the mathematical fact that the
original linear-mass definition of $J$ is a valid invariant characterization of
the CP phase.  If this were the physically correct definition, then
(\ref{jeq2}) would be revised to read $J\sim 10^{-10}$, which is in
the right ballpark.  

A more specific criticism of the argument was given in \cite{FG}, who
noted that the argument cannot be applied  to $K\bar K$
mixing in the neutral kaon system, coming from the box diagram\\
\centerline{\includegraphics[width=0.3\textwidth]{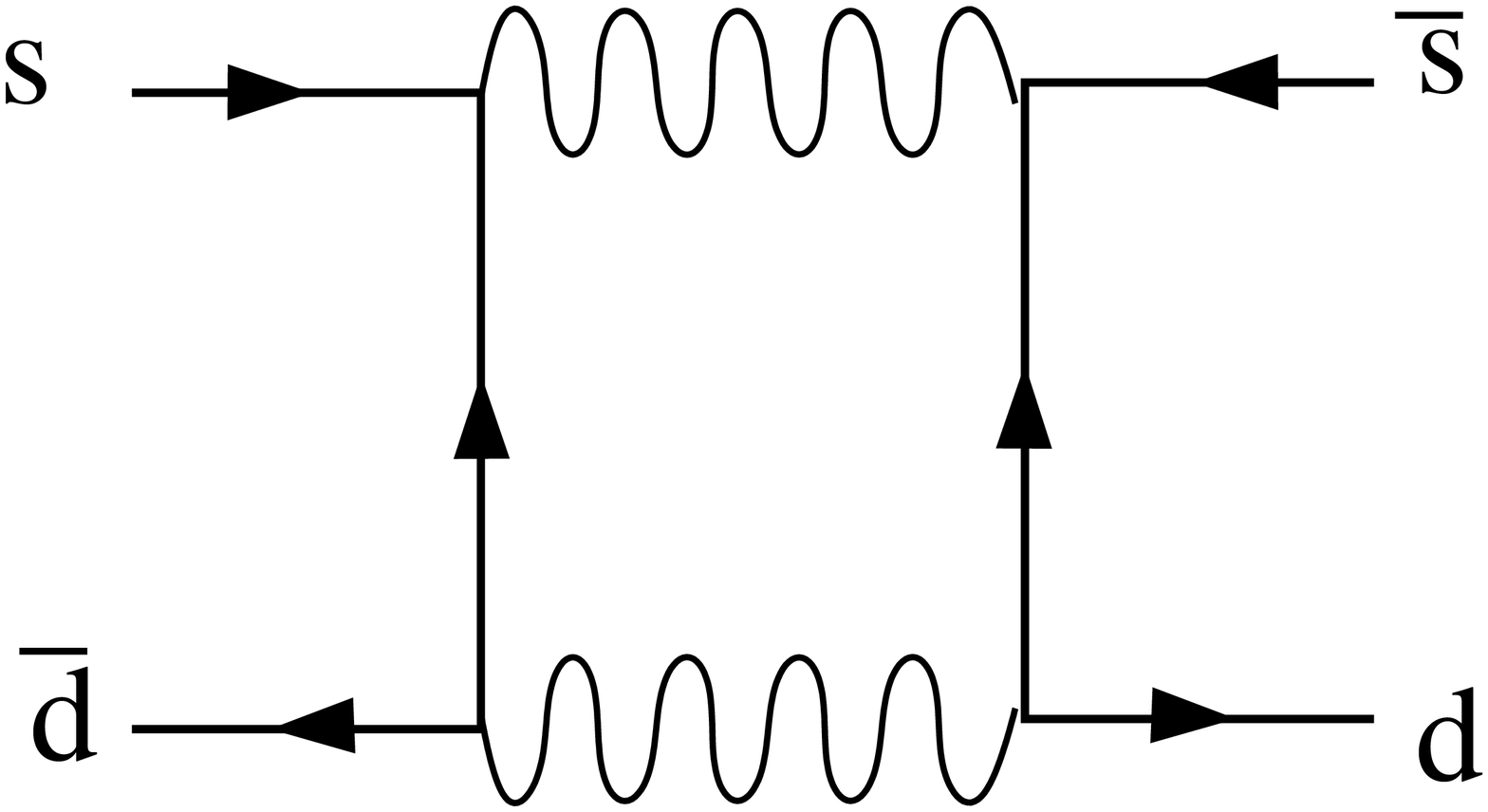}} Farrar
and Shaposhnikov point out that this CP-violating effect is not
proportional to $J$, and that the relevant scale is much  smaller
than 100 GeV---it is the mass of $K^0$.  The idea that $J/T^{12}$ is
the correct measure of CP violation only makes sense if all the
ratios of mass to temperature can be treated as perturbatively
small.  This is clearly not the case for the ratio of the top quark
mass to the $K^0$ mass.  Could there not also be some scale lower
than 100 GeV playing a role in the baryogenesis mechanism?  Ref.\
\cite{FG} attempted to construct such a mechanism within the SM,
which was refuted by \cite{Gavela}.  There is no theorem proving that
it is impossible to find some other mechanism that does work, but so
far there are no convincing demonstrations, and most practitioners of
baryogenesis agree that CP violation in the SM is too weak; one needs
new sources of CP violation and hence new physics beyond the SM. 
We will see that it is rather easy to find such new sources; for
example the MSSM has many new phases, such as in the term
$W = \mu H_1 H_2$ in the superpotential, and in the gaugino masses.

\section{Electroweak Phase Transition and Electroweak Baryogenesis}
In the previous section we showed that B violation is present in the
standard model, and new sources of CP violation are present in
low-energy extensions of the SM.  But the remaining requirement, 
going out of thermal equilibrium, is not so easy to achieve at low
energies.  Recall the condition for out-of-equilibrium decay,
(\ref{ooe}), 
\beq
	\alpha \ll {m\over M_p}
\eeq 
If $m\sim 100$ GeV, we require extraordinarily weak couplings which
are hard to justify theoretically, and which would also be hard to
verify experimentally.

But the EWPT can provide a departure from thermal equilibrium if it is
sufficiently strongly first order.  The difference between first and
second order is determined by the behavior of the Higgs potential
at finite temperature, as shown in figure \ref{fig10}.  In a first
order transition, the potential develops a bump which separates
the symmetric and broken phases, while in a second order transition
or a smooth cross-over there is no bump, merely a change in sign of
the curvature of the potential at $H=0$.  The critical temperature
$T_c$ is defined to be the temperature at which the two minima
are degenerate in the first order case, or the temperature at which
$V''(0)=0$ in the second order case.

\begin{figure}[h]
\centerline{
\includegraphics[width=1\textwidth]{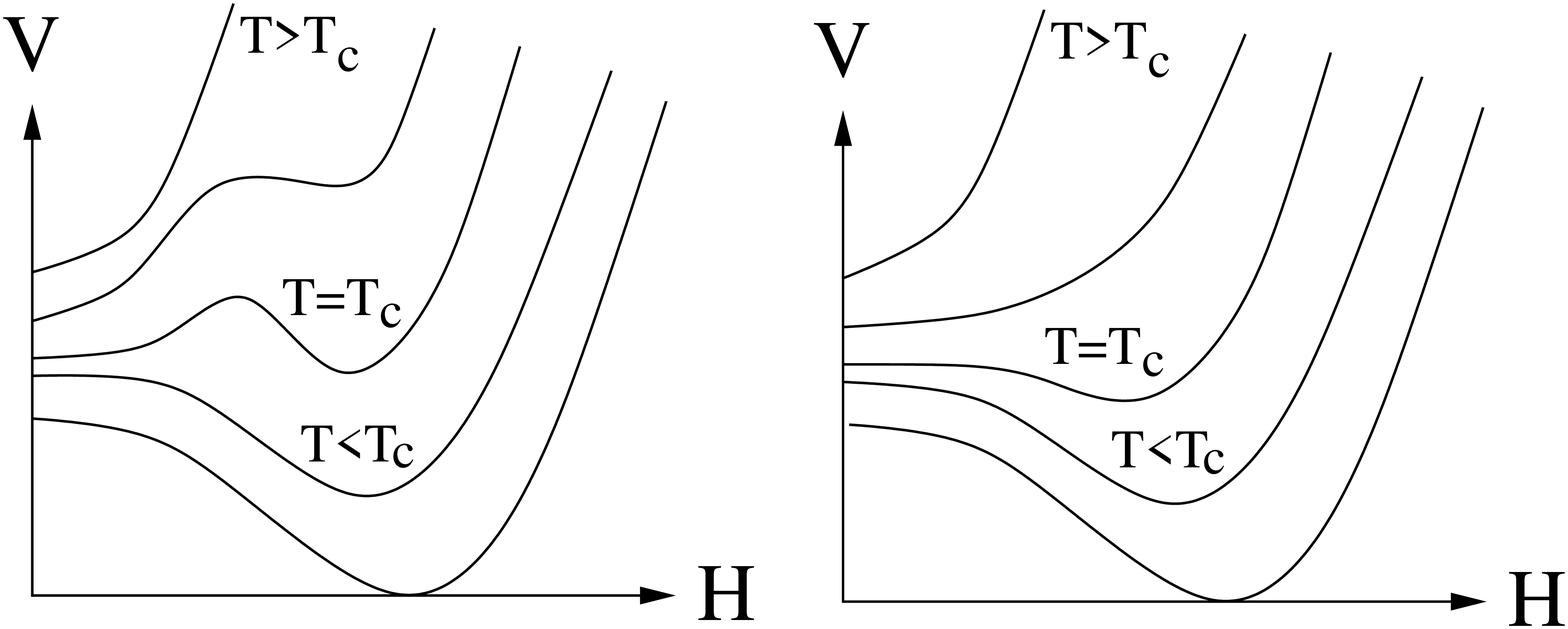}}
\caption{Schematic illustration of Higgs potential evolution
with temperature for first (left) and second (right) order
phase transition.}
\label{fig10}
\end{figure}

A first order transition proceeds by bubble nucleation (fig.\ 
\ref{fig11}), where inside the bubbles the Higgs VEV and
particle masses are nonzero, while they are still vanishing in the
exterior symmetric phase.  The bubbles expand to eventually collide
and fill all of space.  If the Higgs VEV $v$ is large enough inside
the bubbles, sphalerons can be out of equilibrium in the interior
regions, while still in equilibrium outside of the bubbles.  A rough
analogy to GUT baryogenesis is that sphalerons outside the bubbles
correspond to B-violating $Y$ boson decays, which are fast, while
sphalerons inside the bubbles are like the B-violating inverse Y
decays.  The latter should be slow; otherwise they will relax the
baryon asymmetry back to zero.  

\begin{figure}[h]
\centerline{
\includegraphics[width=0.9\textwidth]{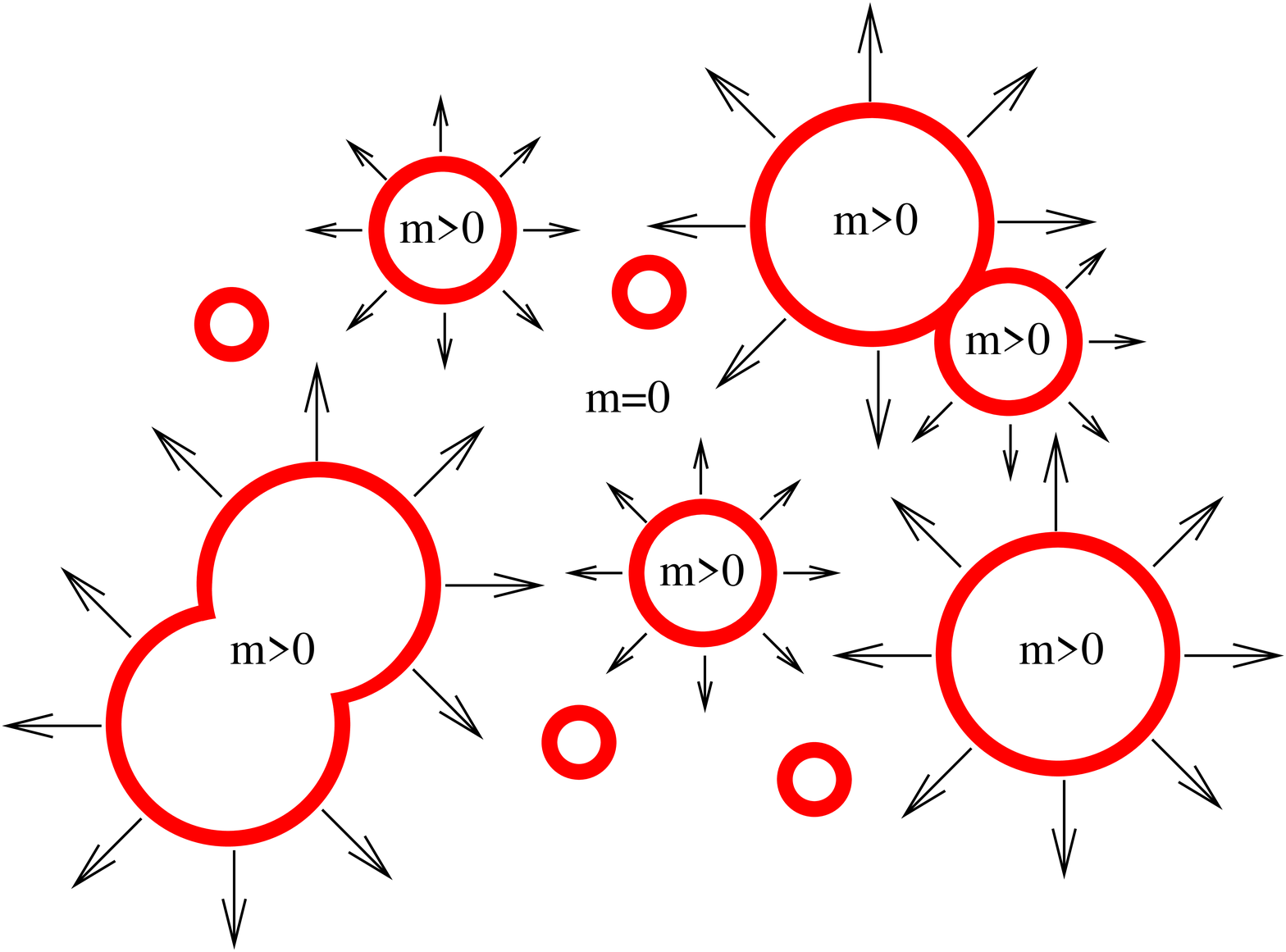}}
\caption{Bubble nucleation during a first-order EWPT.}
\label{fig11}
\end{figure}	

In a second order EWPT, even though the sphalerons go from being
in equilibrium to out of equilibrium, they do so in a continuous way,
and uniformly throughout space.  To see why the difference
between these two situations is important, we can sketch the basic
mechanism of electroweak baryogenesis, due to Cohen, Kaplan and Nelson
\cite{CKN}.  The situation is illustrated in figure \ref{fig12},
which portrays a section of a bubble wall moving to the right.
Because of CP-violating interactions in the bubble wall, we get
different amounts of quantum mechanical reflection of right- and
left-handed quarks (or of quarks and antiquarks).  This leads to
a chiral asymmetry in the vicinity of the wall.  There is an
excess of $q_L + \bar q_R$ relative to $q_R + \bar q_L$ in front of
the wall, and a compensating deficit of this quantity on the other
side of the wall.  This CP asymmetry is schematically shown in 
figure \ref{fig13}.  

\begin{figure}[h]
\centerline{
\includegraphics[width=0.6\textwidth]{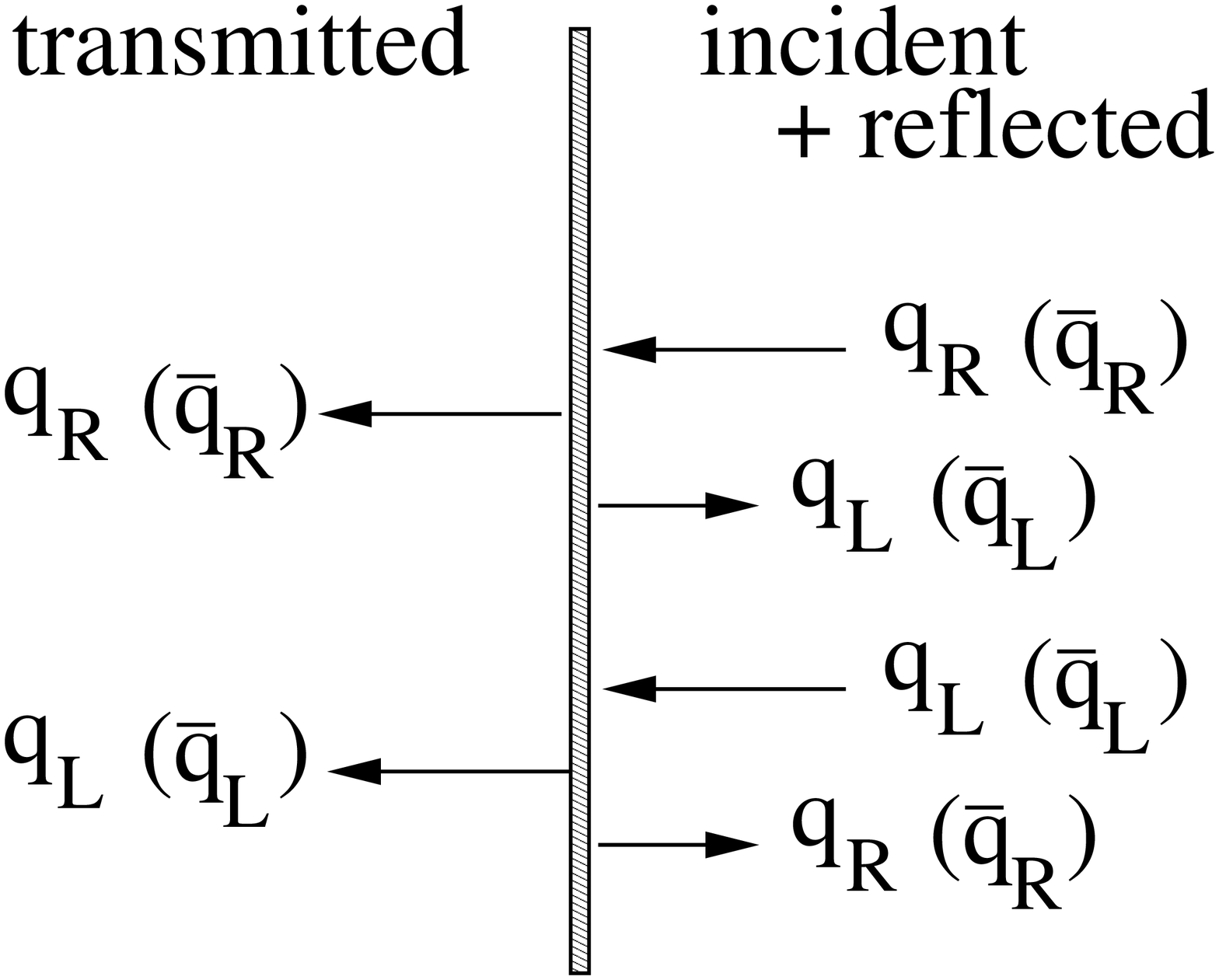}}
\caption{CP-violating reflection and transmission of quarks
at the moving bubble wall.}
\label{fig12}
\end{figure}	

\begin{figure}[h]
\centerline{
\includegraphics[width=0.7\textwidth]{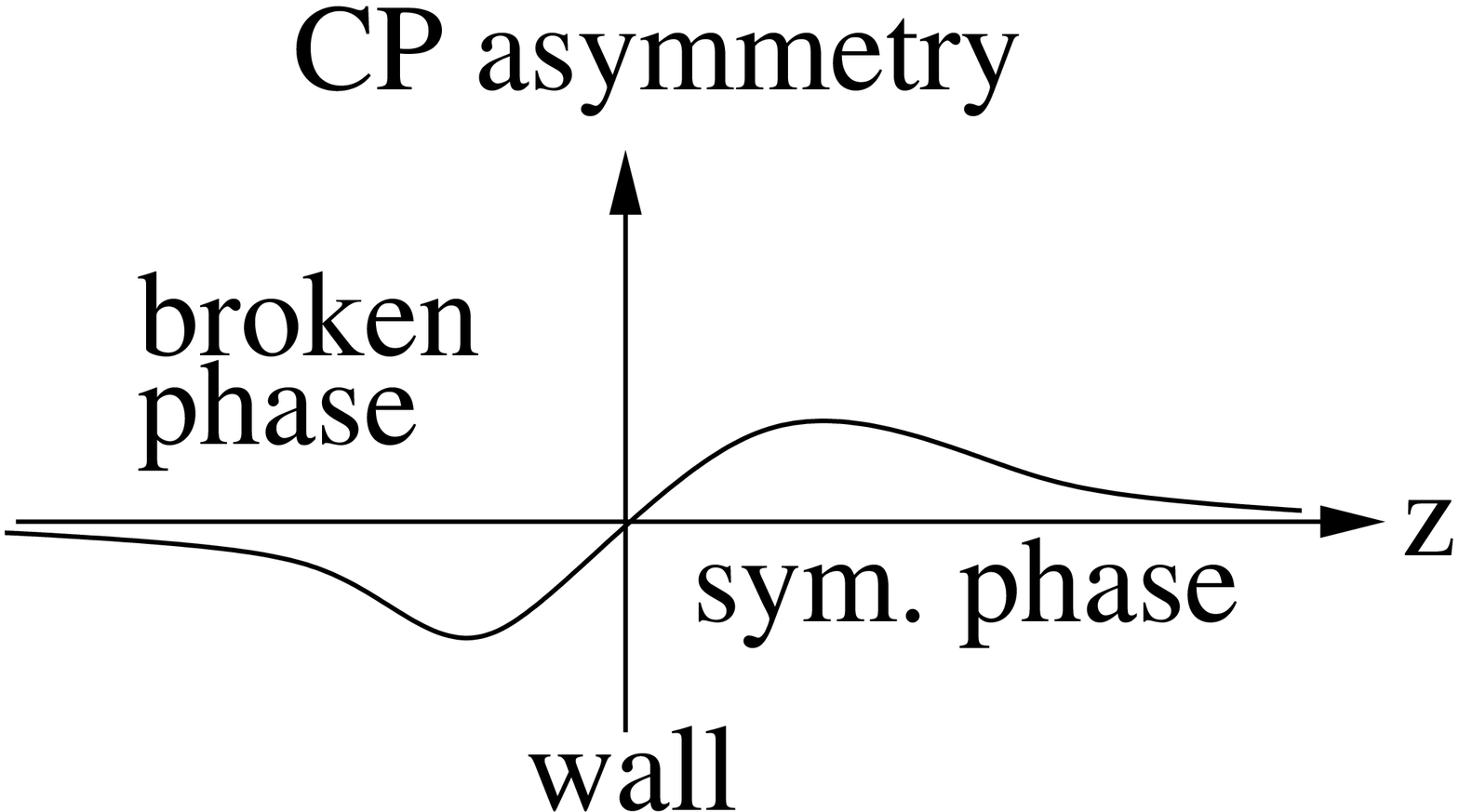}}
\caption{The CP asymmetry which develops near the bubble wall.}
\label{fig13}
\end{figure}

Sphalerons interact only with $q_L$, not $q_R$, and they	try to
relax the CP-asymmetry to zero.  Diagramatically, \\
\includegraphics[width=1\textwidth]{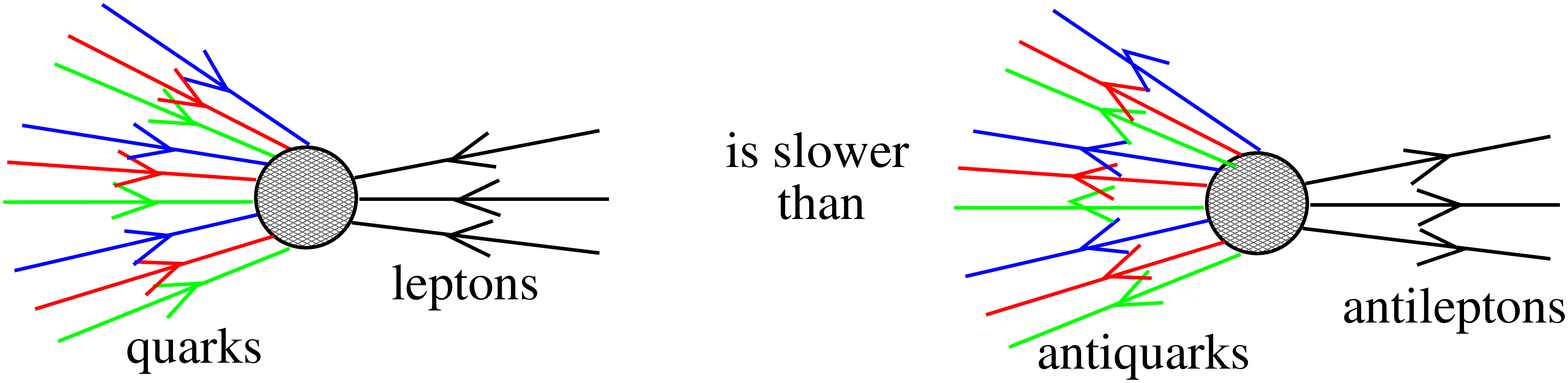}}
simply because there are more $\bar q_L$ than $q_L$ in front of the
wall.  But the first interaction violates baryon number by $-3$
units while the second has $\Delta B = 3$.  Therefore the CP asymmetry
gets converted into a baryon asymmetry in front of the wall
(but not behind, since we presume that sphaleron interactions are
essentially shut off because of the large Higgs VEV). 
Schematically the initial baryon asymmetry takes the form of figure
\ref{fig13a}.

\begin{figure}[h]
\centerline{\includegraphics[width=0.5\textwidth]{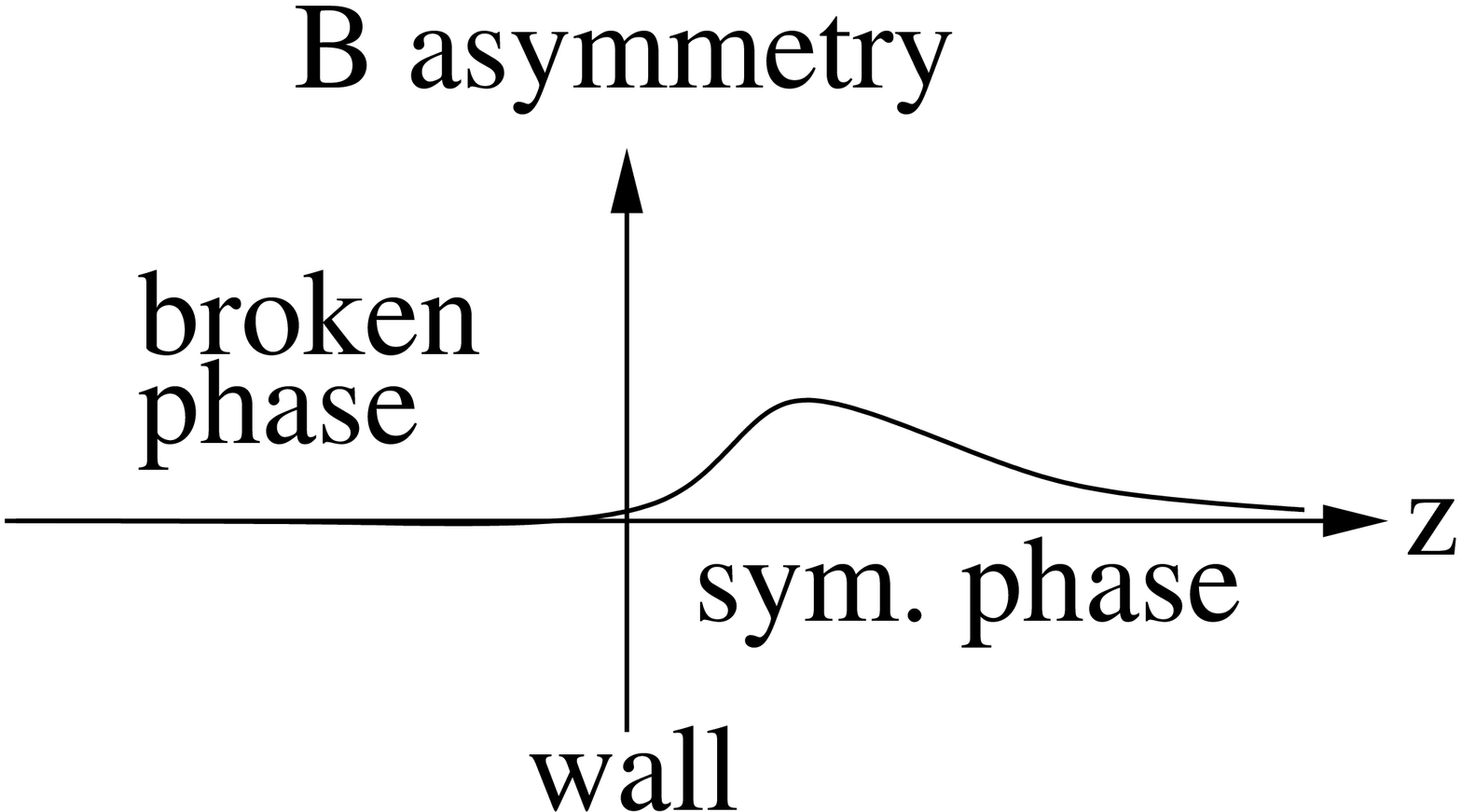}}
\caption{The B asymmetry which initially develops in front of the bubble wall.}
\label{fig13a}
\end{figure}

\begin{figure}[h]
\centerline{
\includegraphics[width=0.7\textwidth]{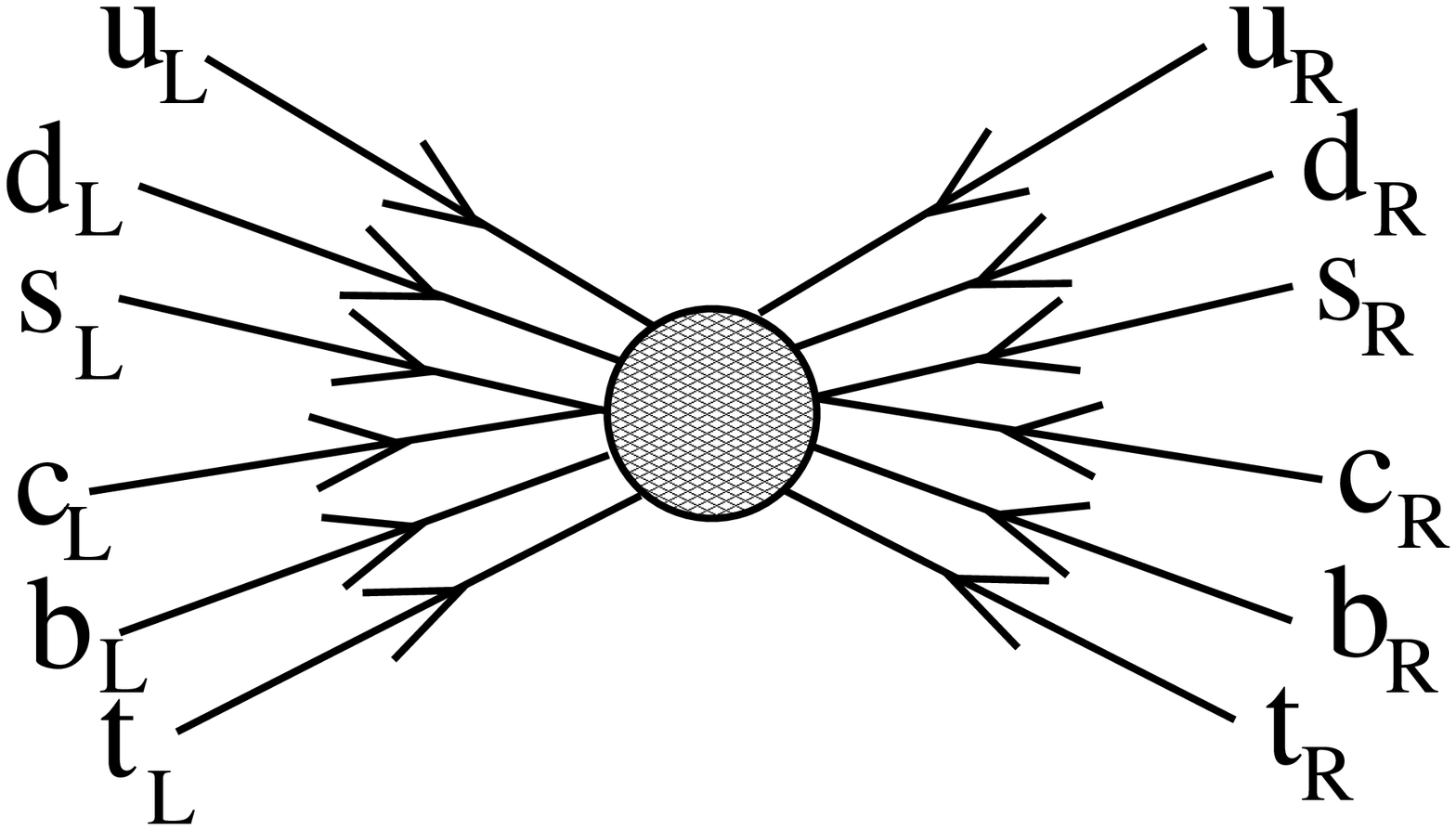}}
\caption{The strong sphaleron.}
\label{fig14}
\end{figure}

If the baryon asymmetry remained in front of the wall, eventually
sphalerons would cause it to relax to zero, because there are other
processes besides sphalerons in the plasma which can relax the CP
asymmetry, for example {\it strong} SU(3) sphalerons which change
chirality $Q_5$ by 12 units, 2 for each flavor, as shown in figure
\ref{fig14}.  (See \cite{GDM} for a lattice computation of the strong
sphaleron rate.)  The combination of weak and strong sphalerons
would relax $Q_5$ and $B+L$ to zero if the wall was not moving.
But due to the wall motion, there is a tendency for baryons to diffuse
into the broken phase, inside the bubble.  If $E_{\rm sph}/T$ is
large enough, $\Gamma_{\rm sph}$ is out of equilibrium and $B$
violation is too slow to relax $B$ to zero.  This is the essence
of electroweak baryogenesis.

\subsection{Strength of the phase transition}
Now that we appreciate the importance of the first order phase
transition, we can try to compute whether it occurs or not.
The basic tool for doing so is the finite-temperature effective
potential of the Higgs field, defined by the path integral
\beq
	e^{-\beta \int d^{\,3}x\, V_{\rm eff}(H)} 	
	= \int \prod_i{\cal D} \phi_i e^{-\int_0^\beta d\tau
	\int d^{\,3}x\, {\cal L}[H,\phi_i]}
\eeq
where $\beta = 1/T$, $H$ is the background Higgs field, and $\phi_i$
are fluctuations of all fields which couple to the Higgs, including
$H$ itself.  Here $\tau$ is imaginary time and the fields are given
periodic (antiperiodic) boundary conditions between $\tau=0$ and $\beta$ if
they are bosons (fermions).  For a compact introduction to field theory
at finite temperature, see ref.\ \cite{Kapusta}.

To evaluate $V{\rm eff}$, we can use perturbation theory.  Since there
are no external legs,  $V_{\rm eff}$ is given by a series of
vacuum bubbles, \\
\centerline{\includegraphics[width=0.7\textwidth]{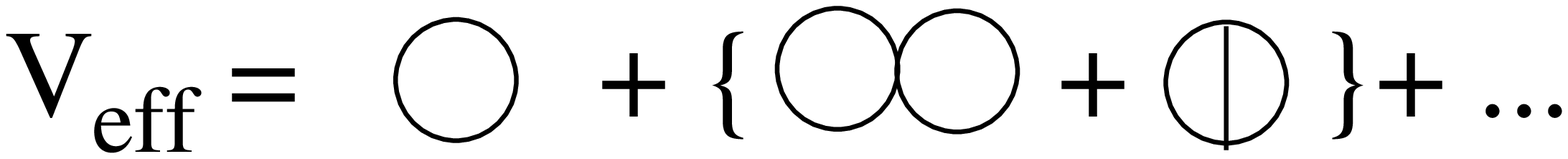}}
The one-loop contribution has a familiar form,
\beq
	V_{\rm 1-loop} = T \sum_i\pm \int {d^{\,3}p\over (2\pi)^3}
	\ln\left(1 \mp e^{-\beta\sqrt{p^2+m^2_i(H)}}\right)
	\ \left\{ {\hbox{bosons}\atop\hbox{fermions}}\right.
\label{v1l}
\eeq
This is the free energy of a relativistic gas of bosons or
fermions.  Recall that the partition function for a free
nonrelativistic gas of $N$ particles in a volume $V$ is
\beqa
F &=& -T\ln Z;  \nonumber\\
Z &=& \left({V\over h} \int {d^{\,3}p}\, e^{-\beta p^2/2m}\right)^N
\eeqa
Eq.\ (\ref{v1l}) is the relativistic generalization, taking into
account that a fermion loop comes with a factor of $-1$.  The
nontrivial aspect is that we must evaluate the particle masses as a
function of the Higgs VEV $H$.  

In the standard model, the important particles contributing to the 
thermal partition function, and their field-dependent masses, are
\beqa
	\hbox{top quark:\ \ } m_t &=& y H \\
	\hbox{gauge bosons,\ \ }W^\pm, W^3, B:\ m^2&=& \frac12 H^2\left(\begin{array}{llll}
	g^2 & & & \\ & g^2 & & \\ & & g^2 & g g' \\
	& & g g' & g^2 \end{array}\right)
\eeqa
(One must of course find the eigenvalues of the gauge boson mass
matrix.)
For the Higgs bosons, we need to refer to the tree-level Higgs
potential
\beq
	V = \lambda\left(|H|^2 -\frac12 v^2\right)^2 =
	\frac14\lambda\left(\phi^2 + \sum_i\chi_i^2 - v^2\right)^2
\eeq
where the real components of the Higgs doublet are
$H = \frac{1}{\sqrt{2}}(\phi + i\chi_1,\ \chi_2 + i\chi_3)$.
$\phi$ is the massive component, while $\chi_i$ are the Goldstone
bosons which get ``eaten'' by the $W$ gauge bosons.  In the effective
potential, we set $\chi_i=0$ so that the VEV is given by 
$H = \frac{1}{\sqrt{2}}\phi$, but to find the masses of the Goldstone
bosons, we must temporarily include their field-dependence in $H$,
take derivatives with respect to $\chi_i$, and then set $\chi_i=0$.
Then
\beqa
	\hbox{Higgs boson,\ \ }H: m^2_H &=& {\partial^2 V\over
	\partial\phi^2}
	= \lambda(3\phi^2-v^2) = \lambda(6H^2 - v^2)\\
	\phantom{.\!\!\!\!\!\!\!\!\!}\hbox{Goldstone bosons,\ \ }\chi_i: m^2_i &=&
	{\partial^2 V\over \partial\chi_i^2} = \lambda(\phi^2-v^2)
	=\lambda (2H^2 - v^2)
\eeqa
We must use these field-dependent masses in the 1-loop thermal
effective potential.  Unfortunately the momentum integrals cannot
be done in any enlightening closed form, except in the high- or
low-temperature limits.  We will be interested in the high-$T$ case,
where $V_{\rm 1-loop}$ can be expanded as 
\beqa
	V_{\rm 1-loop} &=& \sum_{i\in B,F} {m^2_i T^2\over 48}
	\times\left\{ {2,\hbox{\ each real B}\atop 4,\hbox{\ each
	Dirac F}}\right\} - {m^3_i T\over 12\pi} \left\{
	{1,\hbox{\ B}\atop 0,\hbox{\ F}}\right\} \nonumber\\
	&+& {m_i^4\over 64\pi^2}\left(\ln{m_i^2\over T^2} - c_i\right)
	\times \left\{ {-1,\hbox{\ B}\atop +4,\hbox{\ Dirac
	F}}\right\}
	+ O\left(m_i^5\over T\right)
\eeqa
where 
\beq
	c_i = \left\{ {\frac32 + 2\ln 4\pi - 2\gamma_E \cong 5.408,
	\hbox{\ B}\atop c_B - 2\ln 4 \cong 2.635,\hbox{\ F}}\right.
\eeq
Let's examine the effects of these terms, starting with the term
of order $m_i^2 T^2$.  This is the source of a $T$-dependent squared mass
for the Higgs boson, and explains why $V''(0)>0$ at high $T$.
Focusing on just the $H^2$ terms, the contributions to the potential
are
\beqa
	V_{\rm 1-loop} &=& {H^2 T^2}\left[\frac12\lambda
 +\frac3{16}(3g^2+ g'^2) +
	\frac14 y^2\right]  + O(H^3)\nonumber\\
	&\equiv& a T^2 H^2 + O(H^3)\\
	V_{\rm tree} &=& -\lambda v^2 H^2 + O(H^4)
\eeqa
Putting these together, we infer that the $T$-dependent Higgs mass
is given by
\beq
	m_H^2(T) = -\lambda v^2 + a T^2
\eeq
and we can find the critical temperature of the phase transition
(if it was second order) by solving for $m^2_H=0$:
\beq
	T_c = \sqrt{\lambda\over a}\, v \cong 2{\sqrt{\lambda}\over y} v
\eeq
If the transition is first order, it is due to the ``bump'' in the
potential, which can only come about because of the $H^3$ term
in $V_{\rm 1-loop}$.  The cubic term is
\beq
	V_{\rm cubic} \cong - {T H^3\over 12\sqrt{2}\pi}\left
	(3 g^3 + \frac32(g^2+g'^2)^{3/2} + O(\lambda)\right)
	\equiv -E T H^3
\eeq
where the $O(\lambda)$ contributions have been neglected, since we
will see that $\lambda\ll g^2$ is necessary for having a strongly 
first order phase transition.  

With the cubic term, the potential takes the form
\beq
	V_{\rm tot} \cong m_H^2(T) H^2 - E T H^3 + \lambda H^4
\label{pot1}
\eeq
and at the critical temperature it becomes
\beq
	V_{\rm tot} = \lambda H^2 \left( H - {v_c\over \sqrt{2}} 
	\right)^2
\label{pot2}
\eeq
since this is the form which has degenerate minima at $H=0$ and $H=
v_c$.  Notice that the VEV at the critical temperature, $v_c$, is not
the same as the zero-temperature VEV, $v$.   By comparing (\ref{pot1})
and  (\ref{pot2}) we find that
\beq
	m^2(T) = \frac12\lambda v_c^2,\quad E T_c = \sqrt{2}\lambda
	v_c
\eeq
which allows us to solve for $v_c$ and $T_c$.  The result is
\beq
	T_c = {\lambda v\over \sqrt{a\lambda - \frac14 E^2}}
\eeq
and
\beq
	{v_c\over T_c} = {E\over 2\sqrt{\lambda}}
	\cong {3 g^3\over 16\pi\lambda}
\label{vT}
\eeq

The ratio $v_c/T_c$ is a measure of the strength of the phase
transition.  It determines how strongly the sphalerons are suppressed
inside the bubbles.  Recall that 
\beq
	{E_{\rm sph}\over T}\sim {8\pi\over g}\, \left(v\over T\right)
\eeq
appears in $\Gamma_{\rm sph} \sim e^{-E_{\rm sph}/T}$, so the bigger
$v/T$ is at the critical temperature, the less washout of the baryon
asymmetry will be caused by sphalerons.  The total dilution of the
baryon asymmetry inside the bubbles is the factor
\beq
	e^{-\int_{t_c}^\infty \Gamma_{\rm sph} dt}
\eeq
where $t_c$ is the time of the phase transition.  
Requiring this to be not too small gives a bound \cite{Shap}
\beq
	{v_c\over T_c}\gsim 1
\label{sphbnd}
\eeq
which we can use to infer a bound on the Higgs mass \cite{Boch},
by combining (\ref{vT}) and (\ref{sphbnd}).  Given that
\beq
	\alpha_w = {g^2\over 4\pi}\cong {8 M_w^2 G_F\over
4\sqrt{2}\pi} \cong {1\over 29.5}
\eeq
the weak coupling constant is $g = 0.65$, so (\ref{sphbnd}) implies
$\lambda < 3 g^3/16\pi = 0.128$, and therefore
\beq
	m_H = \sqrt{\lambda\over 2}\, v < 32{\rm \ GeV}
\eeq 
This is far below the current LEP limit $m_H> 115$ GeV, so
it is impossible to have a strongly first order electroweak
phase transition in the standard model.

\begin{figure}[h]
\centerline{
\includegraphics[width=0.7\textwidth]{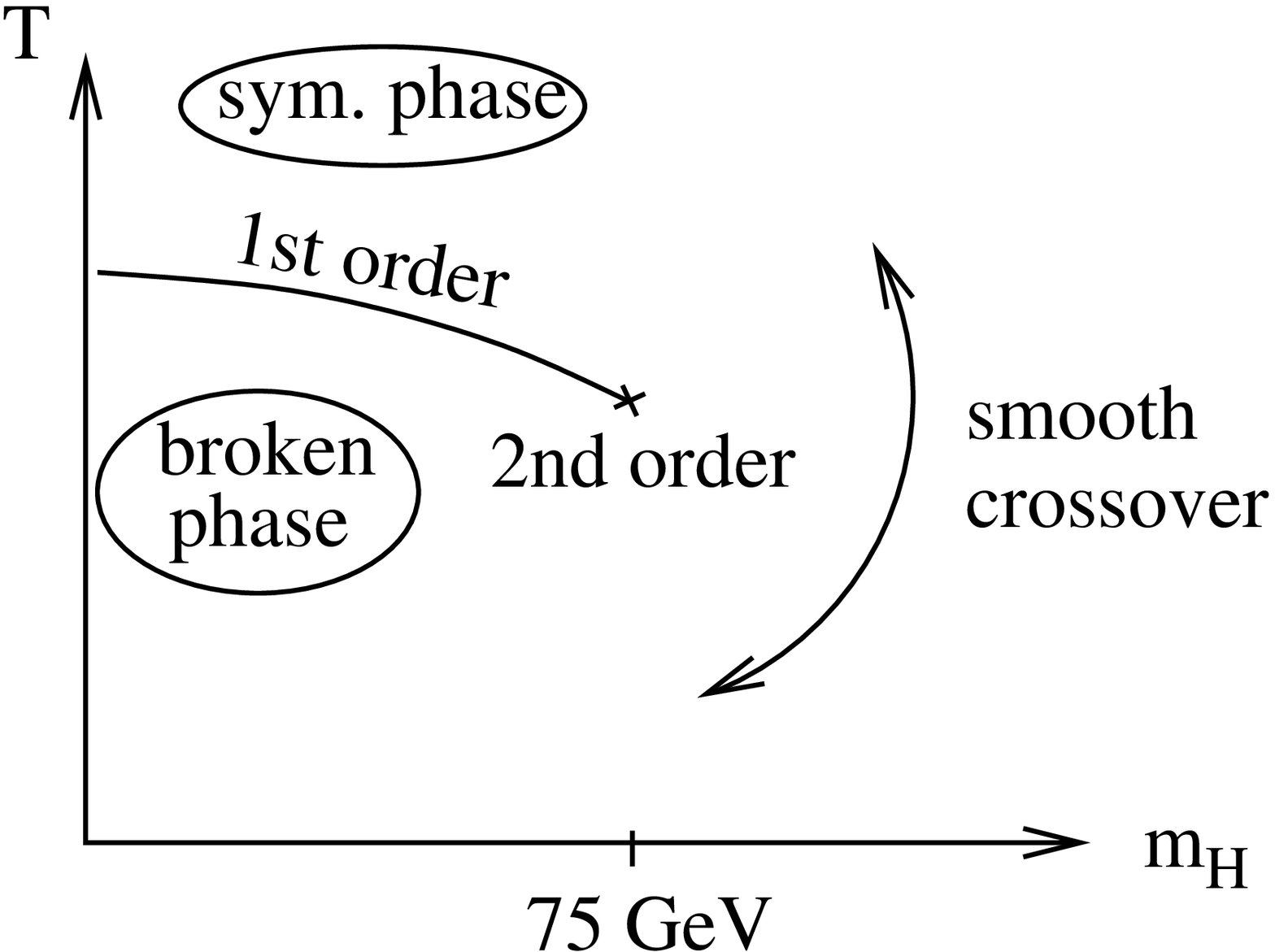}}
\caption{Phase diagram for the EWPT, from ref.\ \cite{KLRS}}
\label{fig15}
\end{figure}

The expressions we have derived give the impression that, regardless of the value of the Higgs mass, the phase transition
is first order, even if very weakly.   This is not the case, as
explained by P.\ Arnold \cite{Arnold}: at high temperatures, the
perturbative expansion parameter is not just made from dimensionless
couplings, but rather 
\beq
	{g^2 T\over m_W}\sim {\lambda\over g^2}\sim \left.{m^2_H\over
	m^2_W}\right|_{T=0}
\eeq
due to the fact that the high-$T$ theory is effectively 3-dimensional,
and is more infrared-sensitive to the gauge boson masses than is the
4D ($T=0$) theory.  In the regime of large $\lambda$, nonperturbative
lattice methods must be used to study the phase transition.  It was
found in \cite{KLRS} that in the $T$-$m_H$ phase diagram, there is a
line of 1st order phase transitions ending at a critical point
with a 2nd order transition at $m_H = 75$ GeV.  For larger $m_H$,
there is no phase transition at all, but only a smooth crossover.  
This is illustrated in figure \ref{fig15}.  

Most attempts to improve this situation have relied on new scalar
particles coupling to the Higgs field.  Ref.\ \cite{AH} considered
a singlet field $S$ with potential
\beq
	{V} = 2\zeta^2|H|^2|S|^2 + \mu^2 |S|^2 
\eeq
including a large coupling to the Higgs field, so 
\beq
	m_s^2 = \mu^2 + 2\zeta^2 |H|^2
\eeq
If $\mu^2$ is not too large ($\mu\lsim 90$ GeV)    then the cubic
term $-\frac{1}{12\pi}T(\mu^2 + 2\zeta^2H^2)^{3/2}$ is sufficiently
enhanced so that $v/T\sim 1$ for heavy Higgs.  If $\mu^2$ is too
large, this does not function as a cubic term, since it can be
Taylor-expanded in $H^2$.  

\subsection{EWPT in the MSSM}

\begin{figure}[h]
\centerline{
\includegraphics[width=0.5\textwidth]{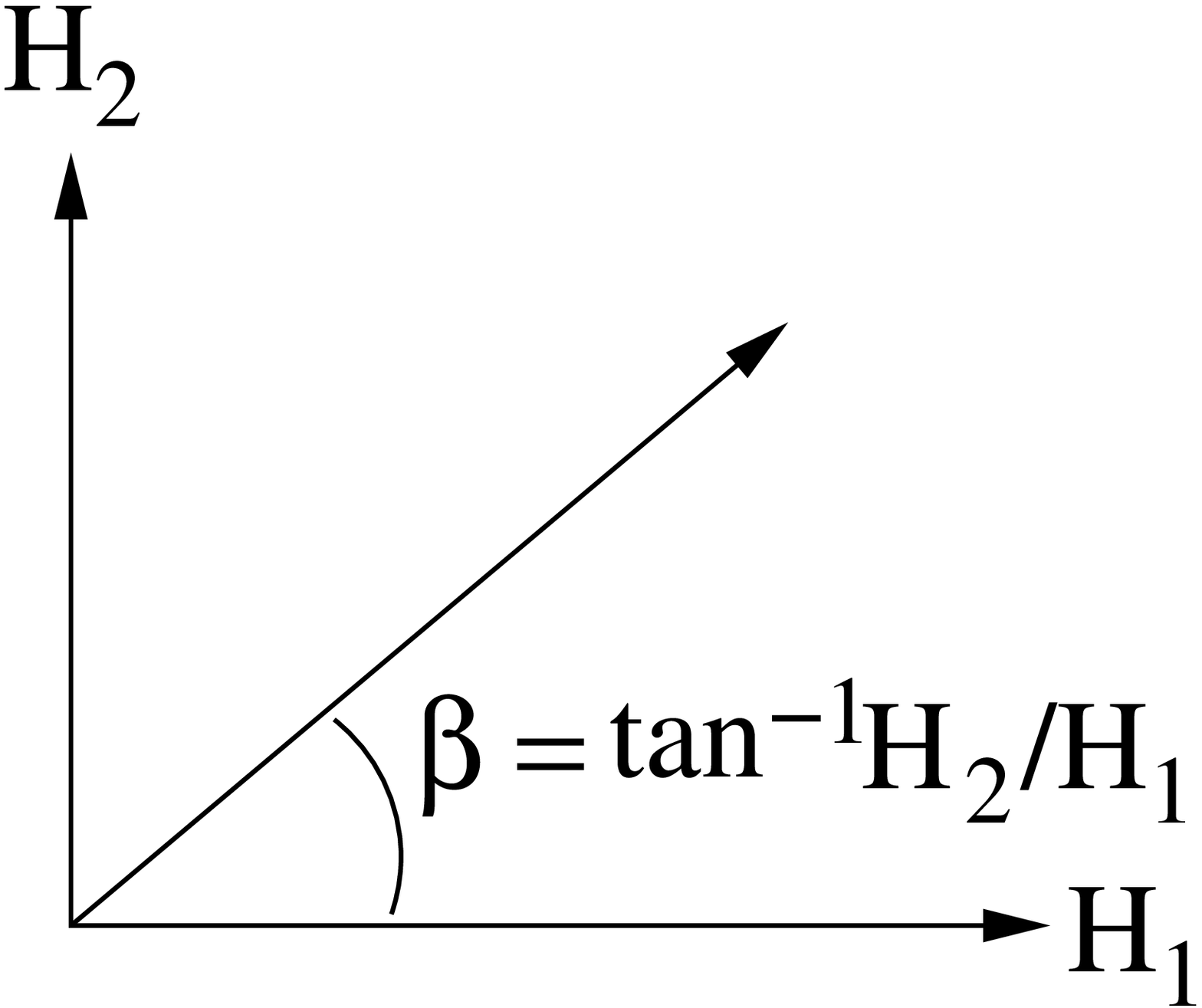}}
\caption{Light effective Higgs direction in the MSSM (in the limit
where the $A^0$ is very heavy).}
\label{fig17}
\end{figure}

It is possible to implement this mechanism in the MSSM because
top squarks (stops) couple strongly to the Higgs \cite{CQW}.
The MSSM is a model   with two Higgs doublets, $H_1$ and $H_2$,
of which one linear combination is relatively light, and plays the
role of the SM Higgs, as illustrated in figure \ref{fig17}.

\begin{figure}[h]
\centerline{
\includegraphics[width=0.5\textwidth]{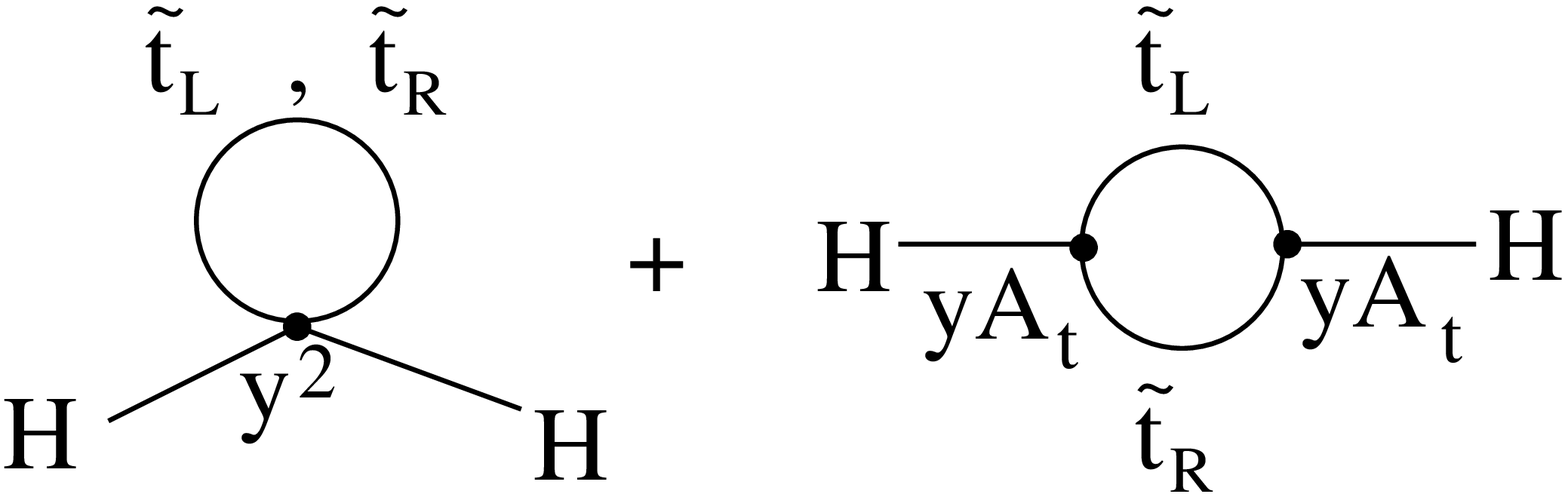}}
\caption{Stop loop contributions to the Higgs mass.}
\label{fig18}
\end{figure}

The stops come in two kinds, $\tilde t_L$ and
$\tilde t_R$, and have mass matrix
\beq
	m^2_{\tilde t} \cong\left(\begin{array}{ll} m^2_Q + y^2 H_2^2
	& y(A^*_t H_2^* + \mu H_1)\\ y(A_t H_2 + \mu^* H_1^*) &
	m^2_U + y^2 H_2^2 \end{array} \right)
\eeq
where $m^2_Q$, $m^2_U$ and $A_t$ are soft SUSY-breaking parameters.
If we ignore the off-diagonal terms (possible if we choose
$A_t\tan\beta = \mu$) it is easier to see what is happening---there
are two unmixed bosons coupling to $H$.  However, we cannot make both
of them light.  This is because $m^2_H$ gets major contributions from
stops at 1 loop, from the diagrams shown in fig.\ \ref{fig18}.  

\noindent These give
\beq
	m^2_H \cong m^2_Z + c {m_t^4\over v^2}\ln\left(
	m_{\tilde t_L} m_{\tilde t_R} \over m^2_t\right)
\label{hmass}
\eeq
If both $\tilde t_L$ and $\tilde t_R$ are light, it is impossible to
make the Higgs heavy enough to satisfy the LEP constraint.  Precision
electroweak constraints dicatate that  $\tilde t_L$ should be the
heavy one.  Otherwise corrections to the $\rho$ parameter,\\
\centerline{\includegraphics[width=\textwidth]{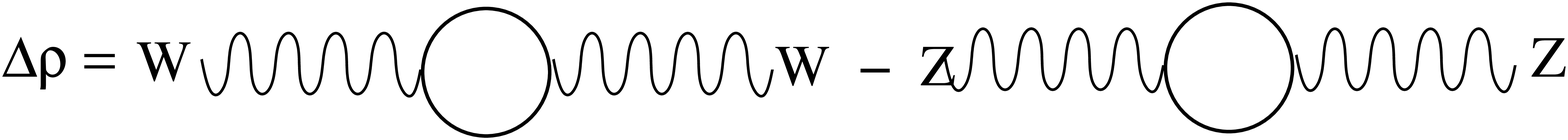}} 
are too large, since $\tilde t_L$ couples more strongly to $W$ and $Z$
than does $\tilde t_R$.

Although we have only discussed thermal field theory at one loop, 
two-loop effects are also important in the MSSM.  The  $\tilde t_R$ 
diagrams\\
\centerline{\includegraphics[width=0.5\textwidth]{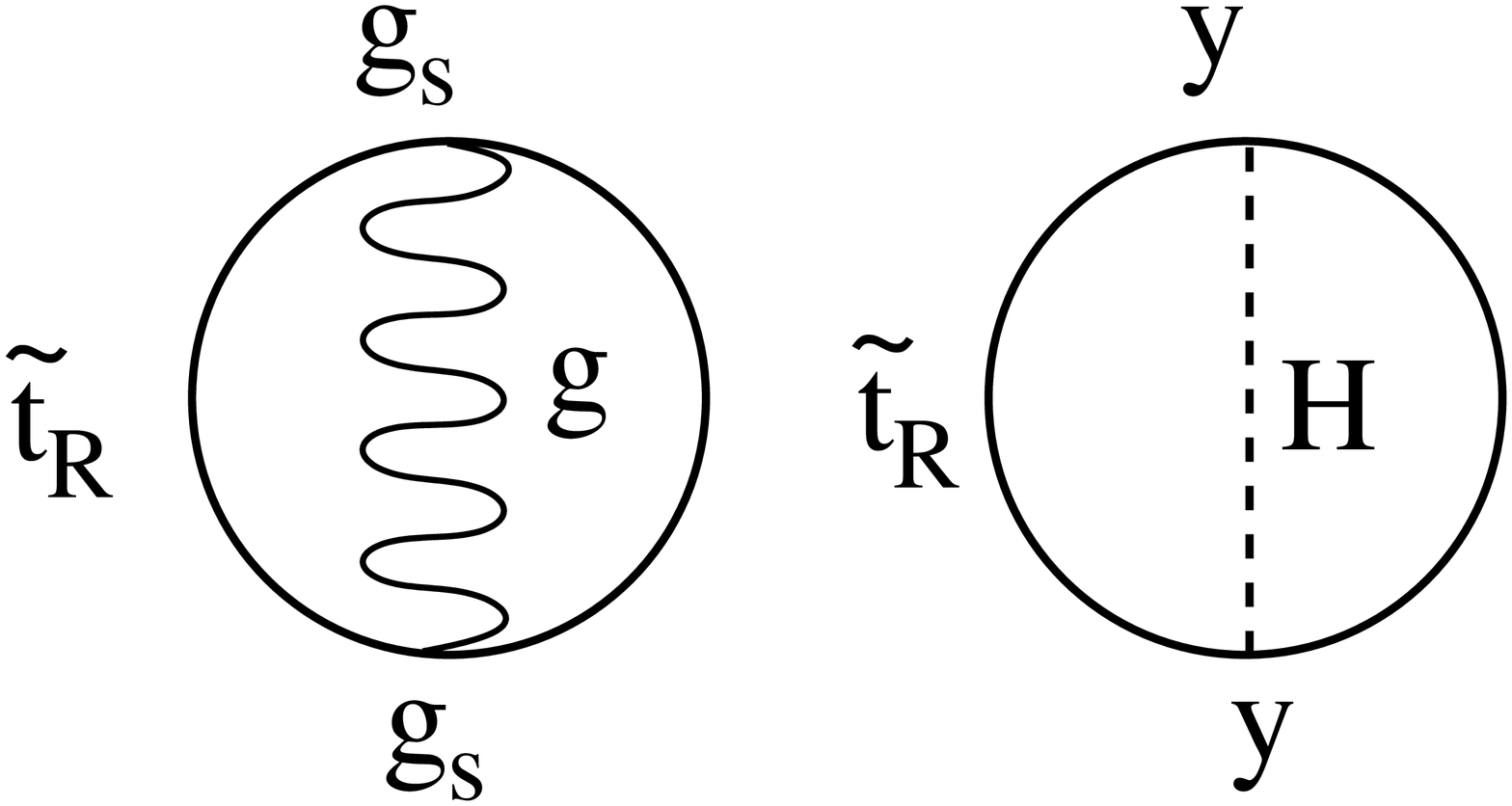}} 
with gluon and Higgs exchange give a contribution of the form
\beq
	\Delta V_{\rm eff} = -c T^2 H^2 \ln{H\over T}
\eeq
which is different from any arising at one loop, and which shift the
value of $v_c/T_c$ by
\beq
	{v_c\over T_c} = {E\over 2\sqrt{2}\lambda} + 
	\sqrt{ {E^2\over 8\lambda^2} + {2c\over \lambda}}
\eeq

Detailed studies of the strength of the phase transition have been
done using dimensional reduction \cite{DR}, 
two-loop perturbation theory (see for example \cite{CM})
and lattice gauge theory \cite{LR} which show that {\it negative}
values of $m^2_U$ (the soft SUSY-breaking mass parameter for $\tilde
t_R$) are required.  This may seem strange; why would we want a cubic term of the
form
\beq
	{\cal L}_{\rm cubic} \sim - {T\over 6\pi}\left(-|m_U^2| 
	+ y^2|H_2|^2\right)^{3/2}\ ?
\eeq
To understand this, we must consider thermal corrections to the {\it
squark} mass as well,
\beq
	m^2_{\tilde t_R} = -|m_U^2| +y^2|H_2|^2 + c_R T^2
\eeq
where $c_R = \frac49 g_s^2 + \frac16 y^2(1+\sin^2\beta)$, which
can be computed similarly to the thermal Higgs mass: one must
calculate
$V_{\rm eff}(H, \tilde t_R)$.  The insertion of 1-loop thermal masses
into the cubic term corresponds to resumming a class of diagrams
called daisies:\\
\centerline{\includegraphics[width=0.7\textwidth]{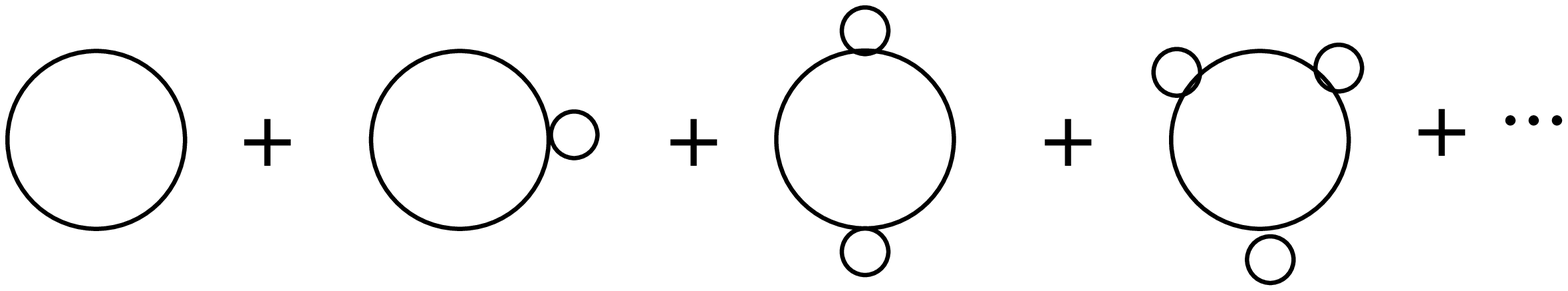}} 
The most prominent effect for strengthening the phase transition
comes when $-|m_U^2| + c_R T^2\cong 0$.  The negative values of
$m_U^2$ correspond to a light stop, $m_{\tilde t_R} < 172$ GeV.  

It is often said that a relatively light Higgs is also needed for a
strong phase transition, but this is just an indirect effect of eq.\
(\ref{hmass}).  If we were willing to make $m_Q$ arbitrarily heavy, we
could push $m_H$ to higher values.  The relation is \cite{CM}
\beq
	m_Q\cong 100 {\rm\ GeV}\, \exp\left[{1\over 9.2}\left({m_H\over{\rm
	GeV}} -  85.9\right)\right]
\eeq
For example, for $m_H = 120$ GeV, $m_Q = 4$ TeV.  This is unnaturally high since
SUSY should be broken near the 100 GeV scale to avoid fine-tuning. 
And it looks strange to have $m_Q \gg |m_U|$.  

The conclusion is that it is possible, but not natural, to get a
strong enough phase transition in the MSSM.  However it is easy in the
NMSSM, which includes an additional singlet Higgs, with superpotential
\beq
	W = \mu H_1 H_2 + \lambda S H_1 H_2 + {k\over 3} S^3 + r S
\label{nmssm}
\eeq
giving a potential
\beq
	V = \sum_i\left|{\partial W\over \partial H_i}\right|^2 + 
\left|{\partial W\over \partial S}\right|^2 + \hbox{\ soft
SUSY-breaking terms}
\label{nmssmpot}
\eeq
Ref.\ \cite{HS} finds large regions of parameter space where the phase
transition is strong enough.

Similar studies have been done for the general 2-Higgs doublet model
(2HDM) \cite{CL,FHS}, with
\beqa
	V &=& -\mu_1^2|H_1|^2 - \mu_2^2|H_2|^2 - (\mu_3^2 H_1^\dagger
	H_2+ {\rm h.c.}) + \frac12\lambda_1|H_1|^4 + 
	 \frac12\lambda_2|H_2|^4  \nonumber\\
	 &+&  \lambda_3|H_1|^2|H_2|^2 + 
	\lambda_4|H_1^\dagger H_2|^2 + 
	\frac12\lambda_5\left[ (H_1^\dagger H_2)^2 +  {\rm h.c.}
	\right]
\label{thdm}
\eeqa
There are several extra scalar degrees of freedom coupling to the
light Higgs in this model, giving a large region of parameter space
where the EWPT is strong.

\begin{figure}[h]
\centerline{
\includegraphics[width=0.75\textwidth]{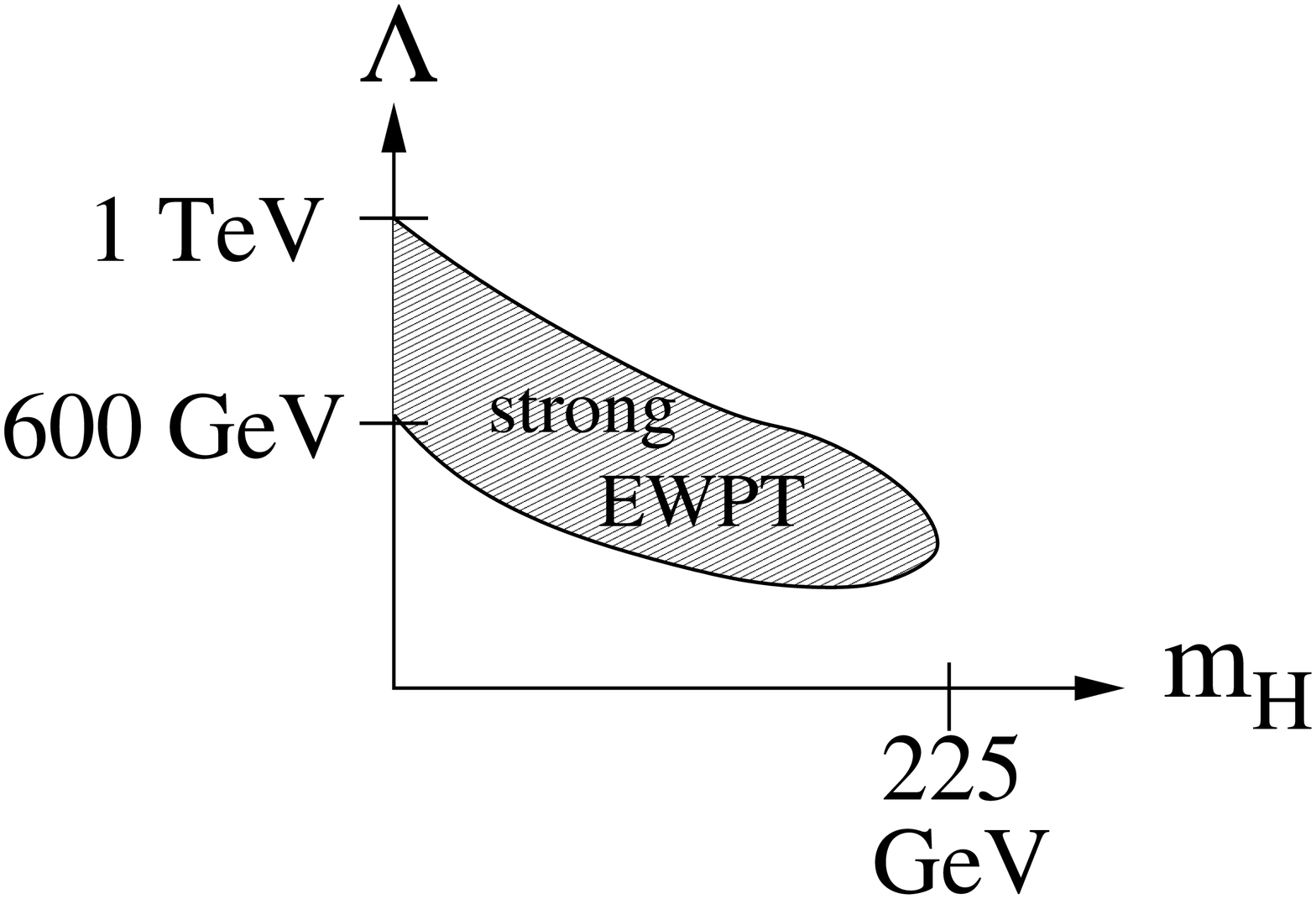}}
\caption{Region of strong EWPT with a $|H|^6/\Lambda^2$ operator.}
\label{fig19}
\end{figure}

A model-independent approach to increasing the strength of the EWPT
has been presented in \cite{GSW}, where the effects of heavy particles
interacting with the Higgs are parametrized by a low-energy effective
potential,
\beq
	V = \lambda\left(|H|^2 - {v^2\over 2}\right)^2 + 
	{1\over \Lambda^2}|H|^6
\eeq 
It is found that $v/T>1$ is satisfied in a sizeable region of
parameter space, as roughly depicted in figure \ref{fig19}. 
This could provide another way of understanding the enhancement of the
strength of the phase transition in models like the NMSSM or
the 2HDM, in cases where the new particles coupling to $H$ are not as
light as one might have thought, based on the idea of increasing the
cubic coupling.

\section{A model of Electroweak Baryogenesis: 2HDM}
Let us now go through a detailed example of how electroweak
baryogenesis could work.  The simplest example is the two Higgs
doublet model, just mentioned in the previous section, where it is
assumed that the top quark couples only to one of the Higgs fields.
This assumption is useful for avoiding unwanted new contributions
to flavor-changing neutral currents (FCNC's). 

\begin{figure}[h]
\centerline{
\includegraphics[width=0.75\textwidth]{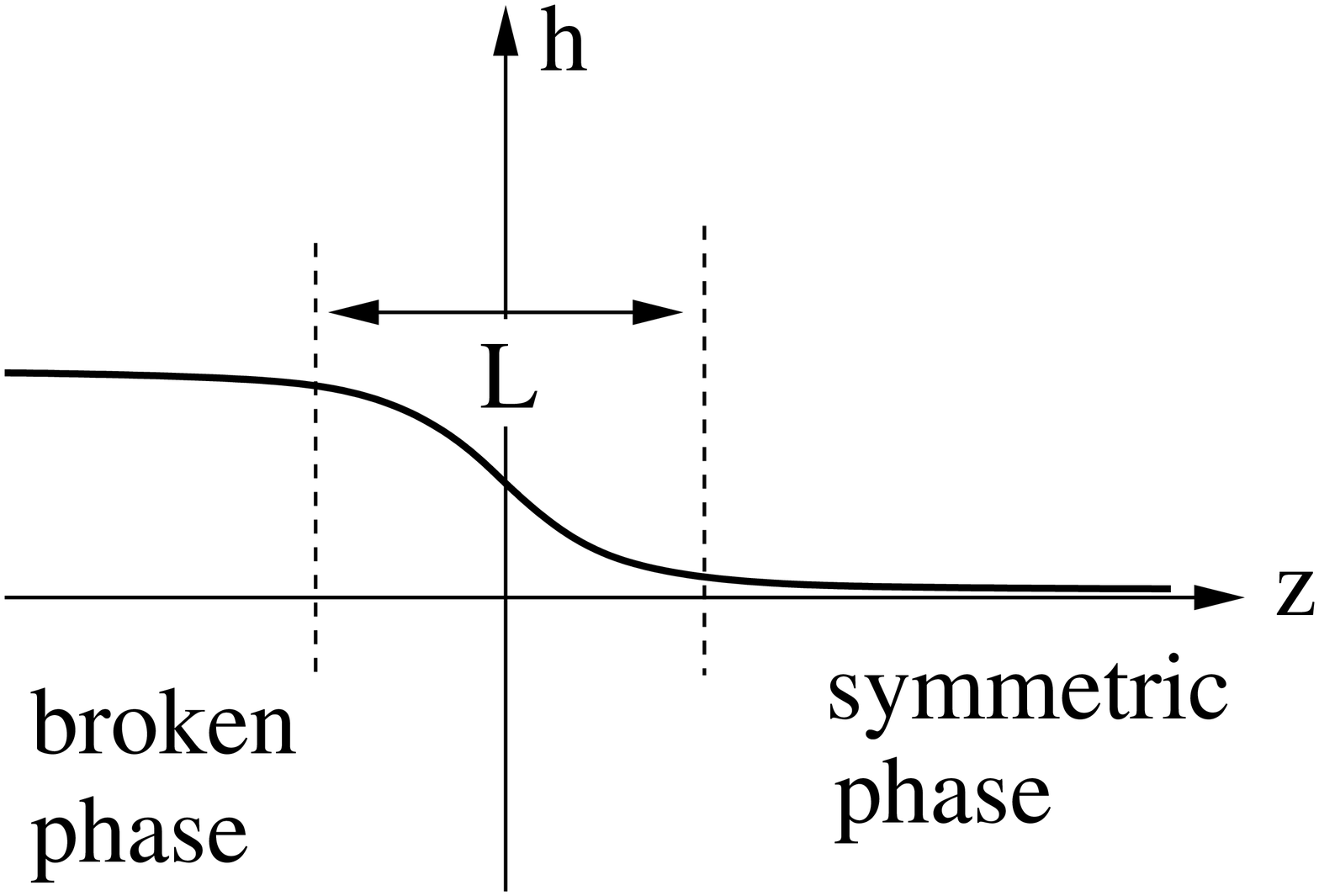}}
\caption{Shape of Higgs field in the bubble wall.}
\label{fig20}
\end{figure}

The basic idea is as follows.  There are phases in the
potential (\ref{thdm}), in the complex parameters $\mu_3^2$ and
$\lambda_5$.  One of these can be removed by a field redefinition,
leaving an invariant phase
\beq
	\phi\equiv {\rm arg}\left(\mu_3^2\over \lambda_5^{1/2}\right)
\eeq
The bubble wall will be described by a Higgs field profile
sketched in figure \ref{fig20},
\beq
	H_i(z) = {1\over\sqrt{2}} h_i(z) e^{i\theta_i(z)}
\eeq
where
\beqa
	h_i(z) &\cong& \left({\cos\beta\atop\sin\beta}\right)
	h(z),\nonumber\\
	h(z) &\cong& {v_c\over 2}\left(1 - \tanh\left(z\over
	L\right)\right)
\label{hsoln}
\eeqa
This shape can be understood by considering the SM, where 
\beqa
	{\cal L} &=& |\partial H|^2 - \lambda\left(|H|^2-{v^2\over
2}\right)^2\nonumber\\
	E &=& \frac12(\partial_z h)^2 + \left\{ 
	{ {\lambda\over 4}(h^2-v^2)^2 \hbox{\ at \ } T=0 \atop
	 {\lambda\over 4}h^2(h-v_c)^2 \hbox{\ at \ } T=T_c}
	\right.
\eeqa
At the critical temperature, the energy is minimized when 
\beq
	\partial_z^2 h = 2\lambda h (h-v_c)(2h-v_c)
\label{heom}
\eeq
Substituting the ansatz (\ref{hsoln}) for $h$ into (\ref{heom}),
one finds that it is a solution provided that
\beq
	L = {1\over\sqrt{\lambda} v_c} \sim {1\over m_h}
\eeq

We can also solve for the phases $\theta_i$.  In fact, only the
difference
\beq
	\theta = -\theta_1 + \theta_2
\eeq
appears in the potential (\ref{thdm}), due to U(1)$_y$ gauge
invariance; it can be rewritten as
\beqa
	V(h_1,h_2,\theta) &=& -\frac12\sum_i \mu_i^2 h_i^2 -
	\mu_3^2\cos(\theta+\phi) h_1 h_2 \nonumber\\
	&=& \frac18 \sum_i\lambda_i h_i^4 + 
	\frac14\left( \lambda_3 + \lambda_4 + \lambda_5\cos 2\theta
	\right) h_1^2 h_2^2
\eeqa
where now all parameters are real, and the invariant CP phase $\phi$
has been shown explicitly.  (Of course, we could move it into the
$\lambda_5$ term if we wanted to, by shifting $\theta\to\theta-\phi$.)
If $h_i = \left(c_\beta\atop s_\beta\right) h$, then
\beqa
	c^2_\beta \partial_z(h^2\partial_z\theta_1) &=& {\partial V\over
	\partial\theta_1}\nonumber\\
	s^2_\beta \partial_z(h^2\partial_z\theta_2) &=& {\partial V\over
	\partial\theta_2} = - {\partial V\over
	\partial\theta_1}
\eeqa
and $c^2_\beta \theta_1 + s^2_\beta \theta_2 = 0$, while
$\partial_z(h^2\partial_z\theta) = (c_\beta^{-2} + s_\beta^{-2})
{\partial V\over\partial\theta}$.  These equations allow us to solve
for $\theta_1$ and $\theta_2$ in the vicinity of the bubble wall.
Notice that the trivial solution $\theta=$ constant is prevented if
$\phi\neq 0$ since 
\beq
	{\partial V\over\partial\theta} = \mu_3^2 h_1 h_2
	\sin(\theta+\phi) - \frac12\lambda_5\sin 2\theta h_1^2 h_2^2
\eeq
vanishes at different values of $\theta$ depending on the value of
$z$.  In fact, as $z\to \infty$, $\theta\to-\phi$, while for
$z\to-\infty$, $\theta$ approaches a different value due to the
nonvanishing of the quartic term in the broken phase.   
 This guarantees that there will be a nontrivial profile for
$\theta$ as well as $h$ in the bubble wall.  
	
The result is a complex phase for the masses of the quarks inside the
wall.  Focusing on the top quark since it couples most strongly to the
Higgs,
this mass term is given by
\beq
	{y\over \sqrt{2}}h_2(z) \bar t e^{i\theta_2\gamma_5} t
	= {y\over \sqrt{2}}h_2(z) \bar t\left( \cos\theta
	+i\gamma_5\sin\theta_2\right) t
\eeq
This means that the propagation of the quark inside the wall will
have CP violating effects.  We can no longer remove $\theta_2$ by
a field redefinition, as we did in section 2, because $\theta_2$ is
a function of $z$, not just a constant.  If we try to remove it
using $t\to e^{-i\theta_2\gamma_5/2}t$, although this removes 
$\theta_2$ from the mass term, it also generates a new interaction
from the kinetic term,
\beq	
	\bar t i\dsl t\to \bar t i\dsl t + \frac12\bar t\dsl\theta_2
	\gamma_5 t
\eeq
Any physics we derive must the be same using either description, but
we will keep the phase in the mass term as this has a clearer
interpretation.  

Now let's consider the Dirac equation, which can be written with the
help of the chiral projection operators $P_{L,R} =
\frac12(1\mp\gamma_5)$ as
\beq
	(i\dsl - mP_L - m^*P_R)\psi = 0
\eeq
using the complex mass 
\beq
	m = {y\over\sqrt{2}} h_2(z) e^{i\theta_2(z)}
\eeq
Decompose the Dirac spinor into its chiral components as
\beq
	\psi = e^{-iEt} \left({R_s\atop L_s}\right) \otimes\chi_s
\eeq
where $\chi_s$ is a 2-component spinor and $s=\pm 1$ labels spin up
and down along the direction of motion, which we can take to be $\hat
z$.  The Dirac equation becomes two coupled equations,
\beqa
	(E-i s\partial_z)L_s &=& m R_s\nonumber\\
	(E+i s\partial_z)R_s &=& m^* L_s
\eeqa
These can be converted to uncoupled second-order equations,
\beqa
	\left[ (E+i s\partial_z){1\over m} (E-i s\partial_z)\right]
	L_s &=& 0 \nonumber\\
	\left[ (E-i s\partial_z){1\over m^*} (E+i s\partial_z)\right]
	R_s &=& 0
\eeqa
We see that the LH and RH components can propagate differently in the 
bubble wall---a signal of CP violation.  We can think of it as
quantum mechanical scattering,\\
\centerline{
\includegraphics[width=0.75\textwidth]{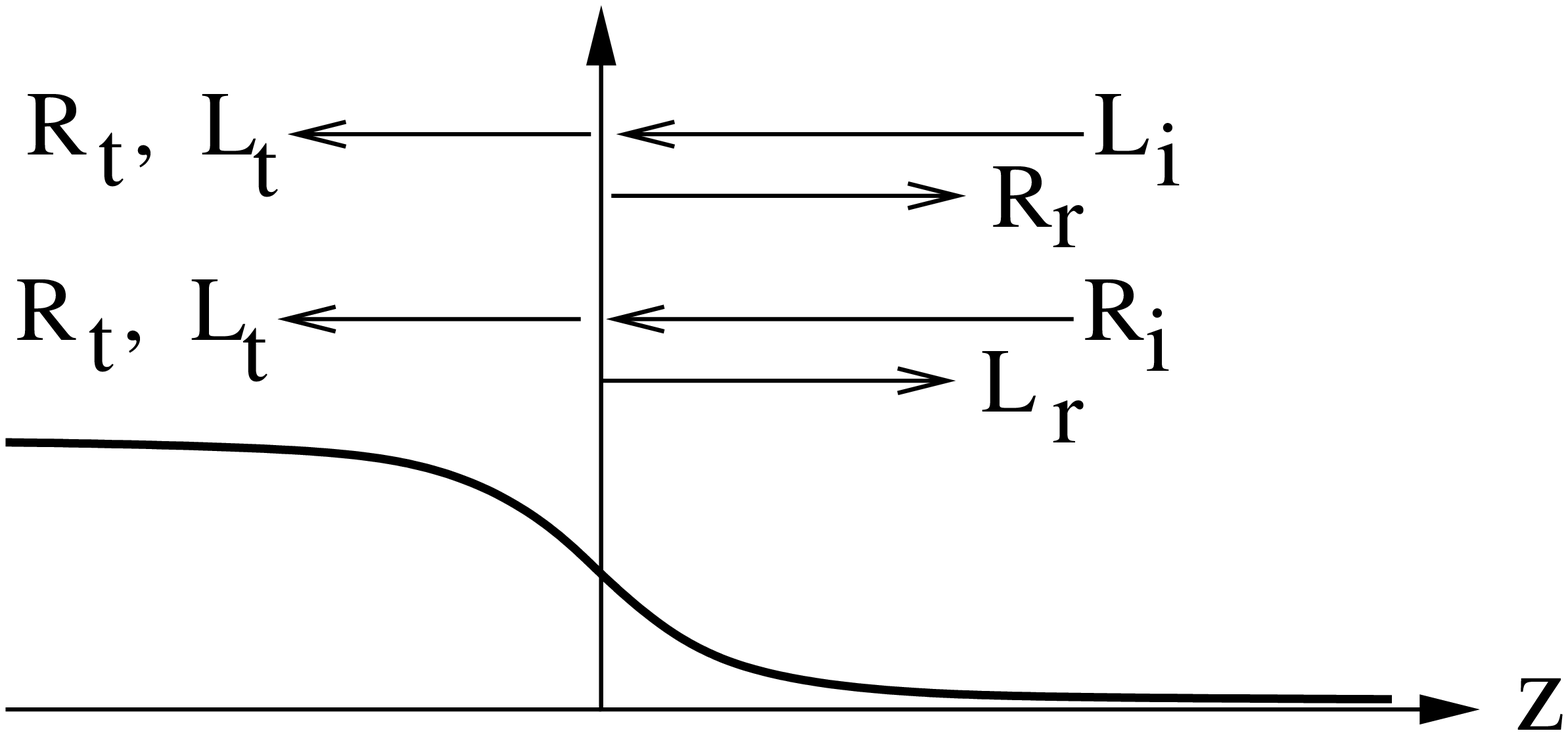}}	
where left-handed components incident from the symmetric phase
reflect into right-handed components and vice-versa.  One can see
that as $z\to\infty$, $m\to 0$, so it is consistent to have RH and LH
components moving in opposite directions.  On the broken-phase side of
the wall, both components must be present because the mass mixes them.

However, quantum mechanical (QM) refelection is only a strong effect if the
potential is sharp compared to the de Broglie wavelength.  In our
case,
\beq
	\lambda \sim {1\over 3T},\quad L\sim {1\over m_h}
\eeq
while for strong scattering we need 
\beq
	{L\over \lambda} = {3T\over m_h} < 1
\eeq
which is not true, except in low-energy part of the particle distribution
functions.  Thus only some low-momentum fraction of the quarks in the
thermal plasma will experience significant QM scattering.  However
there is a classical effect which applies to the dominant momentum components
with $p\sim 3T$: a CP-violating force.  It can be deduced by solving
the Dirac equation in the WKB approximation.  Making the ansatz
\beq
	L_s = A_s(z) e^{i\int^z p(z') dz' - i E t}
\eeq
and substituting it into the equation of motion, we can solve for
$p(z)$ in an expansion in derivatives of the background fields
$|m(z)|$ and $\theta(z)$.  This approach was pioneered in ref.\
\cite{JPT} and refined in \cite{CJK2}.  We find that the canonical
momentum is given by
\beq
	p(z) \cong p_0 + s_c {s E \pm p_0\over 2 p_0}\, \theta'
	+ O\left({d^2\over dz^2}\right)
\label{canp}
\eeq 
where 
\beq
	p_0 = {\rm sign}\, p \sqrt{E^2 - |m^2(z)|}
\eeq
is the kinetic momentum, $s_c = +1$ $(-1)$ for particles
(antiparticles), and the $\pm$ in (\ref{canp}) is $+$ for LH
and $-$ for RH states---but it will be shown that this term has
no physical significance.

The dispersion relation corresponding to (\ref{canp}) is
\beq
	E = \sqrt{(p \pm s_c \theta'/2)^2 + |m|^2} - \frac12 s
	s_c\theta'
\eeq
and the group velocity is
\beq
	v_g = \left({\partial E\over \partial p}\right)_z = 
	{p_0\over E}\left( 1 + s s_c \, {|m|^2\theta'\over 2E^3}
	\right)
\eeq
This is a physically meaningful result: particles with different spin
or $s_c$ are speeded up or slowed down due to the CP-violating effect.
Notice that the effect vanishes if $\theta'=0$ or $m=0$.  Furthermore
the effect of the $\pm \theta'/2$ term in (\ref{canp}) has dropped
out, showing that indeed it is unphysical.  It can be thought of as
a being like a gauge transformation.  One can understand this from
the fact that it has no dependence on $m$.  In the limit that $m\to
0$, the phase $\theta$ has no meaning, so any physical effects of
$\theta$ must be accompanied by powers of $m$.  The same can be said 
concerning derivatives of $\theta$.  We know that if $\theta$ is
constant, it can be removed by a field redefinition, hence any
physical effects can only depend on $\theta'$ or higher derivatives.

We can furthermore identify a CP-violating force which is responsible
for the above effect.  The Hamilton equation which gives the force is
\beq
	\dot p = -\left(\partial E\over \partial z\right)_p 
	= - {(m^2)^2\over 2E} + s s_c {(m^2\theta')'\over 2E^2}
\eeq
The first term on the r.h.s.\ is the CP-conserving force due to the
spatial variation of the mass, while the second term is the
CP-violating force.  This is the main result which allows us to
compute the baryon asymmetry, since $\dot p$ appears in the Boltzmann
equation,
\beq
	{\partial f\over \partial t} + {p\over m}\, {\partial f\over
	\partial z} + \dot p\, {\partial f\over \partial p}
	= \hbox{collision\ terms}
\eeq

To set up the Boltzmann equations we make an ansatz for the
distribution function of particle species $i$, in the rest frame of the bubble wall,
\beq
	f_i = {1\over e^{\beta(\gamma(E_i + v_w p_z) - \mu_i)}
	\pm 1} + \delta f_i
\label{dist}
\eeq
which can be understood as follows.  $\gamma(E_i + v_w p_z)$
is the Lorentz transformation of the energy when we go to the
rest frame of the wall, which is moving at speed $v_w$, and 
$\mu_i$ is the chemical potential.  The first term in (\ref{dist})
tells us about the departure from chemical equilibrium induced by the
CP violating force, while the term $\delta f_i$ is included to
parametrize the departure from kinetic equilibrium.  We are free to
impose the constraint
\beq
	\int d^{\,3}x\, \delta f_i = 0
\eeq
since $\mu_i$ already accounts for the change in the overall
normalization of $f_i$.  

\begin{figure}[h]
\centerline{
\includegraphics[width=0.75\textwidth]{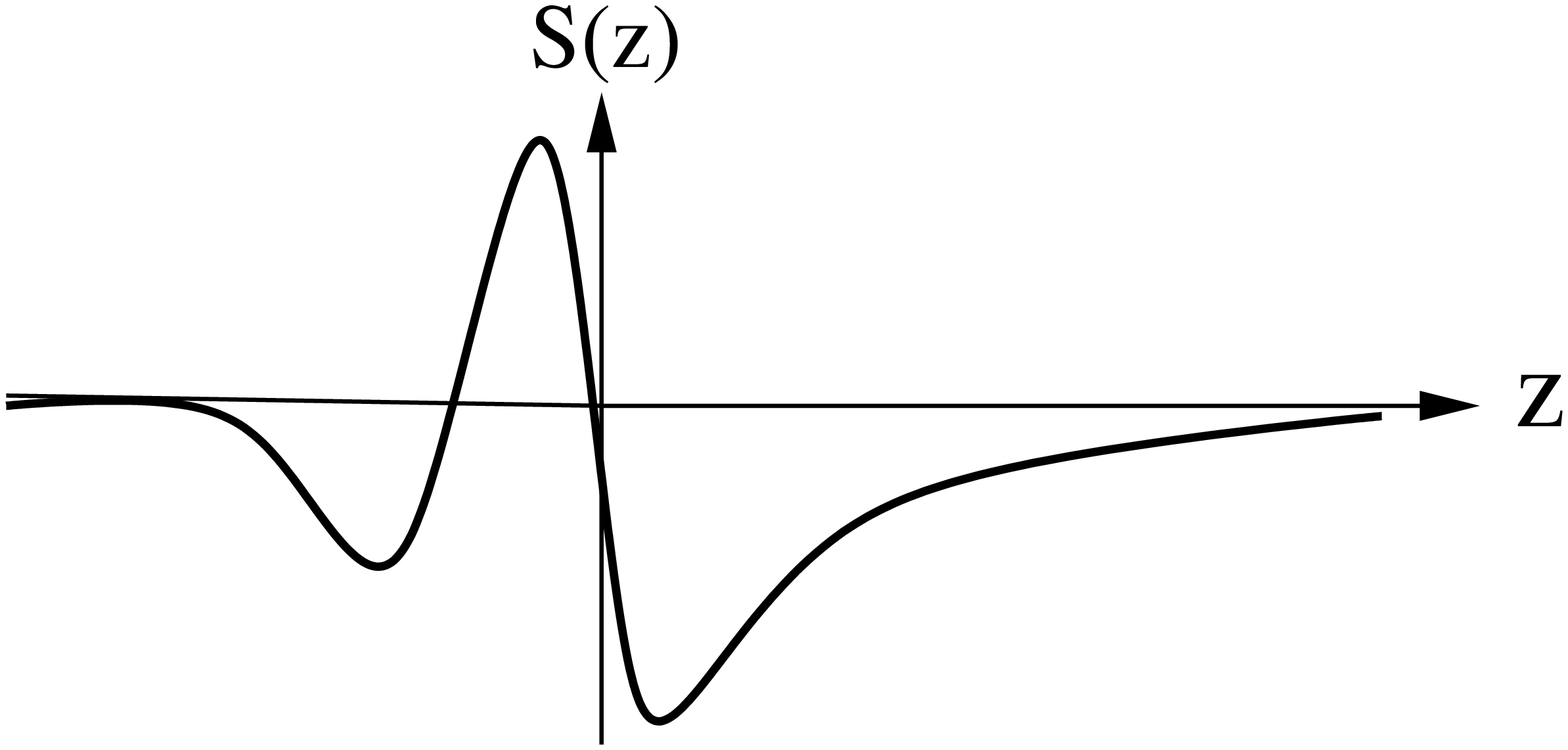}}
\caption{Typical shape of source term in diffusion equation.}
\label{fig21}
\end{figure}

The idea is to make a perturbative expansion where
\beq
	\mu_i \sim \delta f_i \sim m^2\theta'
\eeq
We will thus linearize the Boltzmann equations in $\mu_i$ and $\delta
f_i$, and take the differences between particles and antiparticles.
The next step is to make a truncation of the full Boltzmann equations
(BE) by taking the first two moments,
\beq
	\int d^{\,3}p\, (BE)\quad\hbox{and}\quad \int d^{\,3}p \,
	p_z \, (BE)
\eeq
The reason for taking two moments is that we have two functions
to determine, $\mu(t,z)$ and $\delta f(t,z)$; a more accurate
treatment would involve more parameters in the ansatz for the
distribution functions and taking higher moments of the BE.  
In any case, the truncation is necessary because the full
BE is an integro-differential equation, which we do not know how to
solve, whereas the moments are conventional first-order PDE's.  
Moreover in the rest frame of the wall, the equations no longer depend on
time and therefore become ODE's.  We can solve for $\delta f$ in terms
of $\mu$ and further convert the coupled first-order ODE's into
uncoupled second-order ODE's for the chemical potentials.  These
are known as diffusion equations, and they have the form
\beq
	-(D_i\xi_i'' + v_w \xi_i') + 
	\sum_k \Gamma_k(\pm \xi_i \pm \xi_j \pm \dots) +\dots = S_i
\eeq
where $D_i$ is the diffusion constant, defined in terms of the total
scattering rate $\Gamma_i$ for particle $i$ (in the thermal plasma) as
\beq
	D_i = {\langle v_z^2\rangle\over \Gamma_i}
\eeq
with the thermal average $\langle\cdot\rangle$ applied to the
$z$-component of the particle's squared velocity; $v_w$ is the wall
velocity, usually in the range $0.1-0.01$ depending on details of
the model; $\xi_i$ is the rescaled chemical potential
\beq
	\xi_i = {\mu_i(z)\over T}
\eeq
related to the particle densities by
\beq
	n_i - n_{\bar i} = {T^3\over 6}\, \xi\left\{ {1,\hbox{\ fermions}
	\atop 2,\hbox{\ bosons}} \right.
\eeq
and $\Gamma_k$ is the $k$th reaction rate for particle
in which the particle $i$ is either 
produced or consumed---for example, if $i+j\to
l+m$, then we would write $\Gamma_k(\xi_l + \xi_m - \xi_i - \xi_j)$;
and $S_i$ is the source term due to the CP violating force, which
turns out to be
\beq
	S_i \cong {v_w D_i\over \langle v^2_z\rangle T}
	\left\langle v_z\, {(m^2\theta')'\over 2E^2}\right\rangle'
\eeq
The source term has an asymmetric shape similar to figure
\ref{fig21}, when $m(z)$ and $\theta(z)$ have kink-like profiles.

\begin{figure}[h]
\centerline{
\includegraphics[width=0.75\textwidth]{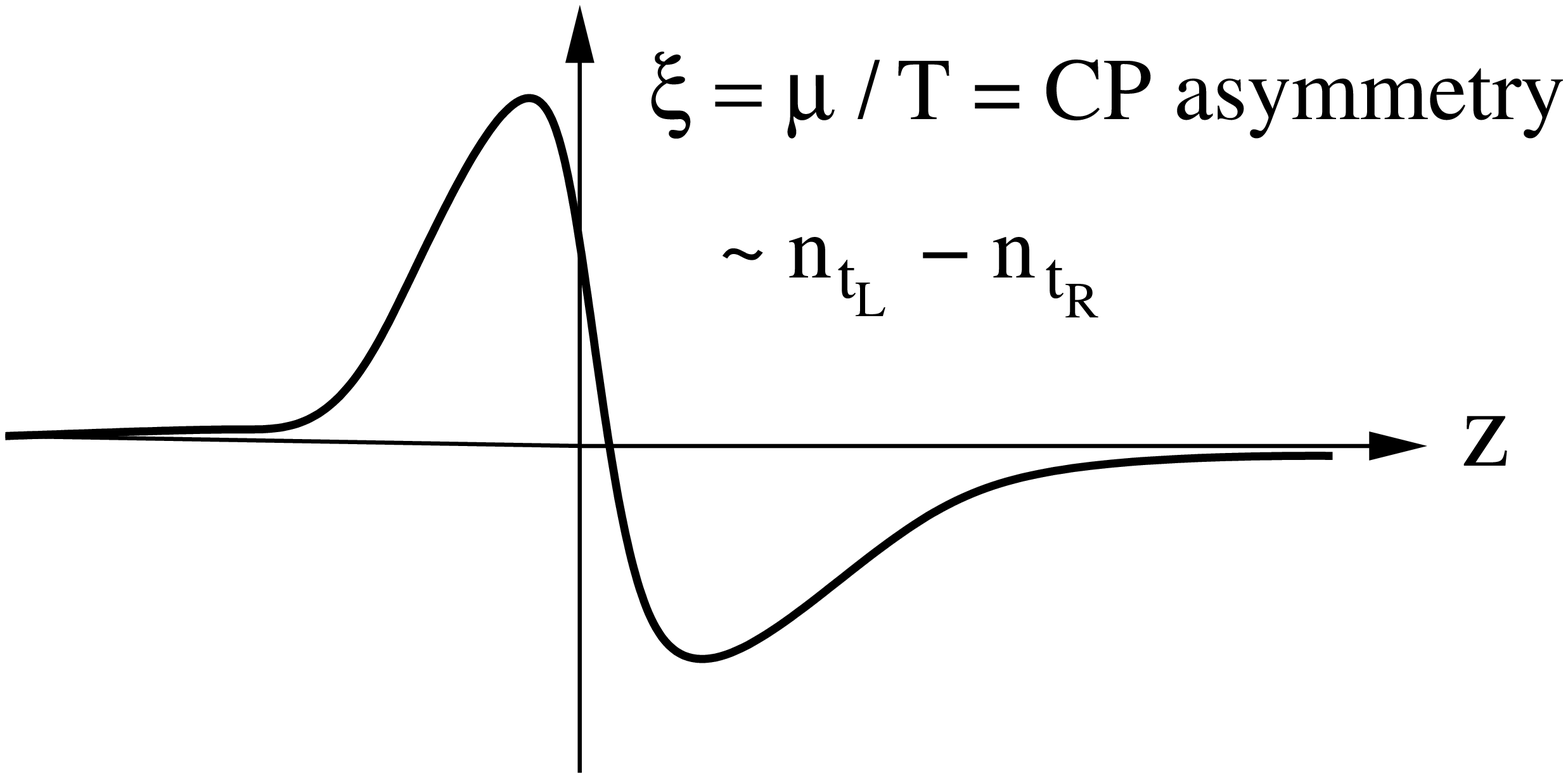}}
\caption{Typical shape of the CP asymmetry from solving the
diffusion equations.}
\label{fig22}
\end{figure}

We can solve the diffusion equation using Green's functions,
\beq
	\left(D {d^2\over dz^2} + v_w {d\over dz} + \Gamma\right)
	G(z-z_0) = \delta(z-z_0)
\eeq
where
\beqa
	G(x) &=& {D^{-1}\over k_+-k_-}\left\{ {e^{-k_+x},\ x>0,\atop
	e^{-k_-x},\ x<0} \right.\nonumber\\
	k_\pm &=& {v_w\over 2D}\left( 1 \pm \sqrt{1 + 
	{2\Gamma D\over 3
	v^2_w}}\right)
\eeqa	
so
\beq
	\xi(z) = \int_{\-\infty}^\infty dz_0\, G(z-z_0) S(z_0)
\eeq
This tells us the asymmetry in one helicity of $t$ quarks, say $(+)$,
which must be equal and opposite to that for the other helicity $(-)$. So far
we have calculated the CP asymmetry near the wall.  It has a shape
similar to that shown in figure \ref{fig22}. 

The final step is to compute the baryon asymmetry.  Since the $q_L$
asymmetry biases sphalerons, we get
\beq
	{d n_B\over dt} \sim 3\Gamma_{\rm sph} \xi - c \Gamma_{\rm
	sph}
	\, {n_B\over T^2}
\label{Bprod}
\eeq
where $3\Gamma_{\rm sph}\xi$ is the term which causes the initial
conversion of the CP asymmetry into a baryon asymmetry, and
$- c \Gamma_{\rm  sph}{n_B/ T^2}$ is the washout term, which would
eventually relax $B$ to zero if sphalerons did not go out of
equilibrium inside the bubbles.  In this equation, we should think
of $\Gamma_{\rm  sph}$ as being a function of $t$ and $z$, 
which at any given position $z$ abruptly
decreases at the time $t$ when the wall passes $z$, so the position of
interest goes from being in the symmetric phase to being in the broken
phase.  (The constant $c$ is approximately 10.)   In the approximation
that the sphaleron rate is zero in the broken phase, and ignoring
the washout term, we should integrate (\ref{Bprod}) at a given point in space until
the moment $t_w$ when the wall passes.  Trading the time integral for an integral
over $z$, we get
\beq
	n_B = \int_{-\infty}^{t_w} dt\, {dn_B\over dt} = 
	\int_0^\infty dz\, {1\over \dot z}\, {dn_B\over dt}
	= {3\Gamma_{\rm sph}\over v_w} \int_0^\infty \xi(z)
\eeq
This can be generalized to the case where the washout term is not
ignored, giving an additional factor $e^{-z(c/v_w)(\Gamma_{\rm
sph}/T^2)}$ in the integrand.  The entire analysis for the 2HDM
has been carried out in \cite{FHS}.

The above derivation of the transport equations gives a clear
intuitive picture of the physics, but more rigorous treatments have
been given in \cite{Schmidt}.
	
\subsection{EDM constraints}
An interesting experimental prediction is that CP violation in EWBG
should also lead to an observable EDM for the neutron and perhaps
other particles.  One needs the invariant CP phase $\phi$ to be 
in the range $10^{-2}-1$ for sufficient baryogenesis, which leads to
mixing between CP-even and odd Higgs bosons (scalars and
pseudoscalars).  Weinberg \cite{WeinEDM} notes that if any of the
propagators $\langle H_2^+ H_1^{+*}\rangle$, $\langle H_2^+ H_1^{+*}\rangle$, 
$\langle H_2^+ H_1^{+*}\rangle$, $\langle H_2^+ H_1^{+*}\rangle$,
$\langle H_2^+ H_1^{+*}\rangle$ have imaginary parts then CP is
violated, giving rise to quark EDM's at one loop, through the
diagram\\
\centerline{
\includegraphics[width=0.5\textwidth]{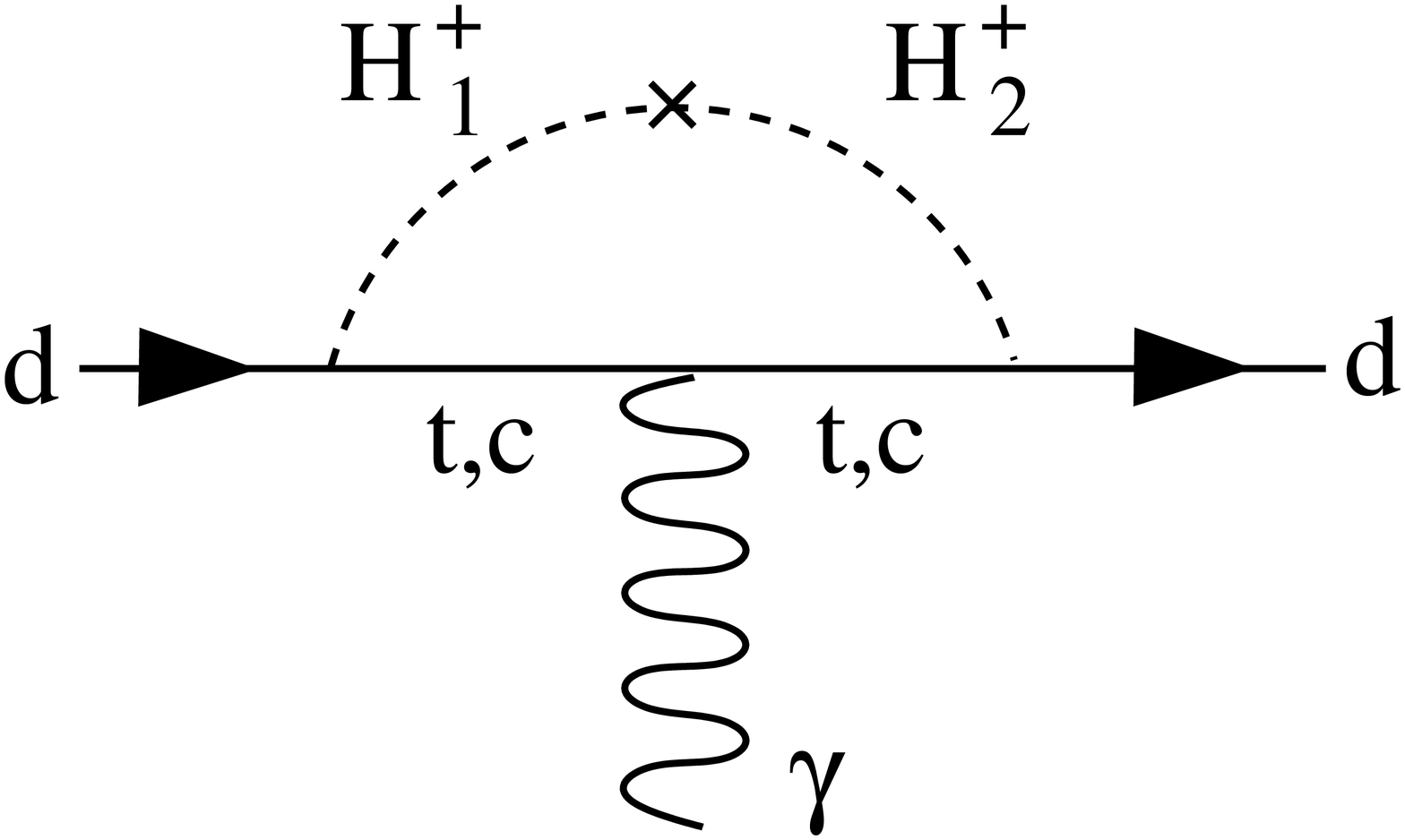}}
The CP-violating propagators occur because the invariant phase $\phi$
appears in the mass matrix of the Higgs bosons.  However this diagram
with charged scalars requires 3 or more Higgs doublets since one of 
the charged fields is a Goldstone boson, and can be set to zero by
going to unitary gauge.  A similar diagram with neutral Higgs exchange
could work, but it is suppressed for the relevant quarks ($u$ and $d$, 
which are valence quarks of the neutron) because the Yukawa couplings
are small.  Moreover we should avoid coupling $H_1$ and $H_2$ to the same
flavor of quarks because this tends to give large FCNC contributions.

\begin{figure}[h]
\centerline{ \includegraphics[width=0.5\textwidth]{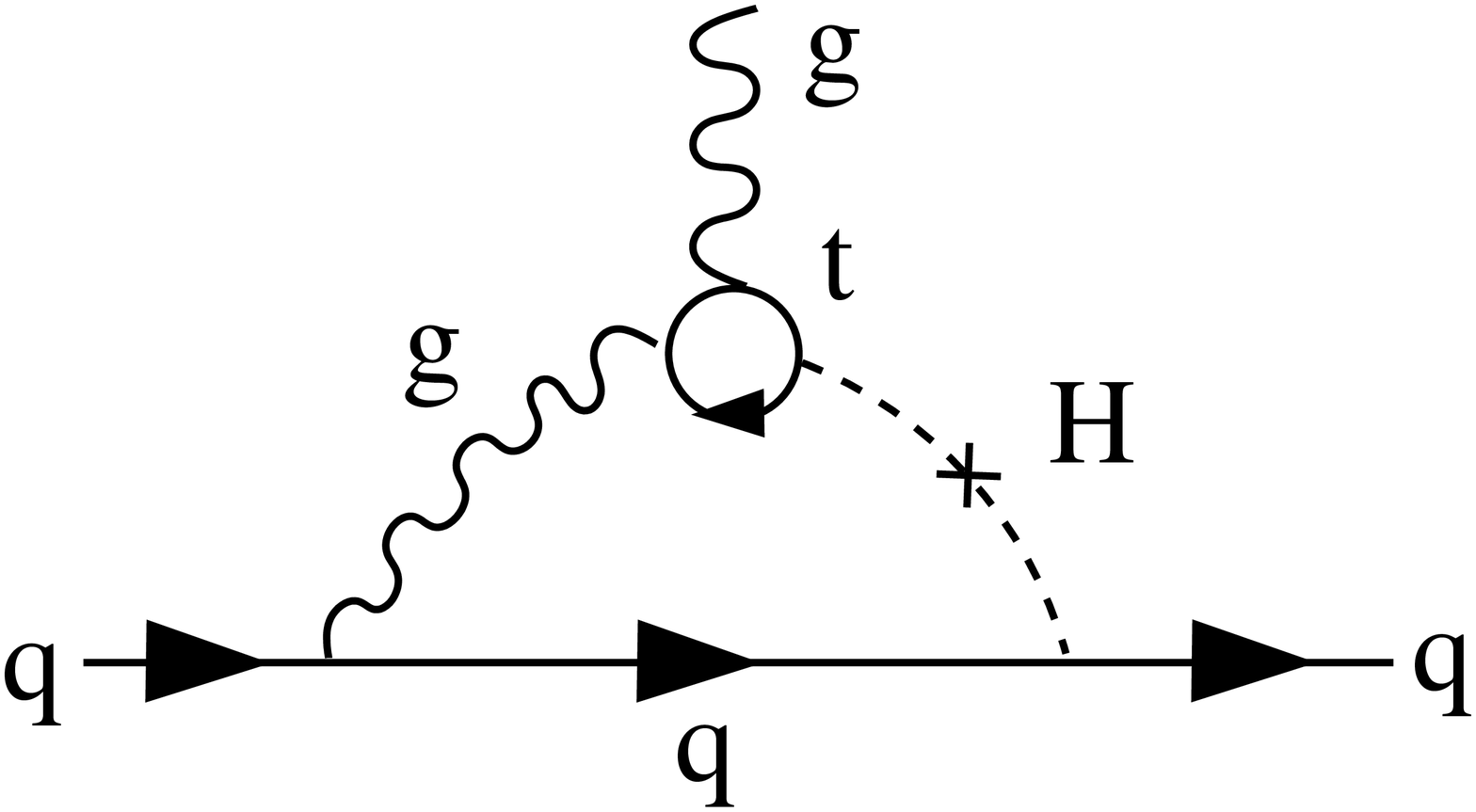}}
\caption{The Barr-Zee \cite{BZ} contribution to the neutron EDM.}
\label{barr-zee}
\end{figure}

It was realized that two-loop diagrams give the largest contribution
(see \cite{GW}, following work of \cite{BZ}), which look like
figure \ref{barr-zee},
where the cross on the Higgs line is the CP-violating mass insertion.
These contribute to the {\it chromo}electric dipole moment of the
quark,
\beq
	{\rm CEDM} = -{i\over 2} g^g_q \bar q\sigma_{\mu\nu} \gamma_5
	G^{\mu\nu} q
\eeq
where $G^{\mu\nu}$ is the gluon field strength.  The quark CEDM's in
turn contribute to the neutron EDM
\beq
	d_n = (1\pm 0.5)(0.55 e\, d^g_u + 1.1 e\, d^g_d) +
	O(d^\gamma_u,\ d^\gamma_d)
\label{nedm}
\eeq
where the factor $(1\pm 0.5)$ is due to hadronic uncertainties.  There
are various ways of estimating the quark contribution to the neutron
EDM; (\ref{nedm}) uses QCD sum rules.  See \cite{PR} for a review.

Ref.\ \cite{FHS} finds that the experimental limit on $d_n$ is close
to being saturated for values of $\phi$ which give enough baryogenesis:
$d_n = (1.2 - 2.2)\times 10^{-26} e\, $cm, which should be compared to
the experimental limit $|d_n| < 3\times 10^{-26} e\, $cm.

There are extra Higgs particles with masses $\sim 300$ GeV
whose strong couplings to the light Higgs give rise to a first order
phase transition, as we have discussed in the previous section.  These
are not CP eigenstates, so they have interactions which could
distinguish them from the Higgs sector of the MSSM.  

\section{EWBG in the MSSM}
We have already discussed how a light $\tilde t_R$ can give a strongly
first order EWPT in the MSSM.  But what is the mechanism for
baryogenesis?  We cannot use the same one as in the 2HDM because in
the MSSM, the Higgs Lagrangian has no $(H_1^\dagger H_2)^2$ coupling
at tree level.  Although such a coupling is generated at one loop, 
its coefficient is too small to get enough baryon production.

The principal mechanism which has been considered is chargino-wino
reflection at the bubble wall.  The first computation of this
effect using the WKB approach, 
 described in the previous section, was done by ref.\ \cite{CJK1}.
(For earlier approaches,  based
on quantum mechanical scattering
or the quantum Boltzmann equation, see \cite{HN} or \cite{CQRVW},
respectively.)  The WKB method has
been carefully scrutinized and verified by \cite{HS} and a number of
other papers \cite{other}.   Charginos/winos have a
$2\times 2$ Dirac mass term
\beq
	\left(\overline{\widetilde W^+_R},\ \overline{\tilde h^+_{1,R}}
	\right)\left(\begin{array}{ll} \tilde m_2 & g H_2(z)\\
	g H_1(z) & \mu\end{array} \right)
	\left({\widetilde W^+_L\atop \tilde h_{2,L}^+}\right)
	+{\rm h.c.}
\eeq
where $\mu$ and $\tilde m_2$ are complex.  Denoting the $2\times 2$
Dirac mass matrix by ${\cal M}$, we diagonalize ${\cal M}$ locally
in the bubble wall using a $z$-dependent similarity transformation
\beq
	U^\dagger {\cal M} V = \left(\begin{array}{ll}
	m_+e^{i\theta_+(z)}  & \\ & m_-e^{i\theta_-(z)}	
	\end{array}\right)
\eeq
so similarly to the top quark in the 2HDM, charginos and winos
experience a CP-violating force in the wall, where
\beq
	m^2_\pm \partial_z\theta_\pm = \pm g^2\, {{\rm Im}(\tilde m_2\mu)
	\over m^2_+ - m^2_-}\, {\partial\over \partial z}
	\left( H_1(z) H_2(z)\right)
\eeq
However the resulting chiral asymmetry in $\tilde h$ or $\widetilde W$
does not have a direct effect on sphalerons.  Instead, we must take
into account decays and scatterings such as\\
\centerline{ \includegraphics[width=0.4\textwidth]{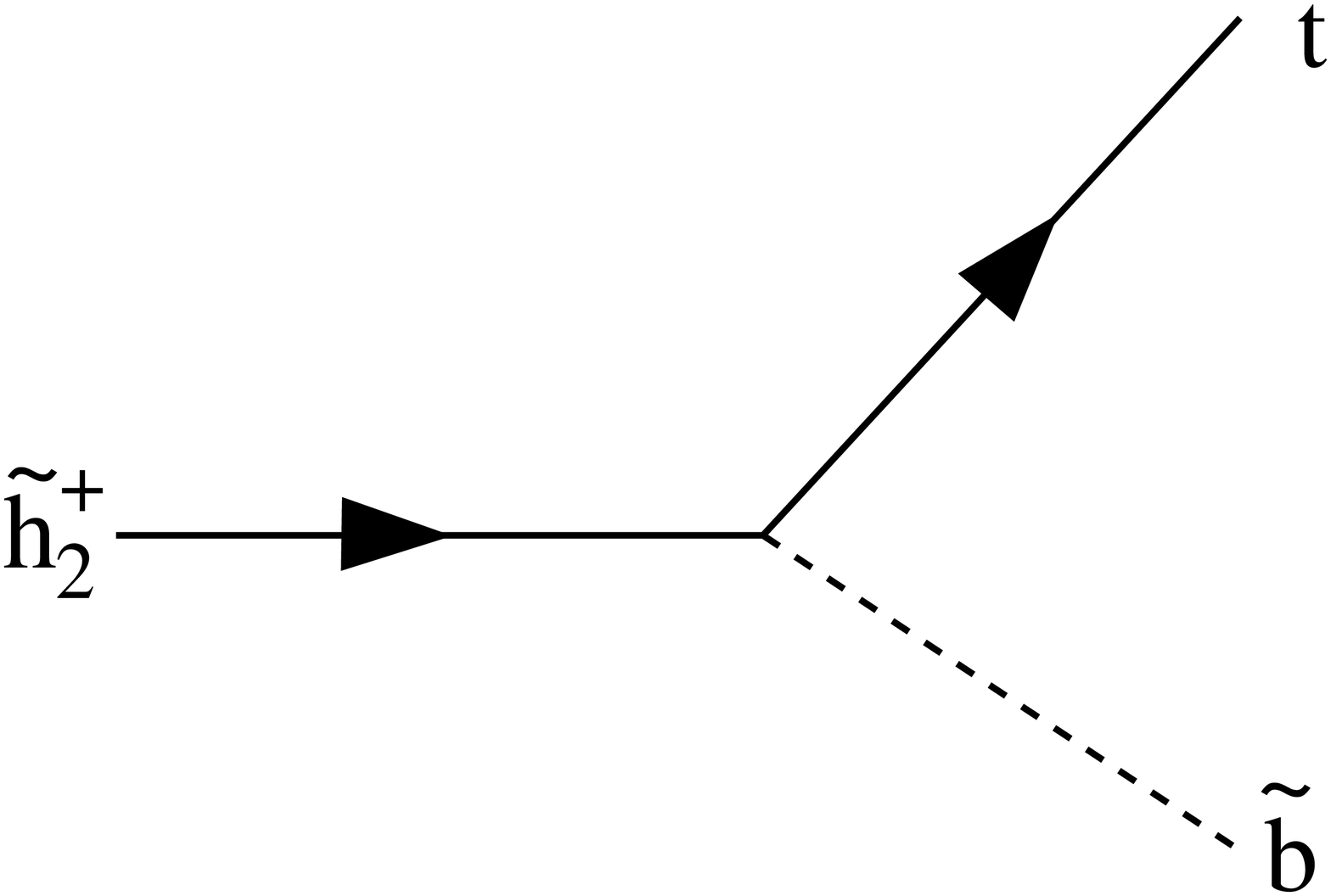}}
which transfer part of the $\tilde h$ CP asymmetry into quarks.
One must solve the coupled Boltzmann equations for all the species
to find the quark CP asymmetry which biases sphalerons.  

\begin{figure}[h]
\centerline{
\includegraphics[width=0.75\textwidth]{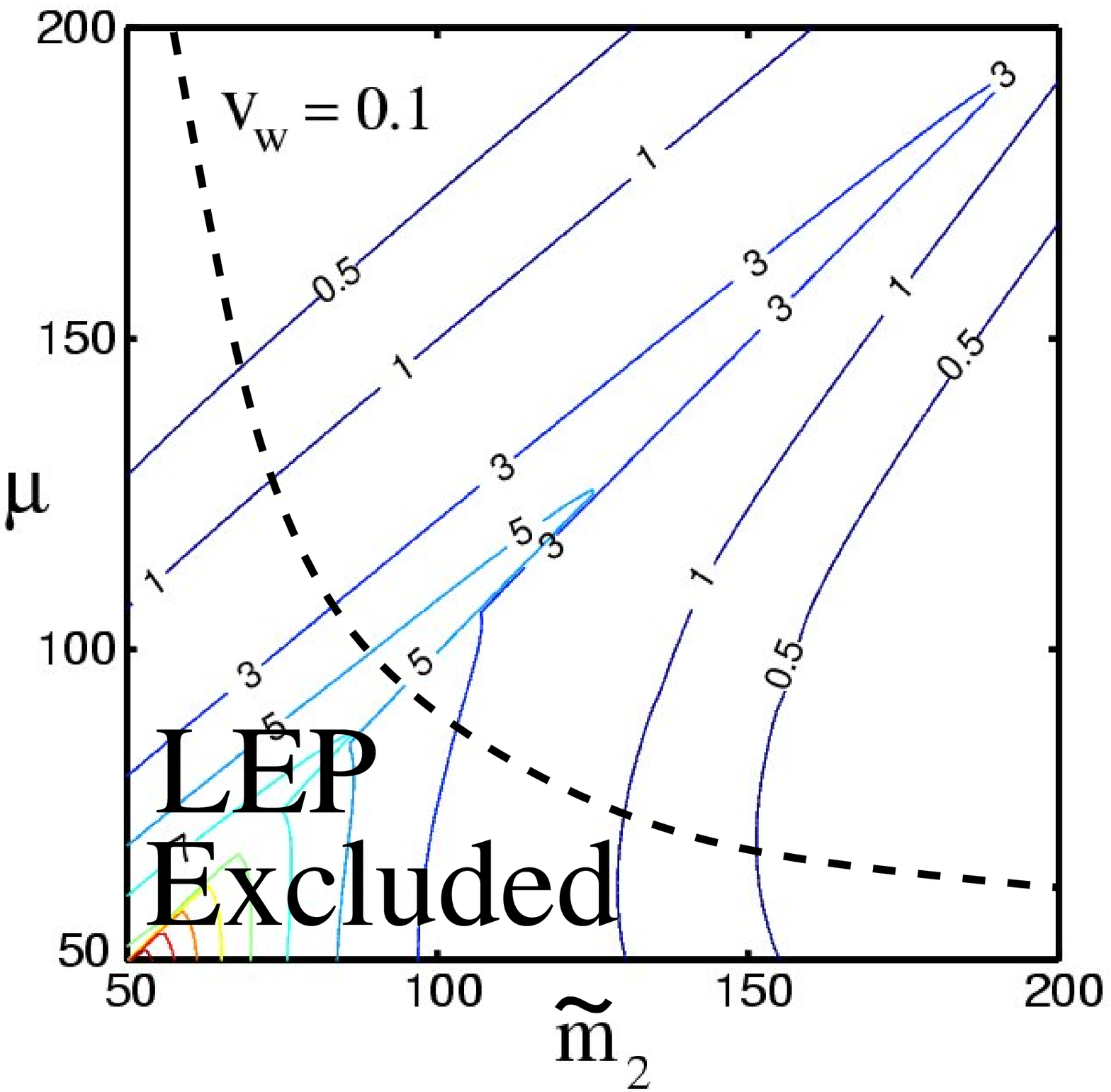}}
\caption{Contours of constant $\eta\times 10^{10}$ from baryogenesis in the MSSM.}
\label{fig23}
\end{figure}

Because of the indirectness of the effect, baryon generation is less
efficient in the MSSM than in the 2HDM.  The effect is largest when
$m^2_+ \cong m^2_-$, which happens when $m_2\cong \mu$, as shown
in figure \ref{fig23}, taken from \cite{CK}.  The figure shows the
contours of $\eta$ in the plane of $\mu$ and $\tilde m_2$, with the
LEP limit on the chargino mass superimposed, $m_{\chi^\pm} >104$
GeV.  This figure was prepared before the WMAP determination of
$\eta$, at which time the preferred value was $3\times 10^{-10}$.
Hence the contours only go up to $5\times 10^{-10}$ in the LEP-allowed
region.  It is possible to choose other values of the MSSM parameters
to slightly increase the baryon asymmetry to the measured value;
for example taking $\tan\beta \lsim 3$ helps, and assuming a wall
thickness (which we have not tried to compute carefully, but is
only parametrized) $L\sim 6/T$ increases $\eta$.  The conclusion is
that all relevant parameters, arg$(\tilde m_2\mu$), $v_w$, $L$,
$\tan\beta$, must be at their optimal values to get a large enough
baryon asymmetry.  The MSSM is nearly ruled out for baryogenesis. 

\begin{figure}[h]
\centerline{
\includegraphics[width=0.75\textwidth]{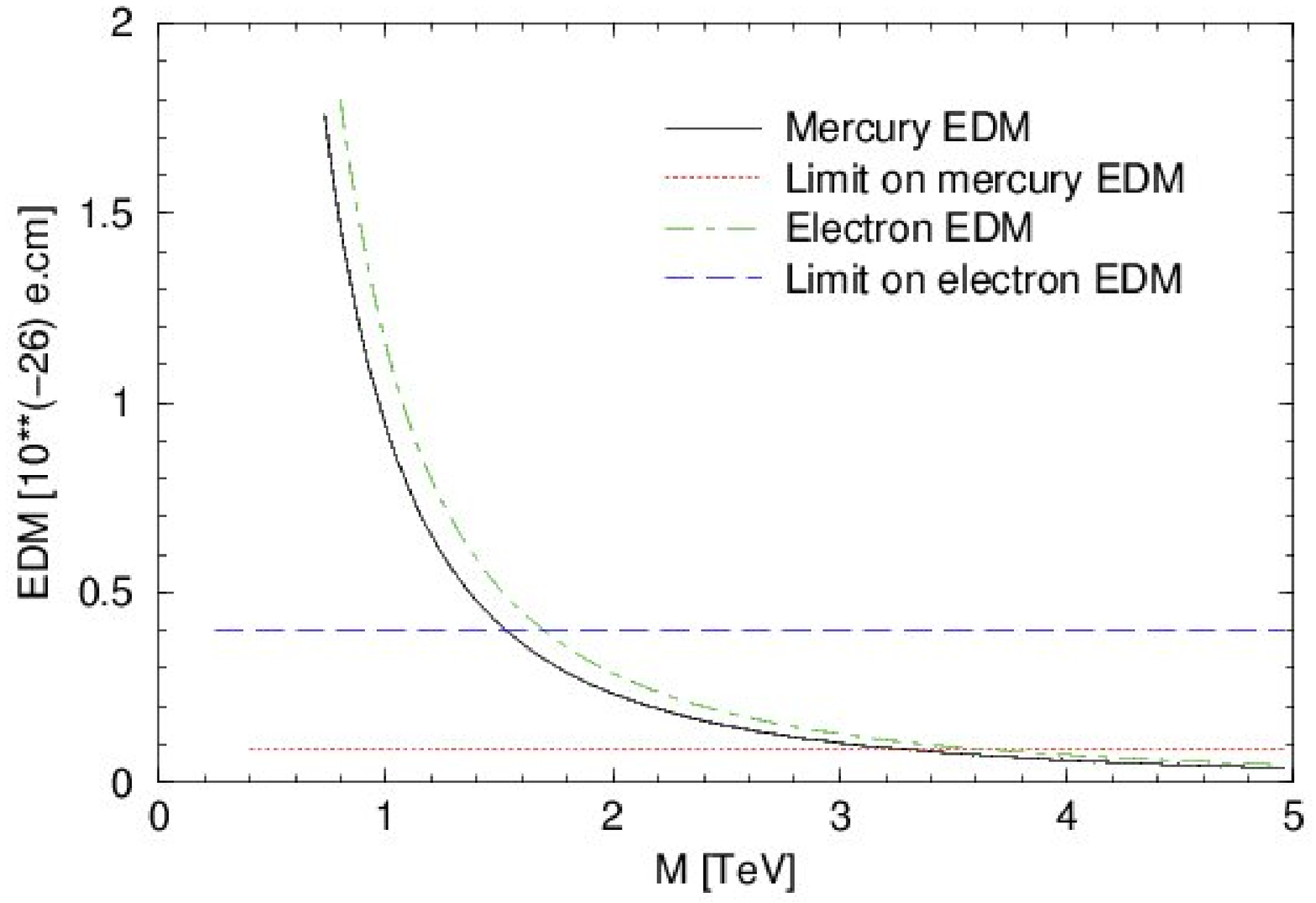}}
\caption{Bound on squark masses from EDM's assuming maximal
phase in the $\mu$ parameter, from \cite{FOPR}.}
\label{fig24}
\end{figure}

In fact some additional tuning is required since the large
CP-violating phases which are needed would lead to large EDM's
through the diagram\\
\centerline{\includegraphics[width=0.5\textwidth]{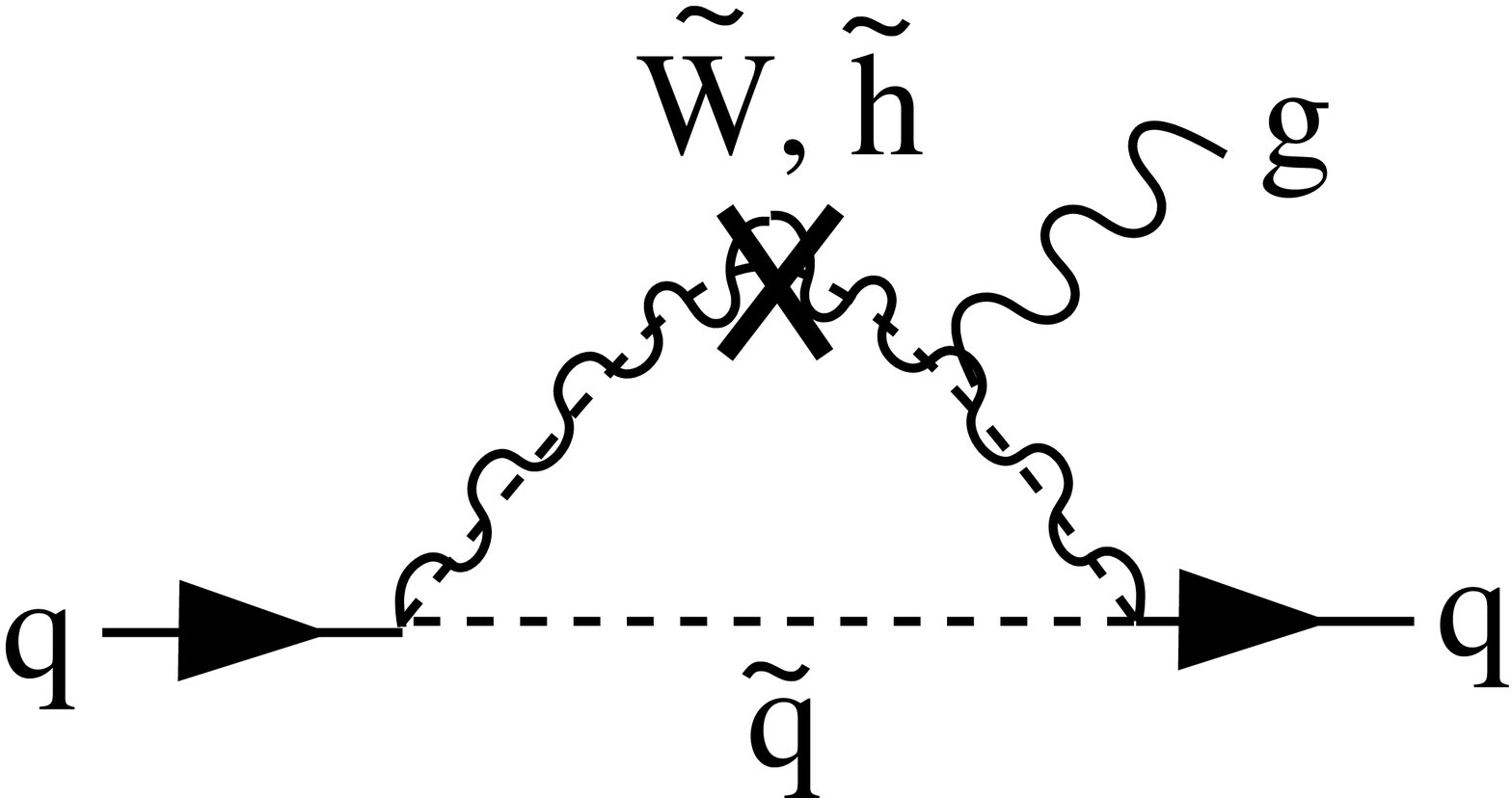}}
If the squark $\tilde q$ is light, the invariant phase (which
we can refer to as the phase of $\mu$) must be small, $\theta_\mu\lsim
10^{-2}$, whereas we need $\theta_\mu\sim 1$.  To suppress the EDM's
it is necessary to take the squarks masses in the range $m_{\tilde q}
\gsim $ TeV.  Roughly, the bound goes like \cite{PR}
\beq
	\theta_\mu \left( 1{\rm \ TeV}\over m_{\tilde q}\right)^2
	< 1
\eeq
In more detail, \cite{FOPR} finds a bound of $m_{\tilde q} > 3$
TeV from the EDM of Hg, if $\theta_\mu$ is maximal, as shown in
figure \ref{fig24}.  
This does not conflict with the need to keep $\tilde b_R$ light
because we only need to make $d^g_u$ and $d^g_d$ (the CEDM's of the
valence quarks of the neutron) small, while the CEDM's of higher
generation quarks might be large.

Assuming heavy first and second generation squarks, the largest
contribution to EDM's can come from two-loop diagrams of the Barr-Zee
type (figure \ref{barr-zee}), where the top loop is replaced by 
a chargino loop \cite{pilafts}.  This contribution does not decouple
when the squarks are heavy, and it gives stringent constraints on 
EWBG in the MSSM through the electron EDM.  A heavy charged Higgs is
needed to suppress this diagram enough to allow for maximal CP
violation in the $\mu$ parameter.

However we can loosen the constraints on the MSSM by adding a singlet
Higgs field $S$ with the superpotential (\ref{nmssm}) and potential
(\ref{nmssmpot}), where the soft SUSY-breaking terms include 
CP violating terms of the form 
$\lambda A S H_1 H_2$ and $k A S^3$.  This model has much more 
flexibility to get a strong EWPT and large CP violation \cite{HS}.  
We no longer need 1-loop effects to generate cubic terms in the Higgs
potential; already at tree level there are cubic terms $H_1 H_2 S$
and $S^3$.  If $S$ participates in the EWPT by getting a VEV, these
cubic terms help to strengthen the phase transition, and it is
possible to have $v_c/T_c > 1$ even with heavy Higgses.

The NMSSM also provides many new CP-violating phases in the Higgs
potential due to complex couplings $k$, $\mu$ and $A_k$, and the model
works similarly to the 2HDM, generating a CP asymmetry directly in the
top quark.

\section{Other mechanisms; Leptogenesis}
Unfortunately there was not enough time to cover other interesting
ideas for baryogenesis, including the very elegant idea of Affleck
and Dine which makes use of flat directions in supersymmetric models
to generate baryon number very efficiently \cite{AD}.

A very popular mechanism for baryogenesis today is leptogenesis, 
invented by Fukugita and Yanagida in 1986 \cite{FY}; see \cite{BDP}
for a recent review.  I am not able to do justice to the subject here,
but it deserves mention.  Leptogenesis is a very natural mechanism,
which ties in with currently observed properties of neutrinos; hence
its popularity.  Unfortunately, the simplest versions of leptogenesis
occur at untestably high energies, similar to GUT baryogenesis. 
Although it is possible to bring leptogenesis down to the
TeV scale \cite{Pilaft}, it requires a corresponding decrease in Yukawa couplings
which renders the theory still untestable by direct laboratory probes.

In leptogenesis, the first step is to create a lepton asymmetry,
which can be done without violating electric charge neutrality of 
the universe since the asymmetry can initially reside in neutrinos.
We know however that sphalerons violate $B+L$, while conserving
$B-L$.  Therefore, an excess in $L$ will bias sphalerons to produce
baryon number, just as an excess in left-handed $B$ did within
electroweak baryogenesis.  Roughly, we can expect that any initial
asymmetry $L_i$ in $L$ will be converted by sphalerons into a final
asymmetry in $B$ and $L$ given by
\beq
	B_f \sim - \frac12 L_i,\quad L_f \sim \frac12 L_i
\eeq
This estimate is not bad.  A more detailed analysis gives
\cite{HT, KS}
\beq
	B_f = -L_i\left\{ { {28\over 79},\ {\rm SM}\atop
	{8\over 23},\ {\rm MSSM}} \right\} \sim -\frac13 L_i
\eeq
since sphalerons couple to $B_L+L_L$ and we must consider the 
conditions of thermal equilibrium between all species.

Leptogenesis is conceptually very similar to GUT baryogenesis, but it
relies on the decays of heavy (Majorana) right-handed neutrinos,
through the diagrams\\
\centerline{
\includegraphics[width=0.75\textwidth]{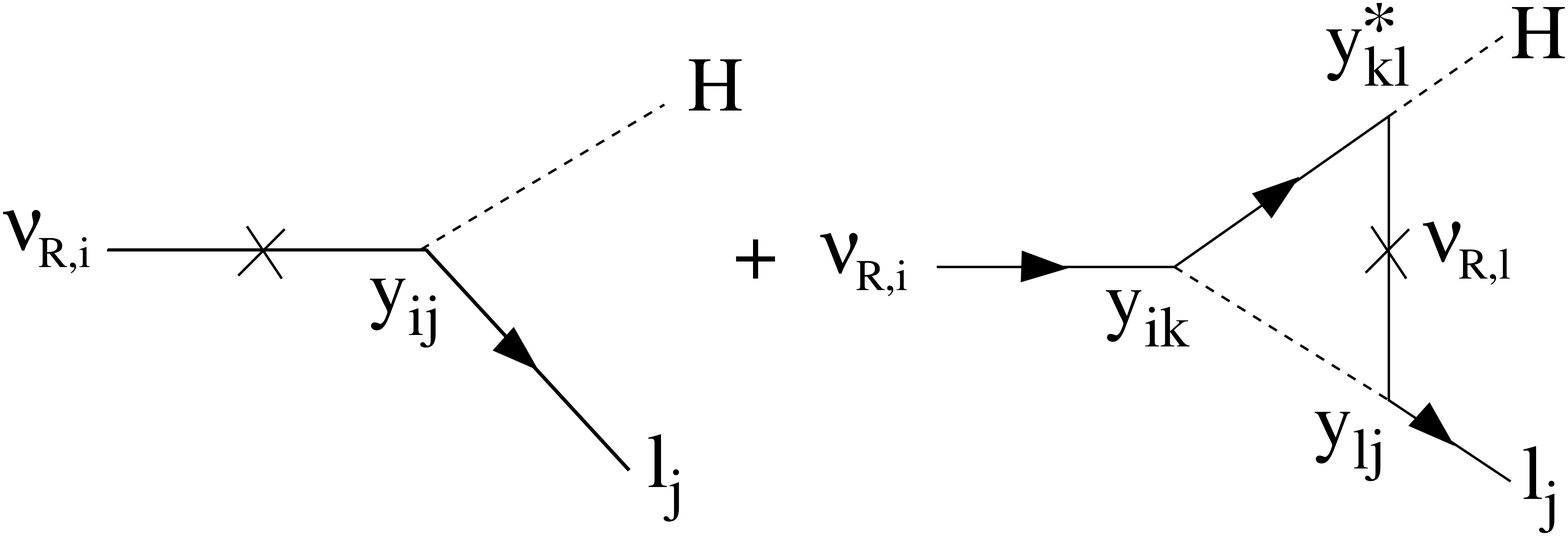}}
The cross represents an insertion of the $L$-violating heavy neutrino
Majorana mass, which makes it impossible to assign a lepton number to
$\nu_R$, and hence makes the decay diagram $L$-violating.
The principles are the same as in GUT baryogenesis.  Physically this
is a highly motivated theory because (1) heavy $\nu_R$ are needed to
explain the observed $\nu_L$ masses, and (2) $\nu_R$'s are predicted
by SO(10) GUT's.  

The first point is well-known; it is the seesaw
mechanism, based on the neutrino mass terms
\beq
	y_{ij}\bar \nu_{R,i} H L_j + {\rm h.c.} -\frac12\left(
	M_{ij} \bar\nu^c_{R,i} \nu_{R,j} + {\rm h.c.}\right)
\label{mm}
\eeq
When the Higgs gets its VEV, the mass matrix in the basis $\nu_L$
and $\nu_R$ is
\beq
	\left(\begin{array}{ll} 0 & m_D\\ m_D^T & M\end{array}\right)	
\eeq
where the Dirac mass matrix is $(m_D)_{ij} = v y_{ij}$ (and we must
take the complex conjugate of (\ref{mm}) to get the mass matrix for the conjugate fields $\nu_R^c$,
$\nu_L^c$).  This can be partially diagonalized to the form
\beq
	\left(\begin{array}{ll} -m_D M^{-1} m_D^T & 0\\ 0 & M\end{array}\right)	
\eeq
so the light neutrino masses are $O(m_D^2/M)\sim y^2 v^2/M$.  More 
exactly, they are the eigenvalues of the matrix $v^2(y M^{-1}
y^T)_{ij}$, up to phases.  We can estimate the size of the $\nu_R$
mass scale:
\beq
	M\sim {y^2 v^2\over m_\nu} \sim y^2\times 10^{14}{\rm\ GeV}
\eeq
where we used the tau neutrino mass, $m_\nu = 0.05$ eV, as measured
through atmospheric neutrino mixing.  Since $y$ can be small, it is
possible to make $M$ smaller than the gravitino bound on the reheat
temperature after inflation.

It is usually assumed that there is a hierarchy such that $M_1\ll
M_{2}, M_3$, since this simplifies the calculations, and it seems like a
natural assumption.  Then $\nu_{R,1}$ is the last heavy neutrino to
decay out of equilibrium, and it requires the lowest reheat
temperature to be originally brought into equilibrium.  

To find the lepton asymmetry, we need the dimensionless measure of CP
violation
\beq
	\epsilon_1 = { \left|{\cal M}_{\nu_{R,1}\to lH}\right|^2
	- \left|{\cal M}_{\nu_{R,1}\to \bar l\bar H}\right|^2
	\over \left| {\cal M}_{\nu_{R,1}\to\hbox{anything}}\right|^2}
	= -{3\over 16\pi}\sum_i {{\rm Im}(y^\dagger y)^2_{i1} 
	\over (y^\dagger y)_{11}} \, {M_1\over M_i} 
\eeq
It can be shown that $\epsilon_1$ has a maximum value of ${3\over
16\pi} {M_1 m_3 \over v^2}$.

The baryon asymmetry can be expressed as
\beq
	\eta = 10^{-2} \epsilon_1 \kappa
\eeq
where $\kappa$ is the efficiency factor, which takes into account the
washout processes\\
\centerline{
\includegraphics[width=\textwidth]{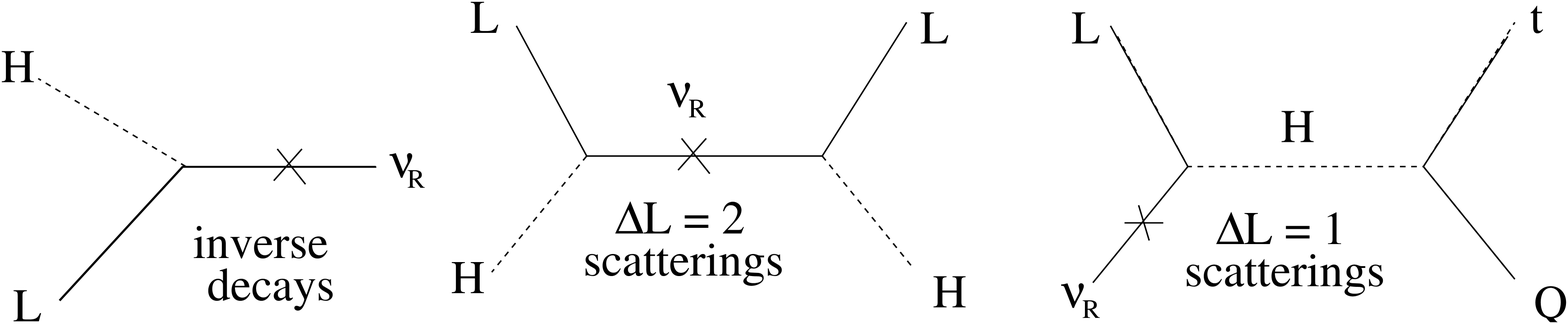}}

It is interesting that leptogenesis can be related to the measured
neutrino masses.  The rate of heavy neutrino decays is given by
\beq
	\Gamma_D = {\tilde m_1 M_1^2\over 8\pi v^2}
\eeq
where
\beq
	\tilde m_1 = { (m_D^\dagger m_D)_{11}\over M_1}
\label{eff}
\eeq
is an ``effective'' neutrino mass; it is not directly related to the
neutrino mass eigenvalues (which requires the transpose of $m_D$
instead of the Hermitian conjugate, and also (\ref{eff}) is
a matrix element rather than an eigenvalue), though it should be of
the same order as the light neutrino masses.  It can be shown that
$m_1 < \tilde m_1 < m_3$, and that the condition for $M_1$ to decay
out of equilbrium is satisfied if
\beq
	\tilde m_1 < m_*\equiv {16\pi^{5/2}\over 2\sqrt{5}}\,
	{v^2\over M_p} = 1.1\times 10^{-3}{\rm eV}
\eeq
The ratio $\tilde m_1/m_*$ enters into the efficiency factor $\kappa$,
and for $\tilde m_1 > m_*$, 
\beq
	\kappa\cong 2\times 10^{-2} \left(0.01 \rm{\ eV}\over \tilde
 	m_1\right)
\eeq
Another interesting connection to the neutrino masses comes through
the $\Delta L=2$ scattering diagram, whose rate does relate to the
actual light masses,
\beq
	\Gamma_{\Delta L =2} = {T^3\over \pi^2 v^4}
	\sum_{i=e,\mu,\tau} m^2_{\nu_i} \equiv  {T^3\over \pi^2 v^4}\,\bar m^2
\eeq
There is a bound on the r.m.s.\ neutrino mass from leptogenesis,
due to this washout process,
\beq
	\tilde m \lsim 0.3\,{\rm eV}
\eeq
which intriguingly is consistent with and close to the bound on the
sum of the neutrino masses from the CMB (see for example \cite{HR}).

It is also possible to derive a lower bound on the lightest
right-handed neutrino mass, $M_1 > 2\times 10^9$ GeV, assuming maximum
value of the efficiency factor \cite{DI}.  This follows from
expressing the maximum value of $\epsilon_1$ in the form
\beq
	\epsilon_1^{\rm max} < {3\over 16\pi}\, {M_1\over v^2}
	\left(\Delta m^2_{\rm atm} + \Delta m^2_{\rm sol}\over 
	m_3\right)
\eeq
and using $\Delta m^2_{\rm atm} = 2.5\times 10^{-3}$ eV$^2$, $m_3\ge
\Delta m^2_{\rm atm}$.

These results which are consistent with the known neutrino masses
(and marginally consistent with the gravitino bound) are suggestive
that leptogenesis could indeed be the right theory.  Unfortunately,
there is no way to find out for sure through independent laboratory
probes of the particle physics, as we hope to do in electroweak
baryogenesis.  On the other hand, it is much easier to quantitatively
predict the baryon asymmetry in leptogenesis than in electroweak
baryogenesis; the Boltzmann equations are much simpler.  One need only
solve two coupled equations, for the heavy decaying neutrino density,
and the produced lepton number,
\beqa
	{dN_{N_1}\over dz} &=& - (D+S) 
	\left(N_{N_1} - N_{N_1}^{\rm eq}\right)\nonumber\\
	{dL\over dz} &=& -\epsilon_1 
	D\left(N_{N_1} - N_{N_1}^{\rm eq}\right) - W L
\eeqa
where $z=M_1/T$, $D$ stands for the rate of decay and inverse decay,
$S$ is the rate of $\Delta=1$ scatterings (which can also produce
part of the lepton asymmetry), and $W$ is the washout terms which 
includes $\Delta L=2$ scatterings; see \cite{Plum}. Care must be
taken when defining the $s$-channel $\Delta L=2$ processes which contribute to
the washout because when the exchanged heavy neutrino goes on shell,
these are the same as inverse decays followed by decays, which have
been separately counted; see \cite{Strumia}.

\bigskip {\bf Acknowledgment}.  I thank the organizers and students
of Les Houches for a very enjoyable and stimulating week.  Thanks
to S.\ Dimopoulos, S.\ Huber, and S.\ Weinberg for useful information
while preparing these lectures, and to R.\ Cyburt for providing figure
1. I am
grateful to B.\ Grzadkowski,
A.\ Pilaftsis, P.\ Mathews, S.\ Menary,  M.\ Schmidt, A.\ Strumia
and H.\ Yamashita
for  pointing out corrections and improvements to the manuscript.  I
also thank K.\ Kainulainen for many years of productive collaboration
on this subject.

\end{document}